\documentclass[usegraphicx]{mn2e}
\usepackage{times}
\usepackage{amssymb}
\usepackage{amsmath}
\begin{document}
%-*-LaTeX-*-
% Copied from gmorris
% Note that some of these call others, eg \kmps uses \km.

% Length
\newcommand{\Mpc}{\rm\thinspace Mpc}
\newcommand{\kpc}{\rm\thinspace kpc}
\newcommand{\pc}{\rm\thinspace pc}
\newcommand{\km}{\rm\thinspace km}
\newcommand{\m}{\rm\thinspace m}
\newcommand{\cm}{\rm\thinspace cm}
\newcommand{\cmps}{\hbox{$\cm\s^{-1}\,$}}
\newcommand{\cmpssq}{\hbox{$\cm\s^{-2}\,$}}
\newcommand{\cmsq}{\hbox{$\cm^2\,$}}
\newcommand{\cmcu}{\hbox{$\cm^3\,$}}
\newcommand{\pcmcu}{\hbox{$\cm^{-3}\,$}}
\newcommand{\pcmcuK}{\hbox{$\cm^{-3}\K\,$}}
% Length

% Time
\newcommand{\yr}{\rm\thinspace yr}
\newcommand{\Gyr}{\rm\thinspace Gyr}
\newcommand{\s}{\rm\thinspace s}
\newcommand{\ks}{\rm\thinspace ks}
% Time

% Frequency
\newcommand{\GHz}{\rm\thinspace GHz}
\newcommand{\MHz}{\rm\thinspace MHz}
\newcommand{\Hz}{\rm\thinspace Hz}
% Frequency

% Temperature
\newcommand{\K}{\rm\thinspace K}
% Temperature

% Pressure
\newcommand{\Kpcmc}{\hbox{$\K\cm^{-3}\,$}}
% Pressure

% Mass
\newcommand{\g}{\rm\thinspace g}
\newcommand{\gpcm}{\hbox{$\g\cm^{-3}\,$}}
\newcommand{\gpcmps}{\hbox{$\g\cm^{-3}\s^{-1}\,$}}
\newcommand{\gps}{\hbox{$\g\s^{-1}\,$}}
\newcommand{\Msun}{\hbox{$\rm\thinspace M_{\odot}$}}
\newcommand{\Msunpc}{\hbox{$\Msun\pc^{-3}\,$}}
\newcommand{\Msunpkpc}{\hbox{$\Msun\kpc^{-1}\,$}}
\newcommand{\Msunppc}{\hbox{$\Msun\pc^{-3}\,$}}
\newcommand{\Msunppcpyr}{\hbox{$\Msun\pc^{-3}\yr^{-1}\,$}}
\newcommand{\Msunpyr}{\hbox{$\Msun\yr^{-1}\,$}}
% Mass

% Energy
\newcommand{\MeV}{\rm\thinspace MeV}
\newcommand{\keV}{\rm\thinspace keV}
\newcommand{\eV}{\rm\thinspace eV}
\newcommand{\erg}{\rm\thinspace erg}
\newcommand{\Jy}{\rm Jy}
\newcommand{\ergpcmc}{\hbox{$\erg\cm^{-3}\,$}}
\newcommand{\ergcmcups}{\hbox{$\erg\cm^3\ps\,$}}
\newcommand{\ergpcmps}{\hbox{$\erg\cm^{-3}\s^{-1}\,$}}
\newcommand{\ergpcmsqps}{\hbox{$\erg\cm^{-2}\s^{-1}\,$}}
\newcommand{\ergpcmsqpspA}{\hbox{$\erg\cm^{-2}\s^{-1}$\AA$^{-1}\,$}}
\newcommand{\ergpcmsqpspsr}{\hbox{$\erg\cm^{-2}\s^{-1}\sr^{-1}\,$}}
\newcommand{\ergpcmcups}{\hbox{$\erg\cm^{-3}\s^{-1}\,$}}
\newcommand{\ergps}{\hbox{$\erg\s^{-1}\,$}}
\newcommand{\ergpspmp}{\hbox{$\erg\s^{-1}\Mpc^{-3}\,$}}
\newcommand{\keVpcmsqpspsr}{\hbox{$\keV\cm^{-2}\s^{-1}\sr^{-1}\,$}}
% Energy

% Force
\newcommand{\dyn}{\rm\thinspace dyn}
\newcommand{\dynpcmsq}{\hbox{$\dyn\cm^{-2}\,$}}
% Force

% Speed
\newcommand{\kmps}{\hbox{$\km\s^{-1}\,$}}
\newcommand{\kmpspmp}{\hbox{$\km\s^{-1}\Mpc{-1}\,$}}
\newcommand{\kmpspMpc}{\hbox{$\kmps\Mpc^{-1}$}}
% Speed

% Luminosity
\newcommand{\Lsun}{\hbox{$\rm\thinspace L_{\odot}$}}
\newcommand{\Lsunppc}{\hbox{$\Lsun\pc^{-3}\,$}}
% Luminosity

% Misc
\newcommand{\Zsun}{\hbox{$\rm\thinspace Z_{\odot}$}}
\newcommand{\gauss}{\rm\thinspace gauss}
\newcommand{\arcsecond}{\rm\thinspace arcsec}
\newcommand{\chisq}{\hbox{$\chi^2$}}
\newcommand{\delchi}{\hbox{$\Delta\chi$}}
\newcommand{\ph}{\rm\thinspace ph}
\newcommand{\sr}{\rm\thinspace sr}
% Misc

% Per something
\newcommand{\pccm}{\hbox{$\cm^{-3}\,$}}
\newcommand{\psqcm}{\hbox{$\cm^{-2}\,$}}
\newcommand{\pcmsq}{\hbox{$\cm^{-2}\,$}}
\newcommand{\pmpc}{\hbox{$\Mpc^{-1}\,$}}
\newcommand{\pmpccu}{\hbox{$\Mpc^{-3}\,$}}
\newcommand{\ps}{\hbox{$\s^{-1}\,$}}
\newcommand{\pHz}{\hbox{$\Hz^{-1}\,$}}
\newcommand{\pcmK}{\hbox{$\cm^{-3}\K$}}
\newcommand{\phpcmsqps}{\hbox{$\ph\cm^{-2}\s^{-1}\,$}}
\newcommand{\psr}{\hbox{$\sr^{-1}\,$}}
% Per something
\newcommand{\pspsqas}{\hbox{$\s^{-1}\,\arcsecond^{-2}\,$}}

\newcommand{\ergpspcmpK}{\hbox{$\erg\s^{-1}\cm^{-1}\K^{-1}\,$}}

\title{Thermal conduction and reduced cooling flows in galaxy clusters}
\author[L.M. Voigt \& A.C. Fabian]
{\parbox[]{6.in} {L.M. Voigt and A.C. Fabian\\
\footnotesize
Institute of Astronomy, Madingley Road, Cambridge CB3 0HA\\
}}

\maketitle

\begin{abstract}
  Conduction may play an important role in reducing cooling flows in
  galaxy clusters. We analyse a sample of sixteen objects using
  \emph{Chandra} data and find that a balance between conduction and
  cooling can exist in the hotter clusters ($T \gtrsim$ 5 keV),
  provided the plasma conductivity is close to the unhindered Spitzer
  value. In the absence of any additional heat sources, a reduced mass
  inflow must develop in the cooler objects in the sample. We fit
  cooling flow models to deprojected data and compare the spectral
  mass deposition rates found to the values required to account for
  the excess luminosity, assuming Spitzer-rate heat transfer over the
  observed temperature gradients. The mass inflow rates found are
  lower than is necessary to maintain energy balance in at least five
  clusters.  However, emission from cooling gas may be partially
  absorbed. We also compute the flux supplied by turbulent heat
  transport (Cho et al.\ 2003) and find conductivity profiles which
  follow a strikingly similar temperature dependence to the
  conductivity values required to prevent cooling. Finally, we show
  that the cluster radio luminosities vary by over five orders of
  magnitude in objects with X-ray luminosities differing by no more
  than a factor of a few.  This suggests that there is unlikely to be
  a straightforward correlation between the mechanical power provided
  by the radio lobes and the rate of energy loss in cooling flow
  clusters.

\end{abstract}

\section{Introduction}
High resolution observations of cooling flow clusters with
\emph{XMM-Newton} and \emph{Chandra} show no evidence for the large
amounts of multiphase gas expected in the inner regions of these
objects (Peterson et al.\ 2001; Tamura et al.\ 2001; Kaastra et al.\ 
2001; Peterson 2002). Several solutions have been proposed to account
for the lack of soft X-ray emission; including models which preserve
the classic mass deposition rates by invoking differential absorption,
mixing or inhomogeneous metallicity distributions (Fabian et al.\ 
2001; Morris \& Fabian 2002; Fabian et al.\ 2002a), and those which
prevent, or significantly reduce, mass dropout by balancing the
radiative losses by some heat source (Bertschinger \& Meiksin 1986,
Tucker \& Rosner 1983; Churazov et al.\ 2001; Br\"uggen \& Kaiser
2001).

Here we discuss the role played by conduction in transporting heat
from the hot gas reservoir outside the cooling radius towards the
centre of the cluster. Narayan \& Medvedev (2001), Gruzinov (2002) and
Fabian et al.\ (2002b) have shown that conduction provides heat fluxes
which are close to those required to stem cooling, assuming the plasma
conductivity is within an order of magnitude of the Spitzer value,
$\kappa_{\rm S}$ (Spitzer 1962).  The actual value for the suppression
of cluster conductivity below $\kappa_{\rm S}$ remains an open
question, and depends on detailed understanding of cluster magnetic
fields. Churazov (2001) estimated the suppression factor to be $\sim$
0.01, whereas Narayan \& Medvedev (2001) argue that it could be as
high as 0.3. If the former scenario were true we could rule out
conduction as a successful heating mechanism.  However, recent support
for heat transfer rates close to the Spitzer rate means that
conduction may have a significant effect on cluster evolution.

In order to test the conduction hypothesis further, detailed spatial
analyses of individual clusters have been carried out by Voigt et al.
(2002) and Zakamska \& Narayan (2002).  In the latter study, a simple
model in which conduction balances cooling was used to generate
theoretical temperature and density profiles.  The conductivity values
required to produce profiles which were reasonable fits `by eye' to
the observed profiles were then used to assess whether or not
conduction can be effective.  Zakamska \& Narayan concluded that half
the clusters in their sample could be prevented from cooling if
conduction is operating at a rate between (0.1--0.4) $\kappa_{\rm S}$.
They suggest that the remaining clusters are heated by the central
radio source.

Voigt et al.\ (2002) calculated the conductivity required to replace
heat loss as a function of radius using the observed temperature and
density profiles of Abell 1835 (Schmidt et al. 2001) and Abell 2199
(Johnstone et al. 2002). We found that whilst conduction at the
Spitzer rate was able to prevent cooling in the outer parts of the
cooling flow region, a factor of 2 or more above the Spitzer rate was
required in the very centre (within about 20 kpc). We suggested that
conduction suppresses cooling, and that in some cases this suppression
is complete, and in others partial. In the latter case a reduced
cooling flow develops.  In this paper we extend the work of Voigt et
al.\ (2002) to a sample of sixteen galaxy clusters using archival
\emph{Chandra} data. In clusters where the required conductivity is
above the Spitzer value we fit cooling flow models to the spectra and
compare the mass deposition rates found to those required to maintain
energy balance with conduction at $\kappa_{\rm S}$.

We consider the possibility that heat is transported by turbulent
diffusion in Section 7.

Throughout the paper we assume a cosmology with $H_{0}$=71$\kmpspMpc$,
$\Lambda$=0.73 and $\Omega_{\rm m}$=0.27.

\section{Clusters and data reduction}
The clusters in our sample are listed in Table~\ref{clusters} and
include objects ranging in redshift from 0.0--0.5 and in temperature
from $\sim$ 2--15 keV. The clusters were observed using chip 7 on the
ACIS-S detector on board \emph{Chandra}, allowing spatially-resolved
analysis of the cooling flow region.  In each case the data were
reduced using the \small{CIAO} \normalsize (\emph{Chandra Interactive
  Analysis of Observations}) software package. We used the Markevitch
script (\small{LC\_CLEAN}\normalsize) to remove flares and strong
point sources were identified by eye and subtracted from the regions
files. ACIS `blank-sky' datasets were used to subtract the background.
Ancillary-response and response matrices were produced using the
\small{CIAO MKWARF} \normalsize and \small{CIAO MKRMF} \normalsize
programs and data were binned to have at least 20 counts per
\small{PHA} \normalsize channel.

\begin{table*}
\centering
\begin{tabular}{lcccccccc} 
\hline\hline
Cluster& Redshift & Obs. date& Exps. time & GTI&
X-ray peak (J2000) &  $d_{\rm L}$ &  $d_{\rm A}$ & Refs.\\
& $z$ & &(ks) &(ks) &RA\hspace{0.5cm} Dec  & (Mpc) &(kpc/$''$) & \\
\hline
\\
2A 0335+096 & 0.0347 & 2000 Sep 06 & 20.0 & 18.1 & (03 38 40.5) (+09 58
11.6) &140.6 & 0.7& [1] \\
A478 & 0.0880  & 2001 Jan 27 &42.9 &38.9 & (04 13 25.4) (+10 27 57.1)
&396.9 &1.6 & [2],[3] \\
PKS 0745-191 & 0.1028 & 2001 Jun 16 &17.9 &14.6 &(07 47 31.2) (-19 17
38.8) &468.5&1.9 & [4],[5] \\
Hydra A & 0.0520   &1999 Nov 02 &24.1 &17.3 &(09 18 05.7) (-12 05
43.3)&228.5 &1.0 & [6],[7]\\
M87$^{\dagger}$ & 0.0043  & 2000 Jul 29 & 38.2     & 33.7  & (12 30 49.4) (+12 23
28.0)& 18.2  & 0.1 & [8],[9],[10],[11]\\
RXJ 1347.5-1145& 0.4510& 2000 Apr 29&  10.1 & 7.2& (13 47 30.7) (-11 45
09.5) &2492.9 &5.7& [12],[13] \\
A1795 & 0.0632 &2000 Mar 21  &19.7 &15.6 &(13 48 52.5) (+26 35 37.8)
&280.0&1.2 & [14],[15] \\
A1835 & 0.2523  & 1999 Dec 11 &19.8 &18.9 &(14 01 02.0) (+02 52 39.7)
&1262.4 &3.9 & [16] \\
PKS 1404-267 & 0.0226 & 2001 Jun 07 &7.3 & 6.0&(14 07 29.8) (-27 01
04.2) &97.1 &0.5&  [17]\\
3C 295 &0.4605 &1999 Aug 30& 19.0& 14.3&(14 11 20.5) (+52 12 10.5)
&2556.0&5.8 & [18]\\
A2029 & 0.0767 & 2000 Apr 12 &19.9 & 19.8 &(15 10 56.1) (+05 44 40.6)
&343.2 &1.1&  [19] \\ 
RXJ 1532.9+3021 & 0.3615 & 2001 Aug 26 &9.5 & 6.2 &(15 32 53.8) (+30
21 00.2) &1916.3&5.0& --  \\
Cygnus A & 0.0562 & 2000 May 21 & 35.2 & 33.3 &(19 59 28.3) (+40 44
02.0) &247.7 &1.1 & [20] \\
A2390 & 0.2301 & 2000 Oct 08 &10.0 &7.4 &(21 53 36.8) (+17 41 44.1)
&1136.8& 3.6& [21],[13] \\ 
Sersic 159-03 (AS 1101)& 0.0564  &2001 Aug 13 &10.1 &9.8 &(23 13 58.5) (-42 43
34.5) &248.7 &1.1 & [22]\\
A2597 & 0.0830 &2000 Jul 28 &39.9 &21.3 &(23 25 19.8) (-12 07
27.6)&362.3&1.5 & [23] \\
\\
\hline
\end{tabular}
\caption{Properties of the sample. Redshift, observation date, 
exposure time, good time interval after data reduction, X-ray
emission peak, luminosity distance and angular
scale. ${}^{\dagger}$ The M87 data
used in this paper were taken from Di Matteo et al.\ (2003). 
Previous papers covering the X-ray spectra of these objects include: 
[1] Mazzotta et al.\ 2003; [2] Johnstone et
al.\ 1992; [3] Sun et al.\ 2002; [4] De Grandi \& Molendi
1999; [5] Hicks et al.\ 2002; [6] David et al.\ 2001; [7] McNamara et
al.\ 2000;  [8] Matsushita et al.\ 2002; [9]
Molendi\ 2002; [10] B\"ohringer et al.\ 2001; [11] Young et al.\ 2002; [12] Allen et al.\ 2002;
[13] Ettori et al.\ 2001;  [14] Ettori et al.\ 2002; [15]
Tamura et al.\ 2001;  [16] Schmidt et al.\ 2001; [17] Johnstone et al.\
1998; [18] Allen at al.\ 2001b;  [19] Lewis et al.\ 2002;  [20] Smith et
al.\ 2002; [21] Allen at
al.\ 2001a;  [22] Kaastra et al.\ 2001;  [23] McNamara et al.\ 2001.}
\label{clusters}
\end{table*}

Spectra were extracted in the 0.5--7.0 keV energy range in circular
annuli around the X-ray emission peak. The following regions were
excluded from the cluster images: strong central sources in PKS
1404--267, Hydra A and 3C 295; enhanced, hot emission south--east of
the X--ray centre in RXJ 1347.5-1145; emission interior to the radio
jets in Cygnus A; X--ray holes in Hydra A and hot emission 180--280
kpc north of the centre in Abell 478. A detailed analysis of the
temperature and density profiles of the sample will be given in future
work (Voigt et al.\ 2003, in prep.).

\section{Spectral Model and Temperature and emission measure profiles}
Deprojected temperature and emission measure profiles were found for
each cluster assuming spherically symmetric emission from shells of
single phase gas. The spectra were fitted in \small{XSPEC} \normalsize
(Arnaud 1996) with the \small{MEKAL} \normalsize (Mewe et al.\ 1985;
Liedahl et al.\ 1995) plasma emission code, absorbed by the
\small{PHABS} \normalsize (Balucinska-Church \& McCammon\ 1992)
photoelectric absorption code.  Deprojection was performed using the
\small{PROJCT} \normalsize routine. The elements were assumed to be
present in the Solar ratios measured by Anders \& Grevesse\ (1989) and
the abundance allowed to vary between shells.  The Galactic absorption
column density was left as a free parameter in the fits, although
linked between shells.  Fixing $N_{\rm H}$ at the value expected along
the line-of-sight to the cluster would lead to spurious results since
the counts are uncertain below $\sim$ 1 keV due to the low energy
quantum efficiency degradation of the detector since launch. We do not
apply the tools provided to correct for this loss in effective area
(\small{ACISABS} \normalsize or \small{CORRARF}\normalsize) since they
appear to over-correct the data, reducing the best-fitting absorption
column density to zero in many objects. (We note that this
over-correction is also found by Takizawa et al.\ (2003) in an
analysis of Abell 3112).  We therefore allow an artificially high
absorption column to account for the loss in low energy counts since
the reduction in effective area is similar in shape to the subtraction
of counts by a foreground screen.  We stress that the temperature
profiles obtained are the same within the 1$\sigma$ error bars both
with and without the correction to the low energy counts when $N_{\rm
  H}$ is left as a free parameter in the fits. We are therefore
confident that the results are robust. The best-fitting $N_{\rm H}$
values and reduced chi-squares of the fits are shown in
Table~\ref{spectra} (model A).

\begin{table*}
\centering
\begin{tabular}{lcccccc} 
\hline\hline
Cluster&Model A& Model A& Model B& Model B & \small{F-test probability} &  Galactic $N_{\rm H}$\\
& \tiny{PROJCT*PHABS(MEKAL)}&Fitted $N_{\rm H}$ &\tiny{PROJCT*PHABS(MEKAL+MKCFLOW)} & Fitted $N_{\rm H}$ &\small{B $\rightarrow$ A} &\\
&$\chi^{2}$/dof & (10$^{20}$ cm$^{-2}$)&$\chi^{2}$/dof &  (10$^{20}$
cm$^{-2}$) & & (10$^{20}$ cm$^{-2}$)\\
\hline
2A 0335+096 & 2227.300/1404 & 27.06$\pm$0.20 & 1888.495/1400  &
28.78$\pm$0.23  & 0.00  & 17.8\\
A478 &  3370.693/2654  & 32.72$\pm$0.17 & 3299.203/2649 & 33.54$\pm$0.20
& 5.5$\times$10$^{-11}$  & 15.2\\
PKS 0745-191 & 2468.181/2193  & 40.73$\pm$0.35  & 2409.976/2187   &
43.14$\pm$0.43  & 1.7$\times$10$^{-9}$  &42.4 \\
Hydra A & 1429.901/1352  & 3.67$\pm$0.19 & 1418.988/1346 &
3.77$\pm$0.20 & 0.11   &  4.90\\
M87  & 3171/1271 & 1.97$\pm$0.12  & --  & -- &-- & 2.5\\
RXJ 1347.5-1145 & 317.3776/312 & 6.63$\pm$0.72 & 317.2172/309  &
6.72$\pm$0.90 & 0.98 &4.85 \\
A1795 & 1753.311/1491   & 1.43$\pm$0.13 &1706.515/1489  &
1.69$\pm$0.14   & 1.8$\times$10$^{-9}$ &1.19 \\
A1835 & 1306.118/1291  & 2.94$\pm$0.24 & 1305.088/1288  &
3.00$\pm$0.27 & 0.78 & 2.32 \\
PKS 1404-267 & 670.9301/625   &  10.06$\pm$0.50  & 649.8231/597   &
10.03$\pm$0.52  &  0.88 &4.52\\
3C 295& 133.3785/137  & 2.49$\pm$1.12  & 133.0522/135  & 2.62$\pm$1.45  & 0.85  &1.33\\
A2029 & 2779.268/2472 & 3.81$\pm$0.12 & 2751.292/2466  &  4.02$\pm$0.13 & 3.5$\times$10$^{-4}$  &3.05\\ 
RXJ 1532.9+3021 & 314.4362/322  & 5.71$\pm$0.68  & 309.5464/319  &
7.04$\pm$1.18   & 0.17   &2.16\\
Cygnus A & 1412.082/1237  & 30.29$\pm$0.32 &1411.647/1235  &
30.42$\pm$0.34  & 0.83  & 34.8\\
A2390 &  589.6790/617 & 11.12$\pm$0.53 & 576.4204/615 &
12.47$\pm$0.62  &  9.2$\times$10$^{-4}$ &6.80 \\
Sersic 159-03 & 867.0667/802   & 6.31$\pm$0.41&  843.1426/798 &
6.60$\pm$0.43  &  1.7$\times$10$^{-4}$  & 1.79 \\
A2597 & 1396.950/1222    & 3.24$\pm$0.22  & 1337.056/1216  &
3.88$\pm$0.24  &  9.9$\times$10$^{-10}$  &2.49 \\
\\
\hline
\end{tabular}
\caption{Chi-squares and best-fitting $N_{\rm H}$ values for models
  A and B. The \small{F-test} probability that model B provides a
  better representation of the data than model A, and the galactic
  absorption column along the line-of-sight to the cluster (from Dickey
  \& Lockman\ 1990) are also tabulated.} 
\label{spectra}
\end{table*}

\begin{figure*}
\vspace{1.0mm}
\includegraphics[angle=-90,width=0.95\columnwidth]{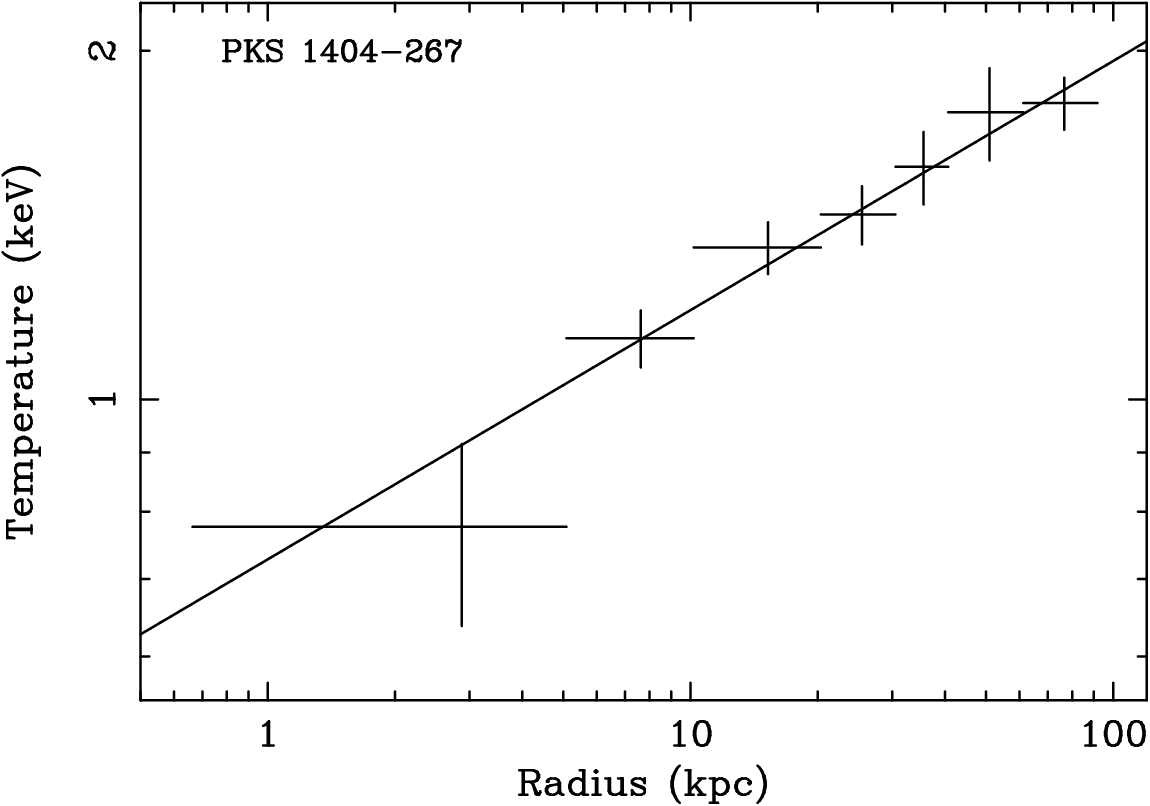}
\vspace{1.0mm}
\includegraphics[angle=-90,width=0.95\columnwidth]{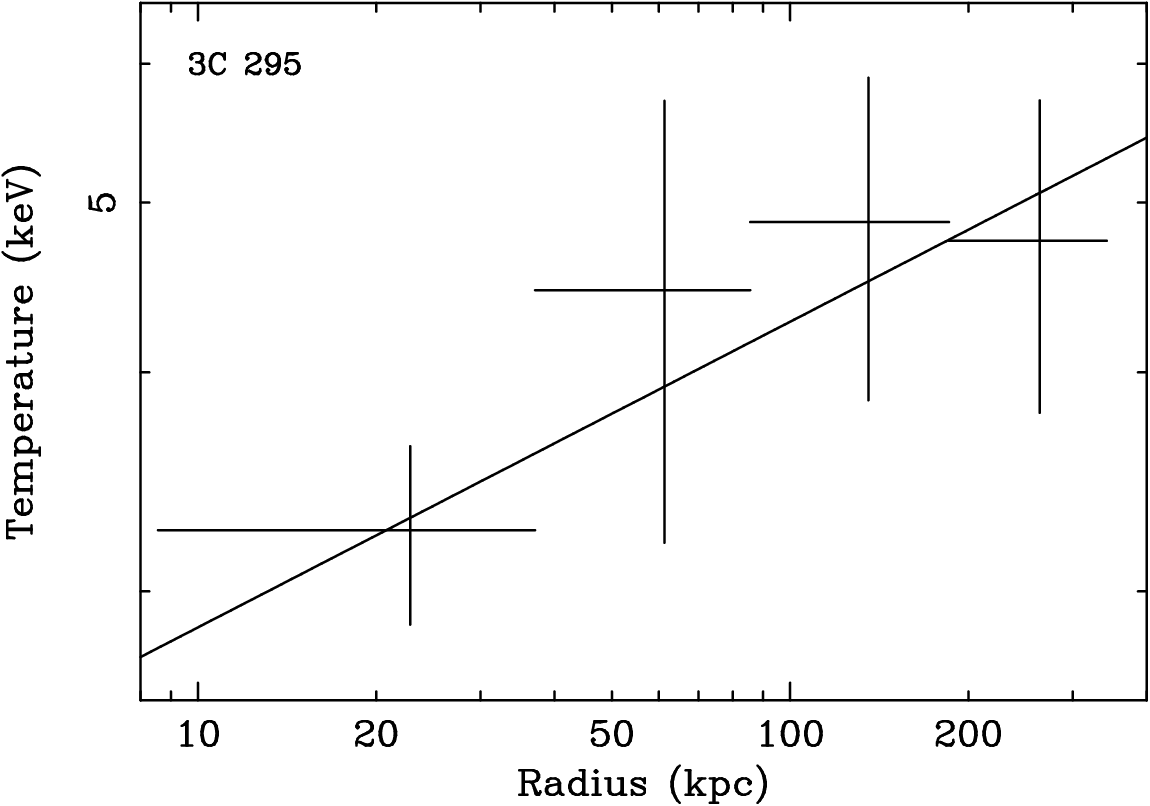}
\vspace{1.0mm}
\includegraphics[angle=-90,width=0.95\columnwidth]{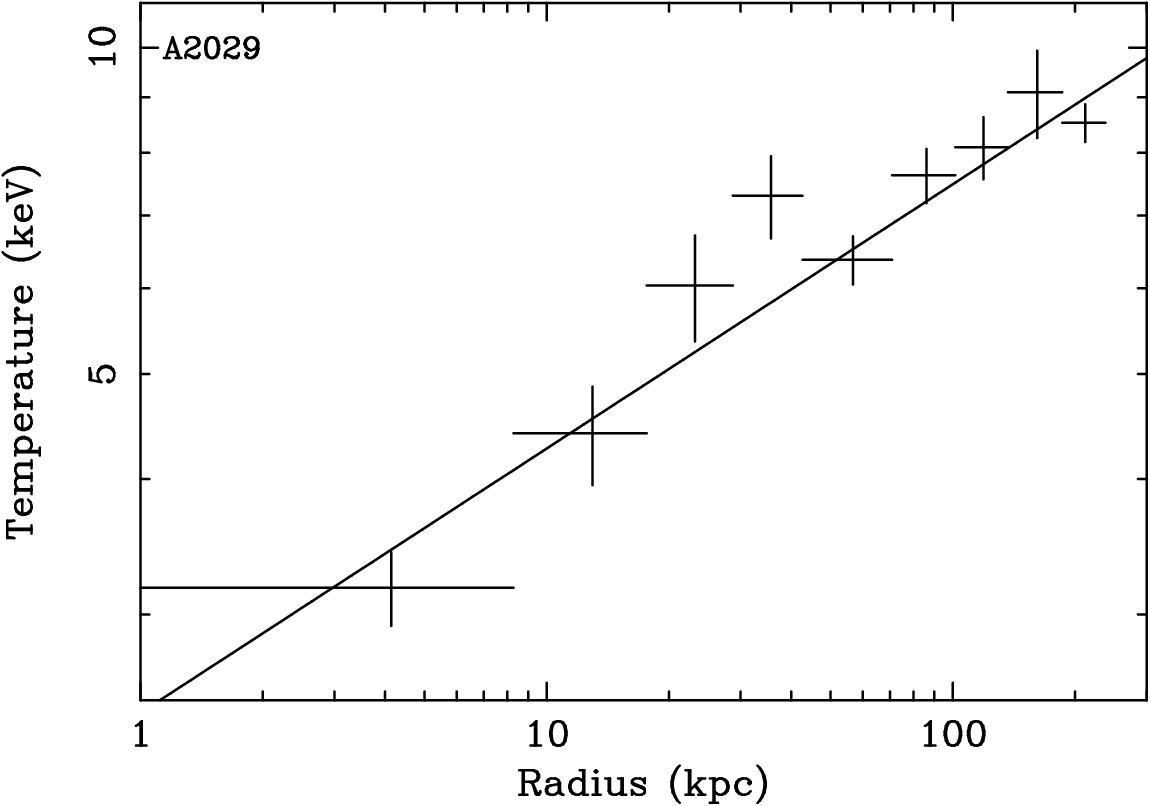}
\vspace{1.0mm}
\includegraphics[angle=-90,width=0.95\columnwidth]{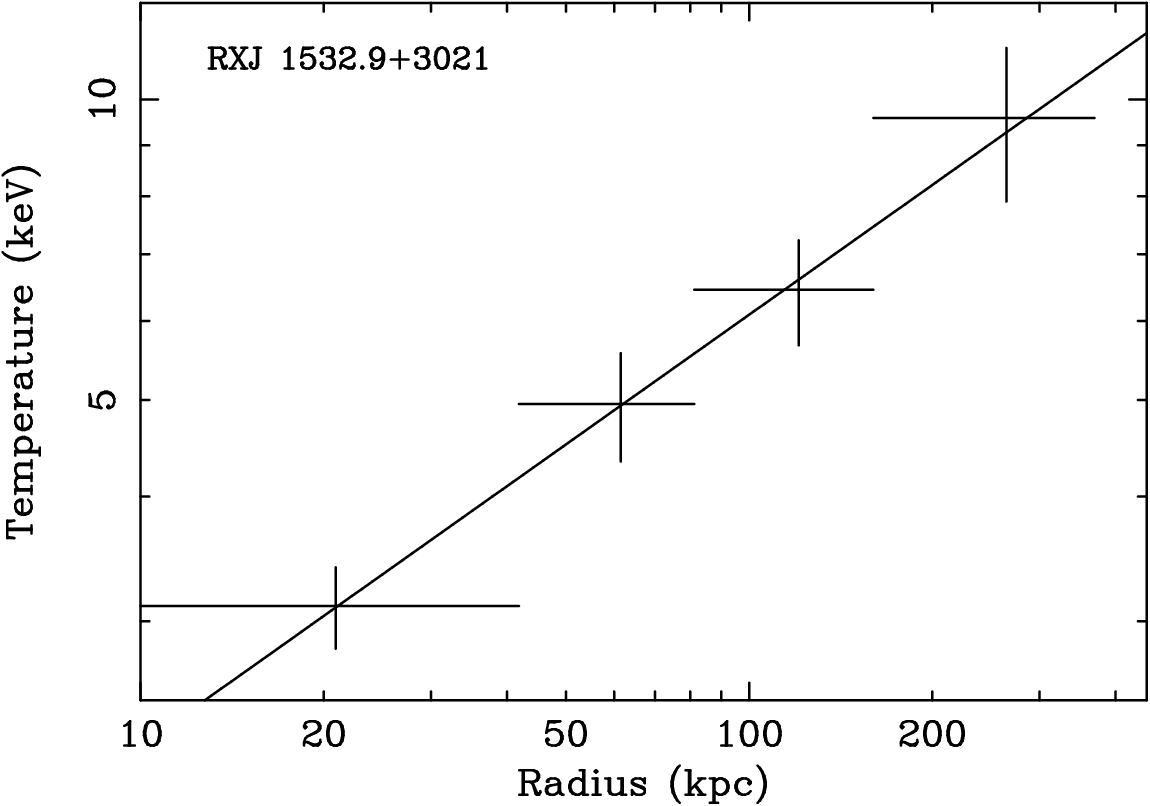}
\vspace{1.0mm}
\includegraphics[angle=-90,width=0.95\columnwidth]{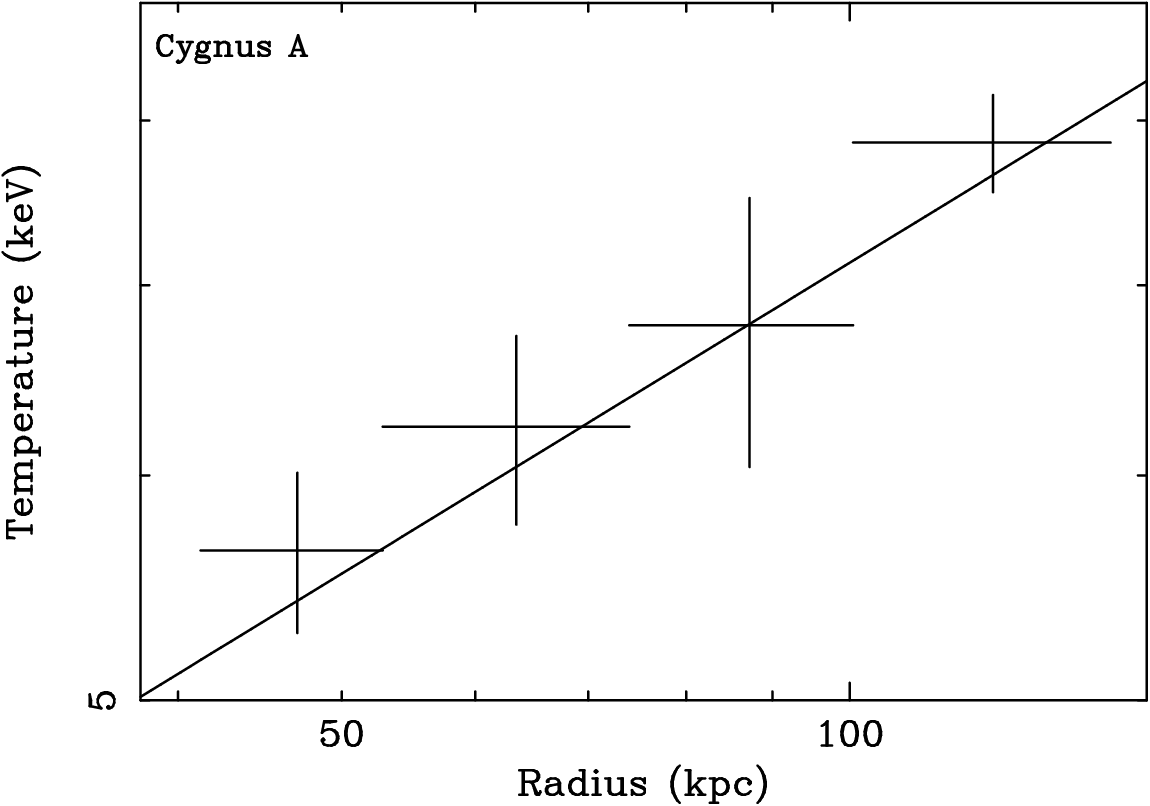}
\vspace{1.0mm}
\includegraphics[angle=-90,width=0.95\columnwidth]{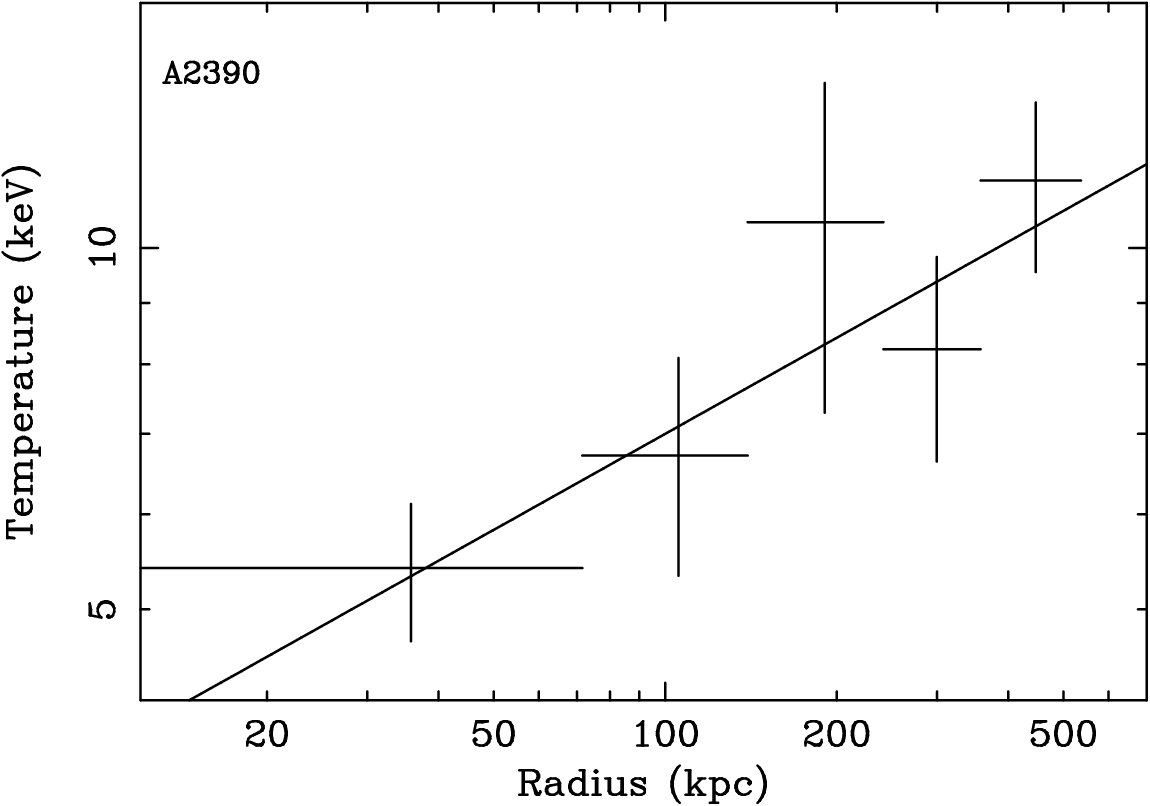}
\vspace{1.0mm}
\includegraphics[angle=-90,width=0.95\columnwidth]{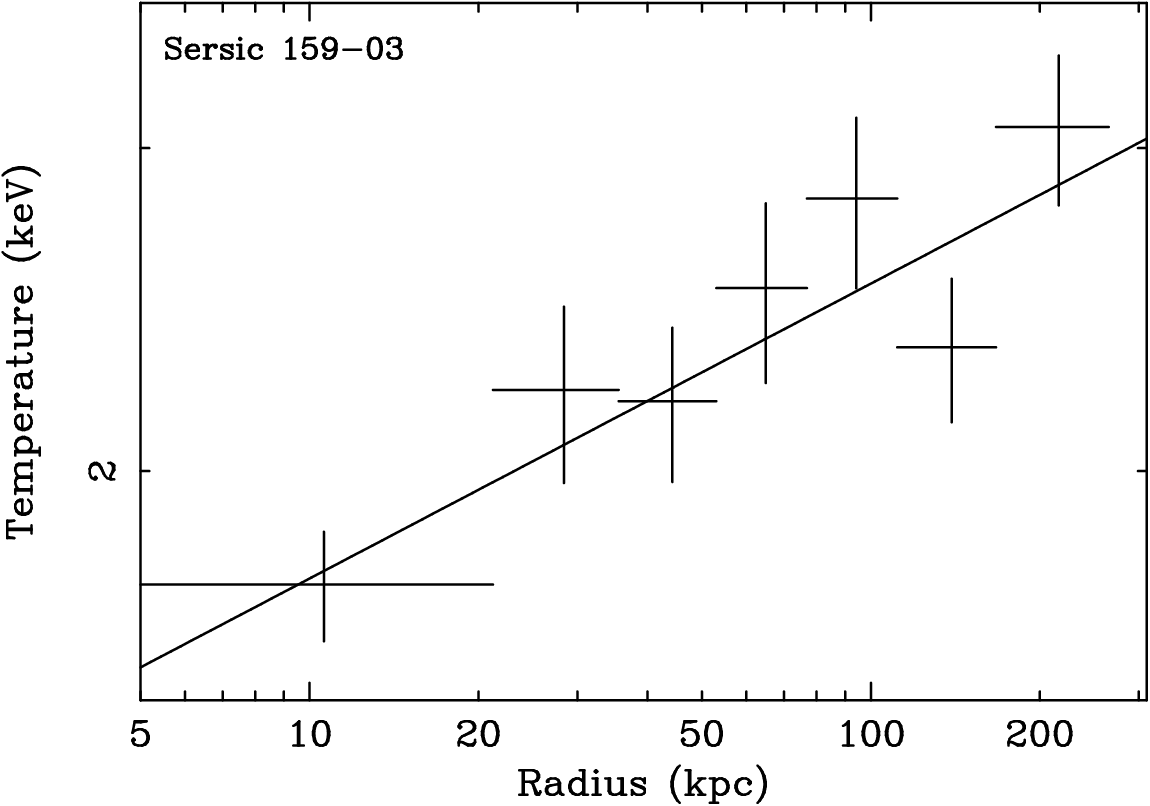}
\vspace{1.0mm}
\includegraphics[angle=-90,width=0.95\columnwidth]{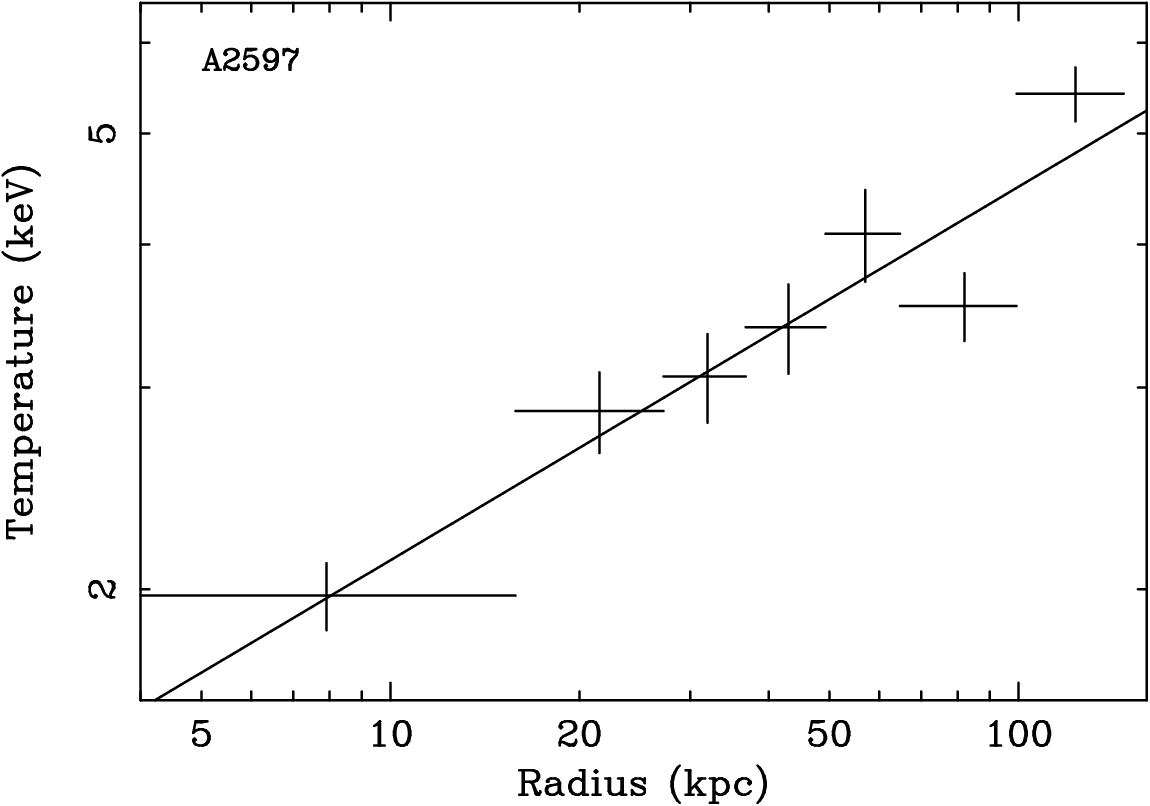}
\caption{Deprojected temperature profiles obtained using the model
  \small{PROJCT*PHABS(MEKAL)} \normalsize in \small{XSPEC}
  \normalsize. The best-fitting power law to the data is shown for
  clusters with over-lapping error bars.}
\label{fig:temp}
\end{figure*}

The temperature profiles obtained are shown for the whole sample in
Fig.~\ref{fig:temp}.  The one sided error bars plotted (necessary
for chi-square calculations) are the root-mean-square of the two sided
1$\sigma$ uncertainties found using the error command in \small{XSPEC}
\normalsize.

\section{Calculation of the effective conductivity}

\subsection{Energy equations}

The luminosity emitted from an isothermal plasma of density $n$,
temperature $T$ and metallicity $Z$ per unit volume is given by

\begin{equation}
L = n^{2} \Lambda(T,Z)  
\end{equation}
where $\Lambda(T,Z)$ is the cooling function for a plasma losing energy
by bremsstrahlung radiation and line emission.

Assuming the shells contain single phase gas, we can write the total
luminosity emitted from shell $j$ with volume $\Delta V_{j}$ as

\begin{equation}
\Delta L^{\mathrm{tot}}_{j} =  EM_{j} \Lambda(T_{j},Z_{j}) 
\label{eqn:cool}
\end{equation}
where $EM_{j}=n_{j}^{2} \Delta V_{j}$ is the emission measure of the
shell.

The net heat transferred to the $j^{\rm th}$ shell by conduction is
given by

\begin{equation}
\Delta L^{\mathrm{cond}}_{j} = 4\pi r_{j}^2 \kappa_{j} \left(\frac{dT}{dr}\right)_{j} - 
4\pi r_{j-1}^2 \kappa_{j-1} \left(\frac{dT}{dr}\right)_{j-1}
\label{eqn:heat}
\end{equation}
where (d$T$/d$r$)$_{j}$ is the temperature gradient and $\kappa_{j}$ the
plasma conductivity across the outer boundary of the $j^{\rm th}$
shell.

We calculate the conductivity required to prevent cooling in the
$j^{\rm th}$ shell by equating Equations~\ref{eqn:cool}
and~\ref{eqn:heat} and summing from the centre of the cluster outwards

\begin{equation}
4\pi r_{j}^2 \kappa^{\rm eff}_{j}  \left(\frac{dT}{dr}\right)_{j} =
 \sum_{i=1}^{j}
EM_{j} \Lambda_{j}
\label{eqn:kappa}
\end{equation}
where $\kappa^{\rm eff}$ is referred to as the effective thermal
conductivity.

\subsection{$\kappa^{\rm eff}$ calculation}

\begin{figure*}
\vspace{1.0mm}
\includegraphics[angle=-90,width=0.95\columnwidth]{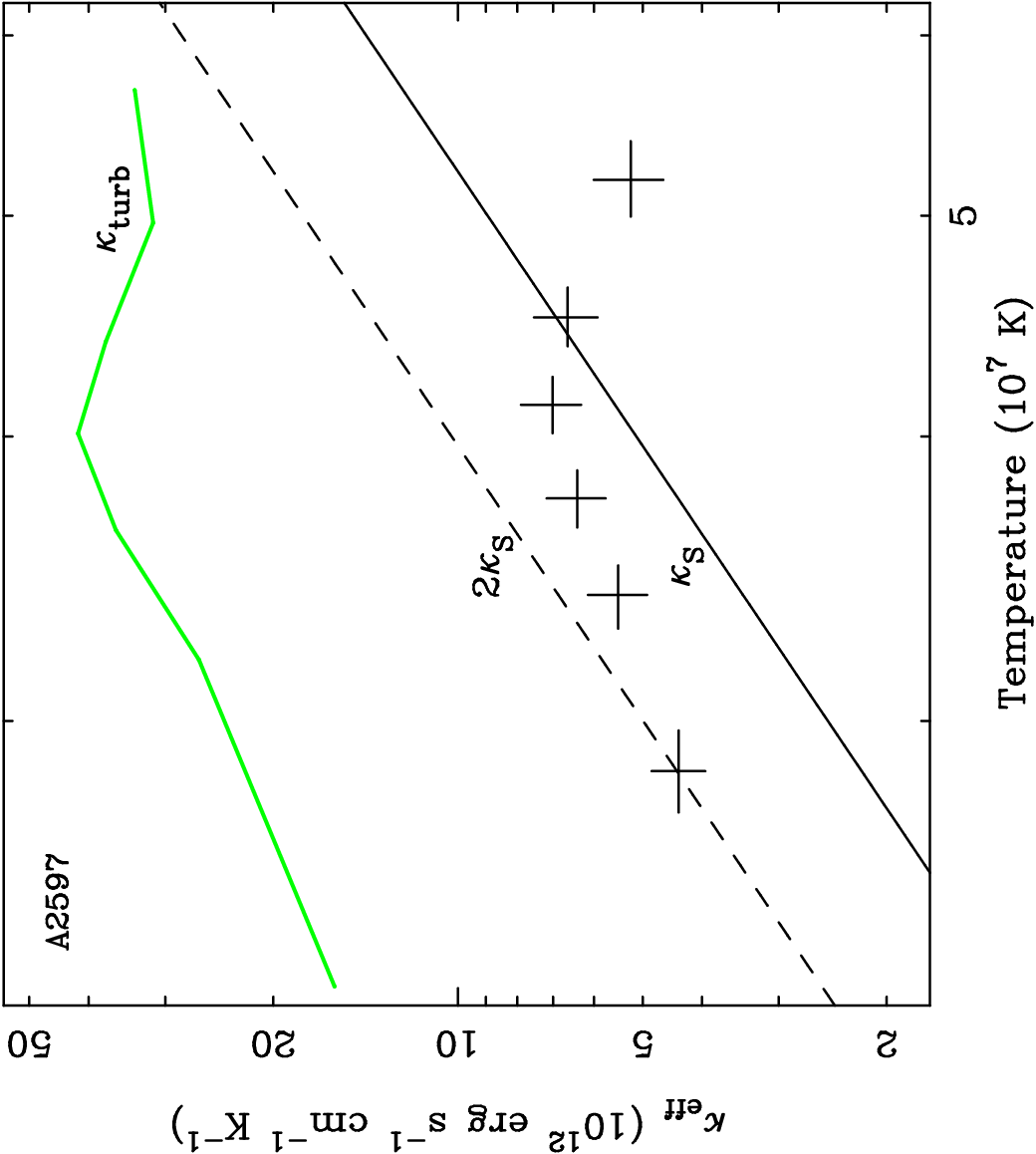}
\vspace{1.0mm}
\includegraphics[angle=-90,width=0.95\columnwidth]{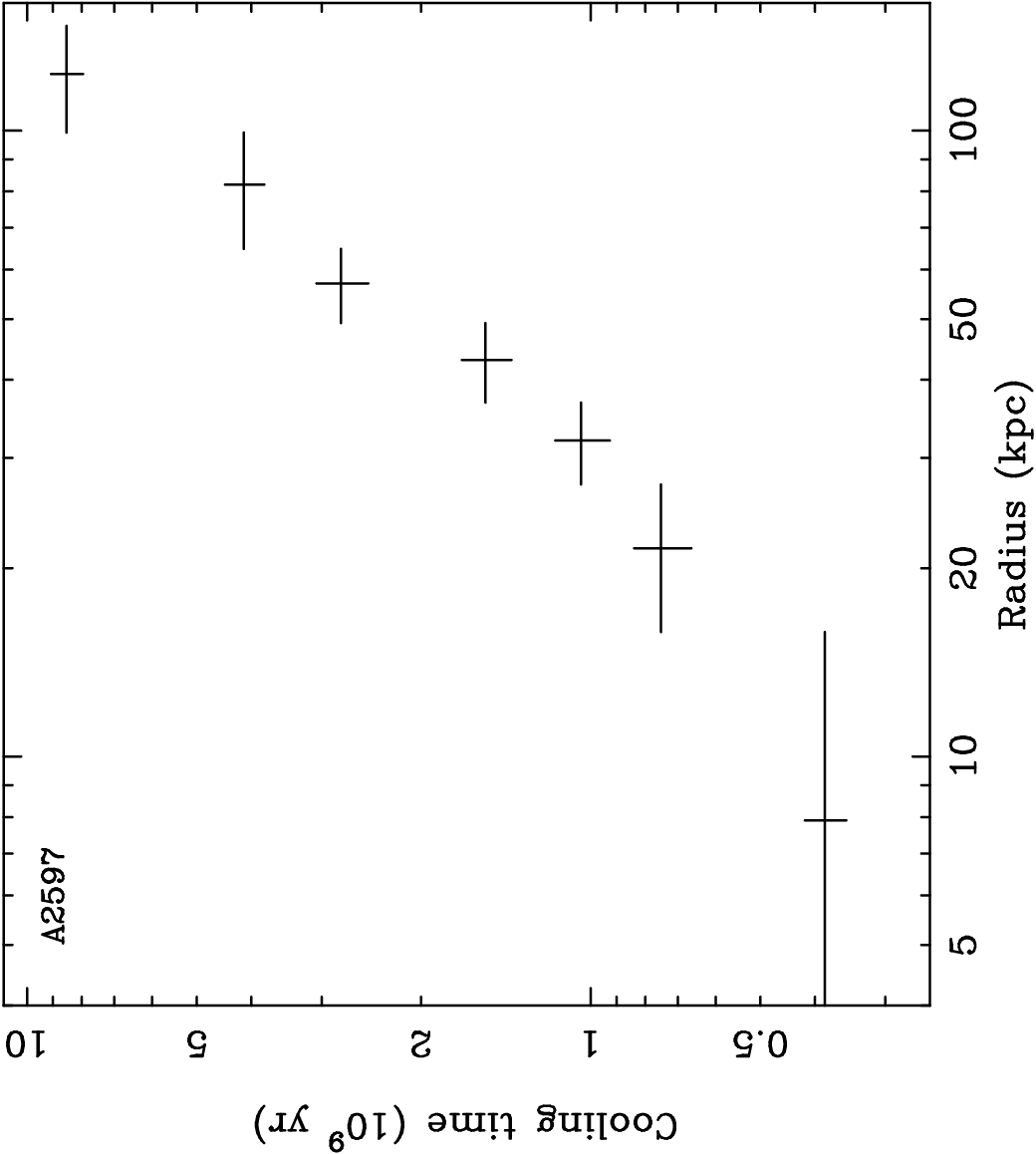}
\caption{Effective conductivity profiles (left) and cooling times
  (right) for each cluster. The effective conductivity at the outer
  boundary of each shell can be compared with the cooling time within
  that shell. $\kappa_{\rm S}$ and $\kappa_{\rm turb}$ are the Spitzer
  and turbulent conductivities, respectively. The age of the Universe
  is 13.7 Gyr for the cosmology used in this paper.}
\label{fig:kappa}
\end{figure*}

The effective conductivity at the outer boundary of each shell was
calculated using Equation~\ref{eqn:kappa}.  We fit power-law models to
temperature profiles where the error bars overlap (see
Table~\ref{fits}). The effective conductivity values and their
1$\sigma$ uncertainties, shown for each cluster individually in
Fig.~\ref{fig:kappa}, were determined using Monte Carlo simulations.
The gas cooling times, $t^{\rm cool}$, within each shell are also
plotted in Fig.~\ref{fig:kappa}.

We calculate $t^{\rm cool}$ using the formula

\begin{equation}
t^{\rm cool} \approx  \frac{\frac{3}{2}nkT}{\epsilon}
\label{eqn:tcool}
\end{equation}
where $\epsilon$ is the total emissivity and $n$ is the total number
density of particles.

Since \(\epsilon \Delta V = EM \Lambda = n_{\rm e}n_{\rm h} \Delta V
\Lambda \), and using \(n \approx 2n_{\rm e}\) and \(n_{\rm e} \approx
(6/5) n_{\rm h}\), then \(\epsilon \approx (5/6) n_{\rm e}^{2}
\Lambda\), where $n_{\rm e}$ and $n_{\rm h}$ are the electron and
hydrogen number densities respectively.  Substituting into
Equation~\ref{eqn:tcool} we have the following expression for the
cooling time of gas in the $j^{\rm th}$ shell

\begin{equation}
t^{\rm cool}_{j} \approx 3.3\frac{kT_{j}}{(EM_{j}/\Delta
  V_{j})^{\frac{1}{2}} \Lambda_{j}}.
\end{equation}

The cooling time profiles (which are mostly $\propto T^{\frac{1}{2}}
n^{-1}$) are plotted for the whole sample in Fig.~\ref{fig:sample}.
They closely track the same radial-dependence curve (a similar result
is found for the entropy profiles $\propto T n^{-\frac{2}{3}}$). We
find no trend in terms of `old' or `young' cooling flows which would
indicate cyclic heating behaviour.

We define the cooling radius as where the cooling time of the gas
drops to 5 Gyr.

\begin{table}
\centering
\begin{tabular}{lcccc} 
\hline\hline
Cluster & $a$ & $b$ & $\chi^{2}$/dof \\
\hline
PKS 0745-191 & 1.18  & 0.38  & 11.83/7\\
Hydra A &1.72  & 0.18  &5.39/6 \\
RXJ 1347.5-1145 &2.84   &  0.31 & 1.21/2 \\
A1795 &1.68  & 0.26  & 6.09/3 \\
A1835 & 1.58  & 0.34 & 2.86/5 \\
PKS 1404-267 & 0.73  & 0.22  & 1.93/5  \\
3C 295 &1.91  & 0.17 & 0.43/2 \\
A2029& 2.43 & 0.24 & 11.86/7\\
RXJ 1532.9+3021 & 0.83  & 0.43  & 0.08/2  \\
Cygnus A & 1.34  & 0.36 & 0.99/2\\ 
A2390 & 2.06  & 0.27 & 1.34/3 \\
Sersic 159-03 & 1.21  & 0.16 & 4.90/5\\
A2597 & 1.00 & 0.33 & 13.42/5 \\
\\
\hline
\end{tabular}
\caption{The parameters and reduced chi-square values for the 
  best-fitting power laws ($T=ar^{b}$) to the temperature profiles. 
A power law is not required for clusters where the error bars are small (M87,
  2A 0335+096 and Abell 478).}
\label{fits}
\end{table}

\section{Thermal conduction}

\subsection{Spitzer conductivity}

The conductivity of a hydrogen plasma completely free from magnetic
fields was derived by Spitzer (1962)

\begin{equation}
\kappa_{\rm S} = \frac{1.84\times10^{-5} T^{\frac{5}{2}}}{\mathrm{ln}\hspace{0.6mm}\Lambda} \ergpspcmpK
\end{equation}
where ln\hspace{0.6mm}$\Lambda$ is the Coulomb logarithm (ratio of the
largest to the smallest impact parameter)

\begin{equation}
\mathrm{ln}\hspace{0.6mm}\Lambda = 37.8 +
\mathrm{ln}\left[\left(\frac{T}{10^{8}\hspace{0.1cm}\mathrm{K}}\right)\left(\frac{n_{\rm
      e}}{10^{-3}\hspace{0.1cm} \mathrm{cm}^{-3}}\right)^{-\frac{1}{2}}\right]. 
\end{equation}
This term depends only weakly on on the plasma temperature and density
and so we can re-write this equation as

\begin{equation}
\kappa_{\rm S} \simeq \kappa_{0} T^{\frac{5}{2}} 
\end{equation}
where \(\kappa_{0} \simeq 5.0\times10^{-7}\) erg s$^{-1}$ cm$^{-1}$
K$^{-\frac{7}{2}}$ for electron densities and temperatures appropriate
to the objects considered.

\subsection{Magnetic field effects}
The presence of magnetic fields in the intracluster medium will reduce
the conductivity below the full Spitzer rate, such that
$\kappa=f\kappa_{\rm S}$, where $f \leq 1$. The gyroradius of charged
particles around magnetic field lines in the intracluster plasma,
$r_{g}$, is much less than the mean free path of charged particles due
to collisions, $\lambda_{e}$. The effective mean free path will be
reduced by a factor \(\sim r_{g}^{2}/\lambda_{e}\) perpendicular to
ordered magnetic field lines. Assuming a radial temperature gradient,
$f$ $\sim$ 0.3 for a radial magnetic field and negligible for a
circumferential field.

However, observations show that the magnetic fields may be tangled
(Carilli \& Taylor\ 2002).  The conductivity then depends on the
coherence length of the magnetic field. The theory was first discussed
by Rechester \& Rosenbluth (1978) and later developed by Chandran \&
Cowley (1998).  These initial studies predicted a suppression factor
$f \sim$ 0.01--0.001. More recently, Narayan \& Medvedev (2001) have
suggested that a magnetic field which is chaotic over a wide range of
scales will lead to a suppression factor in the range $\sim$ 0.1--0.4.
Maron, Chandran \& Blackman\ (2003), on the other hand, support a
value $\sim$ 0.02.

We note that heat conduction at close to the Spitzer rate is observed
in the Solar wind when the electron mean free path is small compared
with the temperature gradient (Salem et al.\ 2003).

The conductive heat flux will be overestimated in the outer regions of
clusters if the mean free path of electrons is comparable to or
greater than the scale length of the temperature gradient (Cowie \&
McKee\ 1977). `Saturation' of the conductivity at large radii may
explain why significant amounts of heat will not be removed from the
cluster outwards (Loeb et al.\ 2002 suggested that conduction at the
Spitzer rate throughout galaxy clusters would result in significant
amounts of thermal energy being leaked to the surrounding intercluster
medium). Also, fields lines may be stretched out radially by mass
inflow, resulting in Spitzer rate conduction being restricted to
cluster centres (Bregman \& David\ 1988; Fabian et al.\ 2002b).

\section{Analysis}

\subsection{Effective conductivity profiles}

\begin{figure}
\includegraphics[angle=-90,width=0.95\columnwidth]{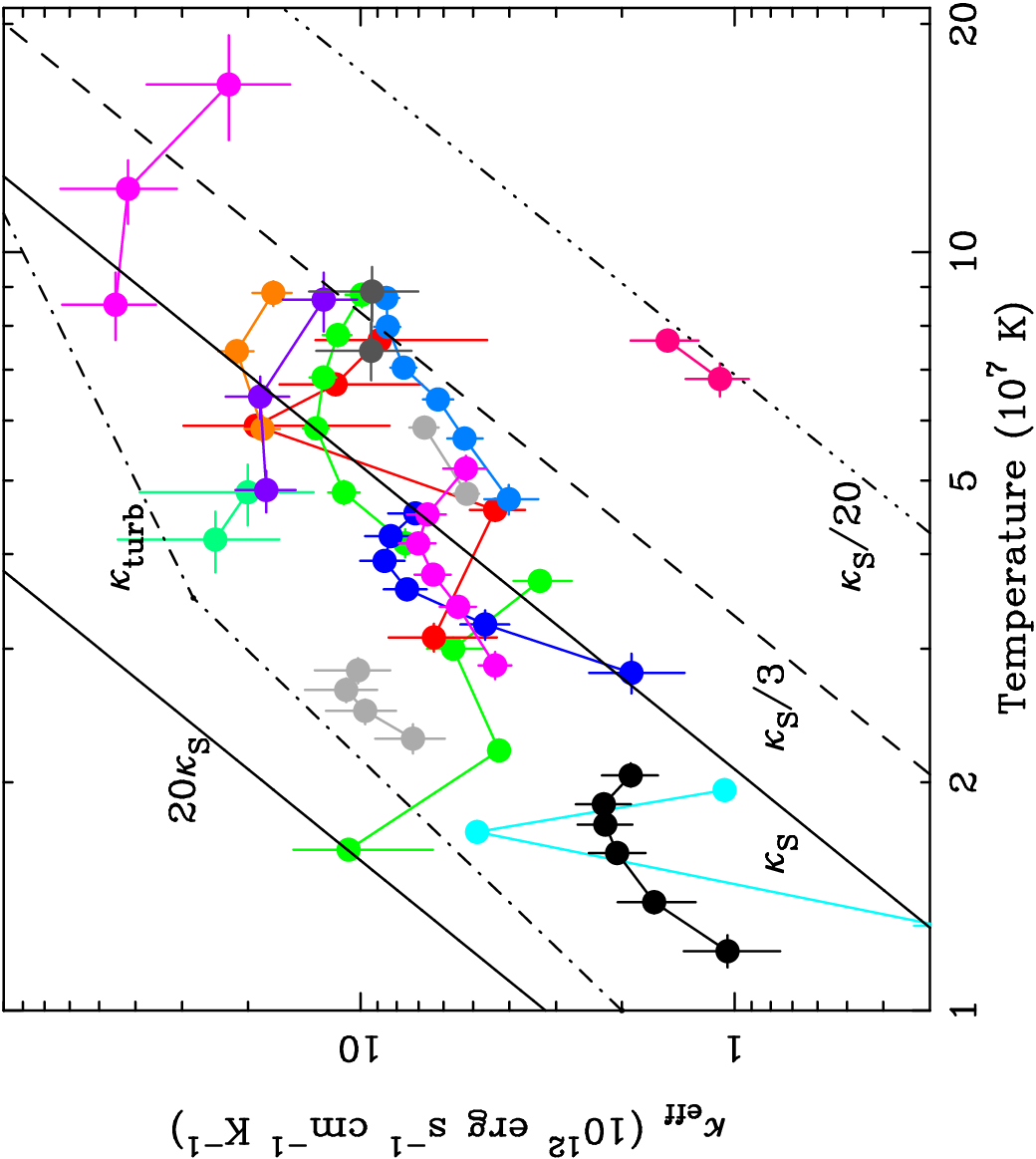}
\vspace{1.0mm}
\includegraphics[angle=-90,width=0.95\columnwidth]{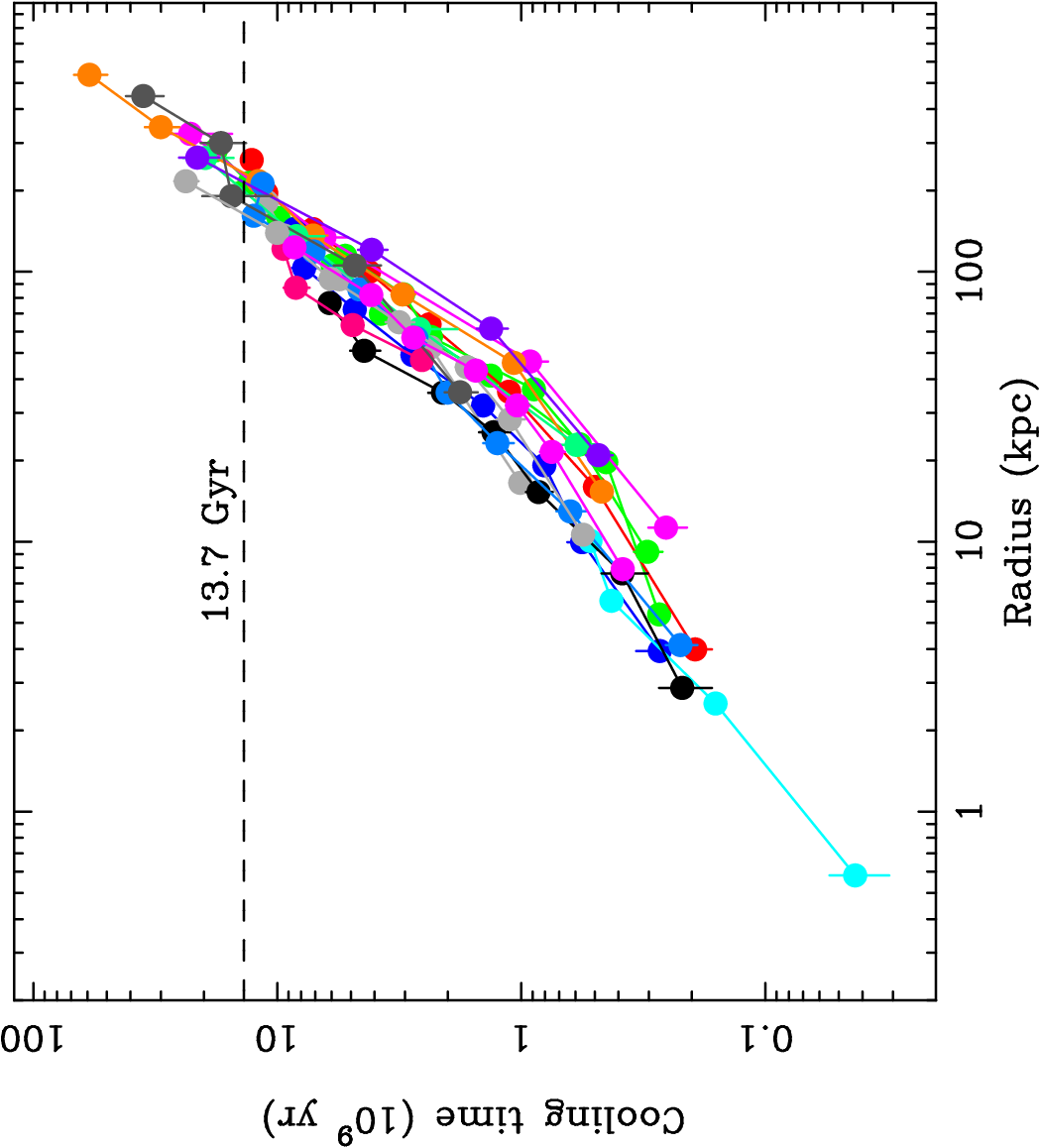}
\caption{Effective conductivity profiles (upper plot) and cooling times
  (lower plot) for the sample. $\kappa_{\rm S}$ is the Spitzer
  conductivity and $\kappa_{\rm turb}$ is the turbulent conductivity.
  Note that $\kappa_{\rm turb}$ is calculated at a fixed cooling time
  (1 Gyr).}
\label{fig:sample}
\end{figure}

The effective conductivity profiles are plotted out to the cooling
radius for the whole sample in Fig.~\ref{fig:sample}. We find that
when the temperature of the cluster drops below $\sim$5 keV,
conductivity values above the Spitzer curve are required.  This
demonstrates the strong dependence of thermal heat transport on
temperature and suggests that even Spitzer-rate conduction will be
unable to prevent cooling in low temperature clusters.

Studying the individual effective conductivity profiles shown in
Fig.~\ref{fig:kappa} in detail, we find that in four clusters (Abell
1795, Abell 2029, Cygnus A and Abell 2390) conduction at or below the
Spitzer rate would prevent cooling from taking place everywhere inside
the cooling radius.  The remaining clusters in the sample require the
plasma conductivity to be greater than $\kappa_{\rm S}$ in at least
one shell. (We note that in studies by Fabian et al.\ (2002) and
Medvedev et al.\ (2003), where the conductivity required to prevent
cooling is calculated at the cooling radius only, clusters such as RXJ
1532.9+3021 and 3C 295 would appear below the Spitzer curve. Detailed
studies of the temperature profiles right into the centres of clusters
are therefore needed to accurately assess the energetic feasibility of
conduction over the entire cooling flow region).

The conductivity values required to prevent cooling at radial
positions within the cluster where the cooling time of the gas is 13,
5 and 1 Gyr are shown in Table~\ref{kappas} as a fraction of
$\kappa_{\rm S}$. Conduction may stop large mass inflows from taking
place in the outer parts of the cooling flow region, but will be
unable to prevent a net heat loss in the inner regions of the majority
of objects in the sample.  Since most of the implied mass deposition
takes place at larger radii, the small mass inflows which must develop
in the centre (in the absence of any additional heat sources) may be
below the spectral constraints imposed by recent observations.

\begin{table*}
\centering
\begin{tabular}{lccccc} 
\hline\hline
Cluster & Cooling time  &Temperature &  $\kappa^{\rm eff}$ & $\dot M^{\mathrm{clas}}$&$\dot M^{\mathrm{spec}}$\\
& (Gyr)& (keV) & (/$\kappa_{\rm S}$)& (M$_{\odot}$ yr$^{-1}$)
& (M$_{\odot}$ yr$^{-1}$)\\
\hline
2A 0335+096  &13.7& -- & -- & &\\
&5.0 & 3.2 & 0.8  &  377$\pm$8 &  80$\pm$5    \\
&1.0 & 2.2 & \bf{2.8} & & \\
A478 & 13.7 &&  & & \\
&5.0 & 6.6  & 0.3  & 892$\pm$22  & 112$\pm^{22}_{19}$ \\
&1.0 & 4.5  & 0.2  &  &  \\
PKS 0745-191 & 13.7 & 9.5  & 0.1  & &\\
&5.0 & 7.6   &  0.3  & 1269$\pm$30  & 317$\pm^{35}_{29}$  \\
&1.0 & 5.1  & 1.0  & &\\
Hydra A & 13.7 & -- & -- & &\\
&5.0 & 3.9 & 1.0 & 239$\pm$7  &  8$\pm^{6}_{4}$ \\
&1.0 & 3.2 & \bf{1.8} &  &\\
RXJ 1347.5-1145 & 13.7 & --  & --  &  &\\
&5.0 & 14.4  & 0.1  & 3518$\pm$304  &  75$\pm^{743}_{72}$ \\
&1.0 & 10.5  & 0.5& &\\
A1795 & 13.7 & -- & -- &&   \\
& 5.0 & 5.4      &   0.4   & 268$\pm$6  & 30$\pm^{5}_{5}$  \\
& 1.0 & 4.1   &   0.6 & & \\
A1835 &13.7 & 10.5  & 0.1  & &  \\
&5.0 & 8.3  & 0.3  & 1453$\pm$64 & 34$\pm^{43}_{34}$  \\
&1.0 & 6.4 & 0.9 & &\\
PKS 1404-267 & 13.7 & -- & -- & &\\
&5.0 & 1.8  &  \bf{2.0} & 54$\pm$4   & 5$\pm$1   \\
&1.0 & 1.4  & \bf{3.9}  & &\\
3C 295 & 13.7 & 4.8   & 1.1 & & \\
&5.0 & 4.5  & \bf{1.7}   & 632$\pm$83  & 16$\pm^{230}_{12}$  \\
&1.0 & 3.9  & \bf{3.2}  & & \\
 A2029 & 13.7 & 8.7   & 0.1  & &  \\
&5.0  & 7.5  & 0.2  &  453$\pm$12 & 30$\pm^{10}_{13}$    \\
&1.0 & 5.2  & 0.4  &&  \\
RXJ 1532.9+3021 & 13.7 & --  & -- & & \\
&5.0 & 7.5  & 0.4   &  2375$\pm$169 & 592$\pm^{357}_{260}$\\
&1.0 & 4.9  & \bf{1.5}   &  & \\
A2390 & 13.7  & 8.9  & 0.1  &  &   \\
&5.0 &  7.7 & 0.2  & 765$\pm$76  & 339$\pm^{42}_{118}$ \\
&1.0 &  -- & --  & &\\
Sersic 159-03 & 13.7  & 2.7  & \bf{2.3}  & & \\
&5.0 & 2.5  & \bf{4.1}  & 245$\pm$13   & 33$\pm^{11}_{9}$ \\
&1.0 &  2.1  & \bf{6.2}  &&  \\
A2597 & 13.7 & -- & --  & &\\
&5.0  & 4.5  & 0.5  & 409$\pm$14   & 57$\pm^{14}_{13}$  \\
&1.0 & 3.1  & \bf{1.5}  & &\\
\\
\hline
\end{tabular}
\caption{Effective conductivity values (shown in bold where
  $\kappa^{\rm eff} > \kappa_{\rm S}$) and temperatures at radial
  positions corresponding  to cooling times of 13.7, 5 and 1 Gyr. 
The integrated classic and
  spectrally-determined mass deposition rates within the cooling
  radius are shown. M87 and Cygnus A are not included in this table since
  the analysis of these clusters does not extend over the entire cooling
  flow region. Data are unavailable (shown by dashed lines) either
  when the inner region is unresolved or the relevant cluster emission
  lies outside the chip boundary.}
\label{kappas}
\end{table*}

\subsection{Conduction + reduced cooling flow?}

For clusters in which conductive heat transfer is insufficient to
prevent cooling, we calculate the rate at which mass must drop out in
each shell, assuming conduction is operating at the Spitzer rate over
the observed temperature gradients. This provides lower limits on the
mass deposition rates required, since the plasma conductivity is
reduced by a factor $f \leq 1$ in the presence of magnetic fields. We
then compare these values with spectral mass deposition rates,
determined by fitting cooling flow models to the spectra.

We note that for any blobs of cool gas to exist, conduction of heat to
the blobs must be highly suppressed. This probably requires them to be
separate magnetic structures.

\subsubsection{Inhomogeneous cooling flows}

The observed bolometric luminosity from a single shell $j$ in a
spherically symmetric, inhomogeneous, steady state, subsonic cooling
flow is given by

\begin{equation}
\Delta L^{\mathrm{tot}}_{j}-\Delta L^{\mathrm{cond}}_{j}= \Delta \dot M_{j} H_{j} + \delta_{j} \Delta \dot M_{j}
 \Delta \Phi_{j}
 + \dot M_{j-1} \left(\Delta H_{j} + \Delta
 \Phi_{j} \right) 
\label{eqn:nulsen}
\end{equation} 
where $\Delta \dot M_{j}$ is the rate at which mass is deposited in
the $j^{\rm th}$ shell and $\dot M_{j-1}$ is the rate at which mass
flows through the shell towards the centre, such that

\begin{equation}
\dot M_{j} = \sum_{i=1}^{j} \Delta \dot M_{i}
\end{equation} 
$H_{j}$, $\Delta H_{j}$, $\Delta \Phi_{j}$ are the enthalpy of the
ambient gas in the $j^{\rm th}$ shell and the enthalpy change and
gravitational potential difference across it. $\delta_{j}$ is a factor
of order unity which takes into account the fact that gas is deposited
throughout the volume of the shell (see e.g. Fabian et al.\ 1985).

The first term represents the luminosity emitted by gas cooling out of
the flow. Since the flow rate is long compared with the cooling rate,
the gas can be assumed to be cooling out at a fixed radius and
therefore, under the condition of hydrostatic equilibrium, at constant
pressure. The spectrum emitted by a perfect gas cooling isobarically
is described by Johnstone et al.\ (1992)

\begin{equation}
{\Delta L^{\mathrm{cool}}_{j}(\nu)={\frac{5k}{2\mu m_{\rm H}}} \dot \Delta M_{j} \int_{0}^{T_{j}}
 \frac{\epsilon_{j}(\nu)}{\Lambda(T)}  \mathrm{d}T}
\label{eqn:rmj}
\end{equation}
where $\epsilon_{j}(\nu)$ is the emissivity at frequency $\nu_{j}$.

The second and fourth terms represent the gravitational potential
energy released due to gas cooling out and gas flowing across the
shell respectively.  The third term is the contribution from the
enthalpy change of gas flowing across the shell. The contribution from
these terms to the total luminosity emitted may be comparable to that
from the first term (Fabian 1994).

We note that in the classic cooling flow scenario cool gas clouds
deposited out of the flow are assumed to be supported against
gravitational infall.  Magnetic fields were invoked to prevent clouds
from falling towards the cluster centre (Fabian\ 1994). Fabian\ 
(2003a) discusses the possibility that blobs of gas fall inwards,
giving up their gravitational potential energy to the surrounding
medium as they descend through the core. The energetics show that this
process could be important in the outer parts of the cooling flow
region.

\subsubsection{Mass deposition rates with and without conduction}
We first calculate the mass deposition rate expected in each shell
assuming the plasma conductivity is zero (i.e. $ \Delta
L^{\mathrm{cond}}_{j}$ = 0); this is the classic mass deposition rate,
$\Delta \dot M_{j}^{\rm clas}$.

\begin{equation}
{\Delta L^{\mathrm{tot}}_{j}={\frac{5k}{2\mu m_{\rm H}}} \Delta \dot  M_{j}^{\mathrm{clas}} T_{j}}
\label{eqn:mdot}
\end{equation}
where we have neglected the contributions from the last three terms in
Equation~\ref{eqn:nulsen}.

We then calculate the mass which must cool out of the flow assuming
conductive heat transfer at the Spitzer rate over the observed
temperature gradients. We refer to this as $\Delta \dot
M_{j}^{\mathrm{cool}}$, given by

\begin{equation}
\Delta L^{\mathrm{tot}}_{j} - \Delta L^{\mathrm{cond}}_{j} =
{\frac{5k}{2\mu m_{\rm H}}} \Delta \dot  M_{j}^{\mathrm{cool}} T_{j}
\label{eqn:mdotcool}
\end{equation}
where we put \(\kappa_{j}=\kappa_{0} T_{j}^{\frac{5}{2}}\) in
Equation~\ref{eqn:heat}.

\subsubsection{Spectral mass deposition rates}

We compare the values calculated for $\Delta \dot
M_{j}^{\mathrm{cool}}$ with spectral mass deposition rates, determined
by fitting a cooling flow models to the data.  The emission from each
shell is represented by an isothermal component (plasma maintained at
a constant temperature) plus a cooling flow component.  The cooling
term is fitted with an \small{MKCFLOW} \normalsize spectrum (see
Equation~\ref{eqn:rmj}) and the isothermal term by a single
\small{MEKAL} \normalsize component.  Since we receive projected X-ray
emission, the model \small{MKCFLOW} \normalsize + \small{MEKAL}
\normalsize is fitted to the deprojected emission from each shell. As
with model A, this is carried out using the \small{PROJCT} \normalsize
routine.  \small{NB} \normalsize The \small{MEKAL} \normalsize
component is not there to account for emission from material outside
the cooling radius and any cooling flow component detected is not a
projection effect (assuming spherical symmetry), as suggested by
Molendi \& Pizzolato\ 2001. The temperatures and abundances in
\small{MKCFLOW} \normalsize are tied to the \small{MEKAL} \normalsize
values within each shell.  The spectral mass deposition rate, $\Delta
\dot M_{j}^{\mathrm{spec}}$, in each shell is the normalization of the
\small{MKCFLOW} \normalsize component. The cooling flow model is fit
to shells only where the cooling time of the gas is less than or equal
to 5 Gyr. As before, the Galactic absorption column is left as a free
parameter in the fits (although linked between shells) and the
abundances are untied. We note that the spectral mass deposition rates
obtained are statistically equivalent both with and without the
\small{CORRARF} \normalsize correction.
 
The best-fitting $N_{\rm H}$ values and reduced chi-squares for the
fits are tabulated in Table~\ref{spectra} (model B), together with the
\small{F}\normalsize-test probabilities that the isothermal model
provides a better representation of the data than the cooling flow
model. The probabilities suggest that adding a cooling flow component
to each shell provides a significantly better fit to the data for the
majority of objects in the sample. For 3C 295, Abell 1835, RXJ
1347.5+1145, PKS 1404-267 and Cygnus A an additional cooling flow
component is not required statistically. There is no relation between
the requirement for a cooling flow component and the need for
super-Spitzer plasma conductivity.

\subsubsection{Comparison between predicted and spectrally-determined
  mass deposition rates} 

\begin{figure*}
\includegraphics[angle=-90,width=0.95\columnwidth]{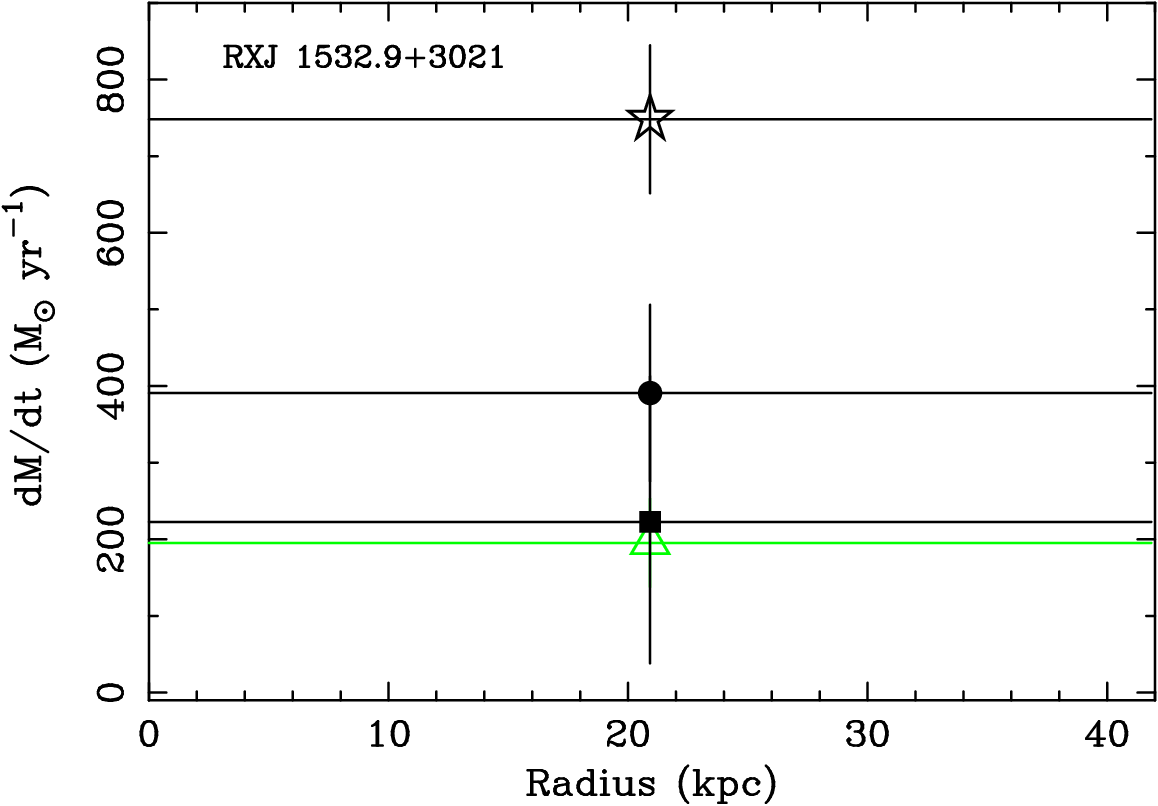}
\vspace{1.0mm}
\includegraphics[angle=-90,width=0.95\columnwidth]{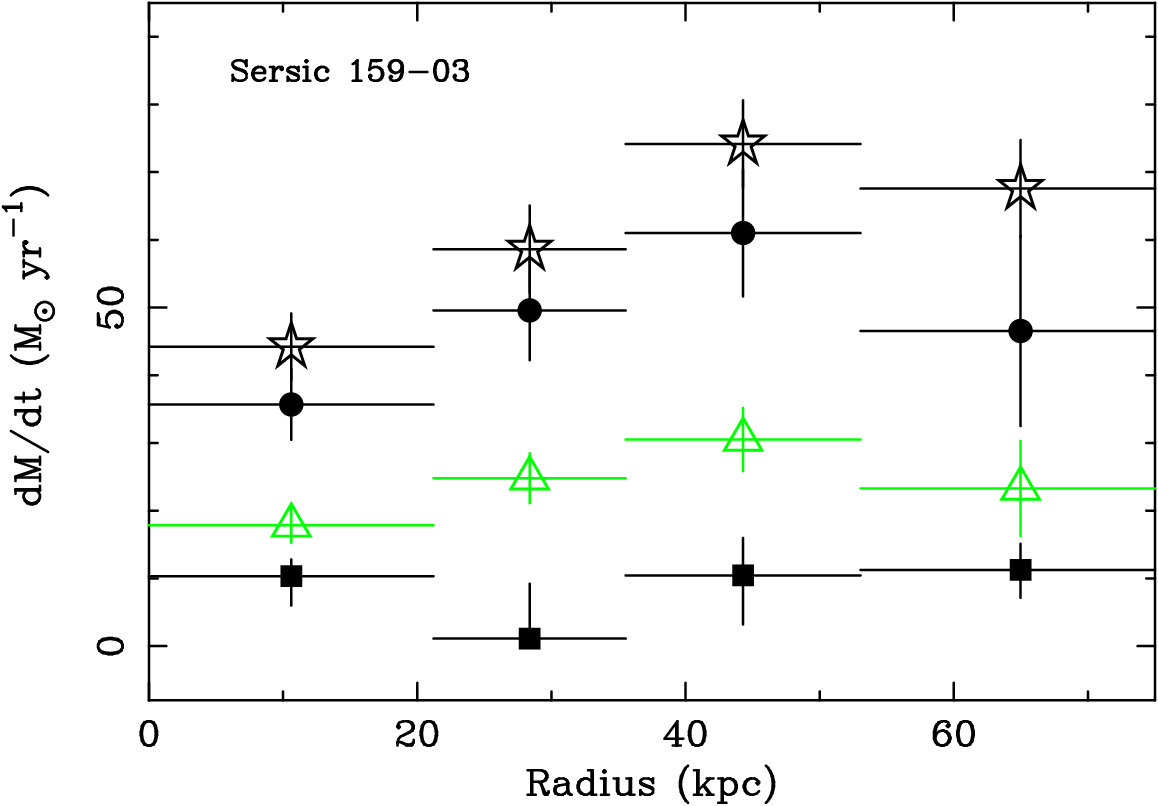}
\vspace{1.0mm}
\includegraphics[angle=-90,width=0.95\columnwidth]{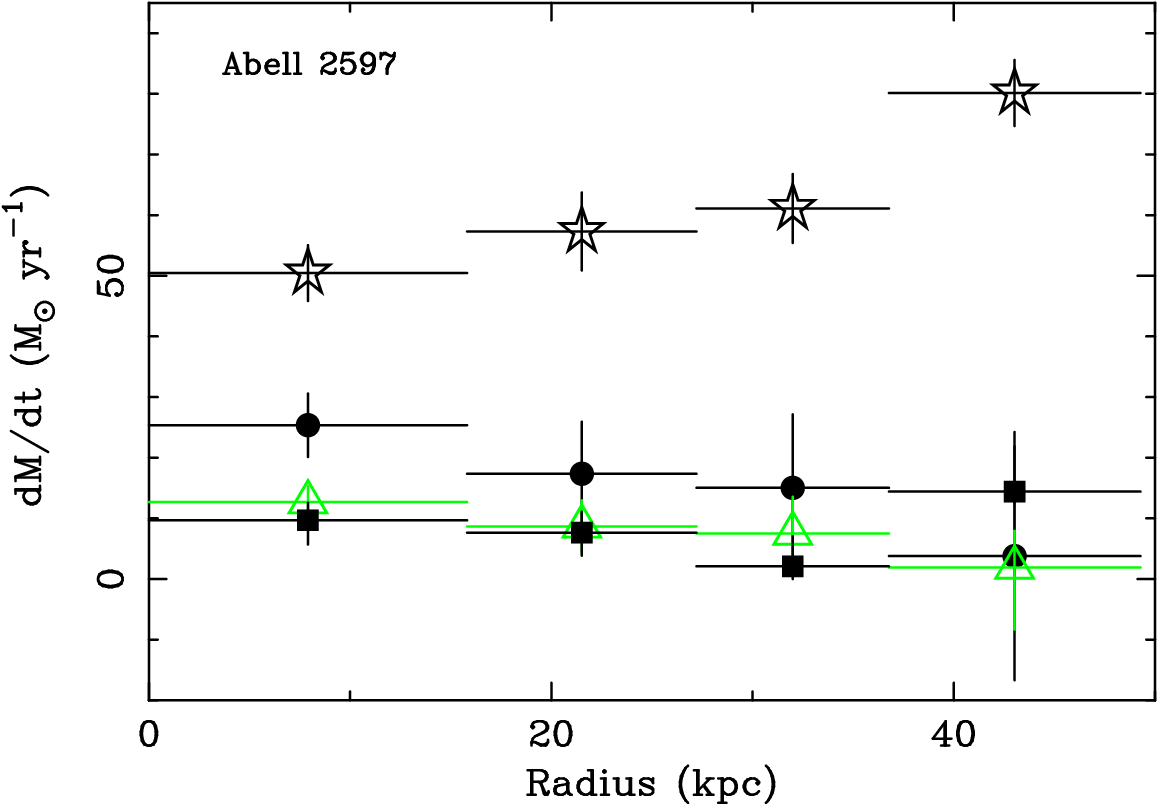}
\caption{Classic mass deposition rates (open stars), 
  mass deposition rates assuming conduction at the Spitzer rate
  (filled circles) and spectral mass deposition rates (filled squares)
  for clusters where $\kappa^{\rm eff} > \kappa_{\rm S}$ (with the
  exception of M87).  The grey open triangles are the cooling rates
  calculated assuming conduction at the Spitzer rate, reduced by a
  factor of 2 to take into account the gravitational contribution to
  the luminosity.}
\label{fig:mdot}
\end{figure*}

In Fig.~\ref{fig:mdot} we plot the classic mass deposition rates,
$\Delta \dot M_{j}^{\mathrm{clas}}$, the mass deposition rates
required assuming conduction at the Spitzer rate, $\Delta \dot
M_{j}^{\mathrm{cool}}$, and the spectral mass deposition rates,
$\Delta \dot M_{j}^{\mathrm{spec}}$, for clusters where the effective
conductivity is above the Spitzer rate.

In five clusters the spectral mass deposition rates are approximately
a factor of two lower than is required for energy balance, even when
$\Delta \dot M_{j}^{\mathrm{cool}}$ is halved to take into account the
gravitational work done on the gas and the enthalpy released by the
gas as it flows across the shell.

\begin{figure*}
\includegraphics[angle=-90,width=0.95\columnwidth]{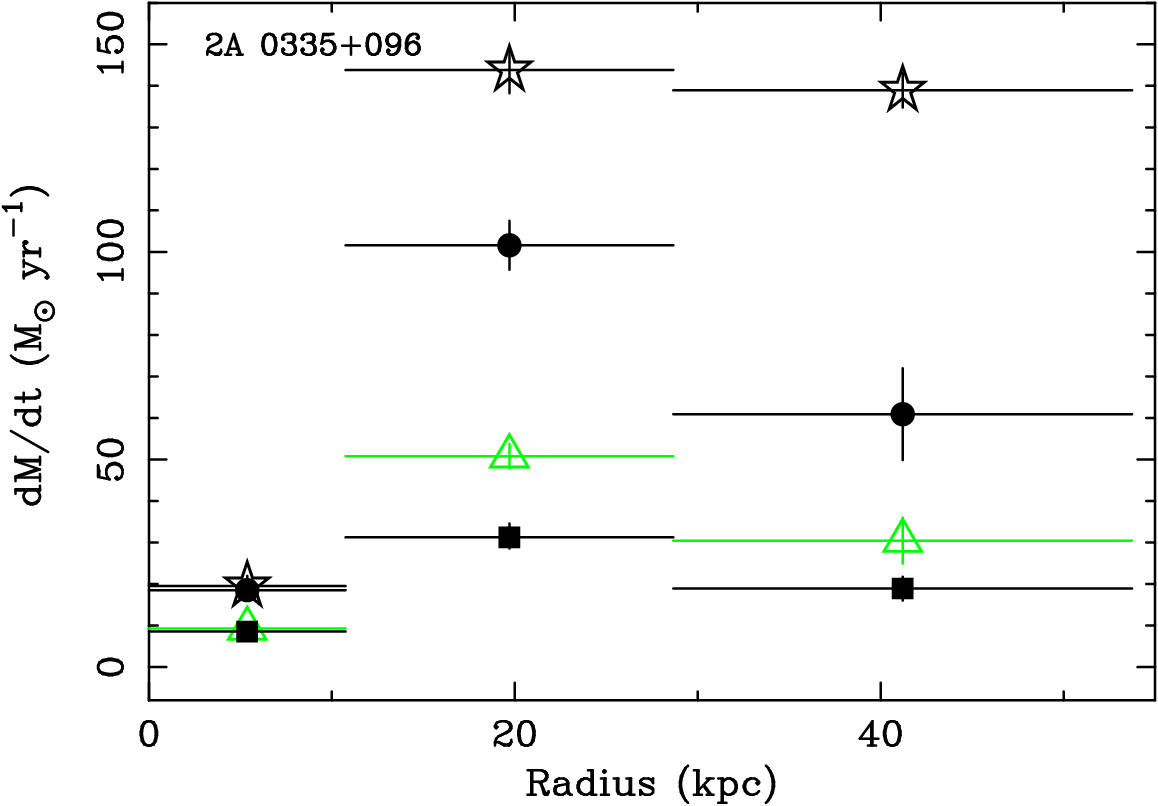}
\vspace{1.0mm}
\includegraphics[angle=-90,width=0.95\columnwidth]{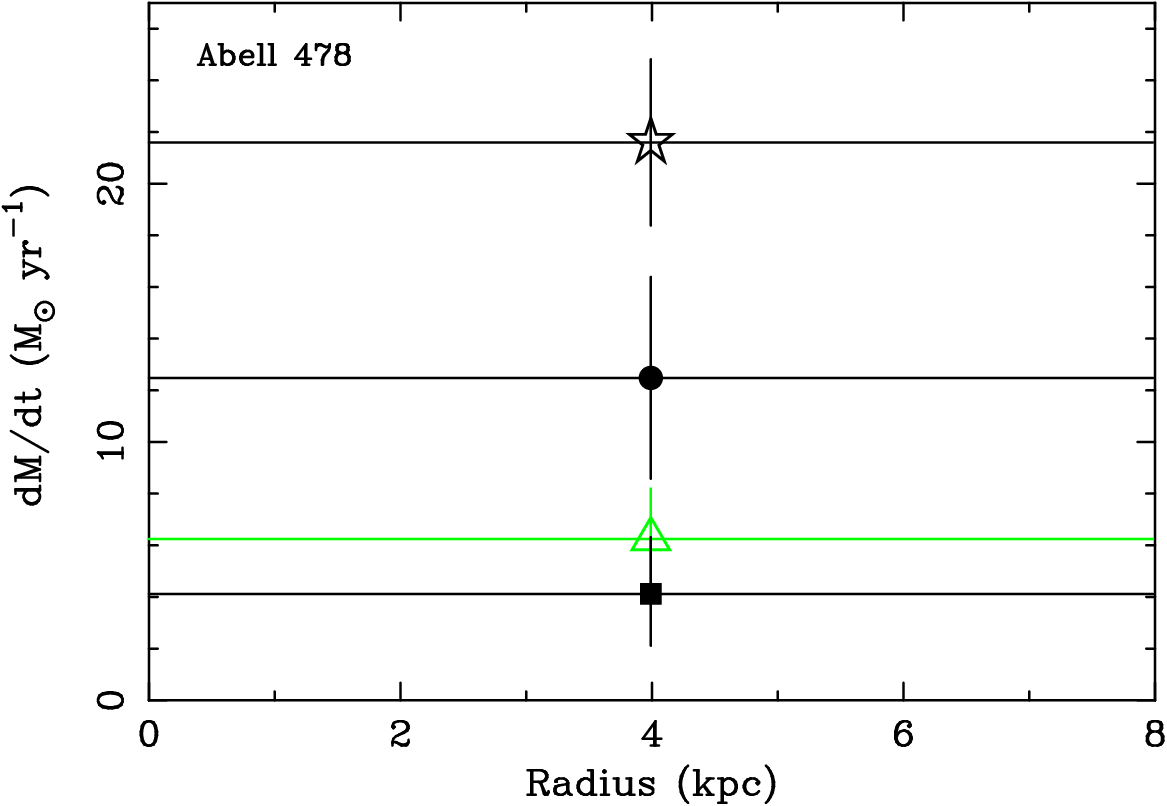}
\vspace{1.0mm}
\includegraphics[angle=-90,width=0.95\columnwidth]{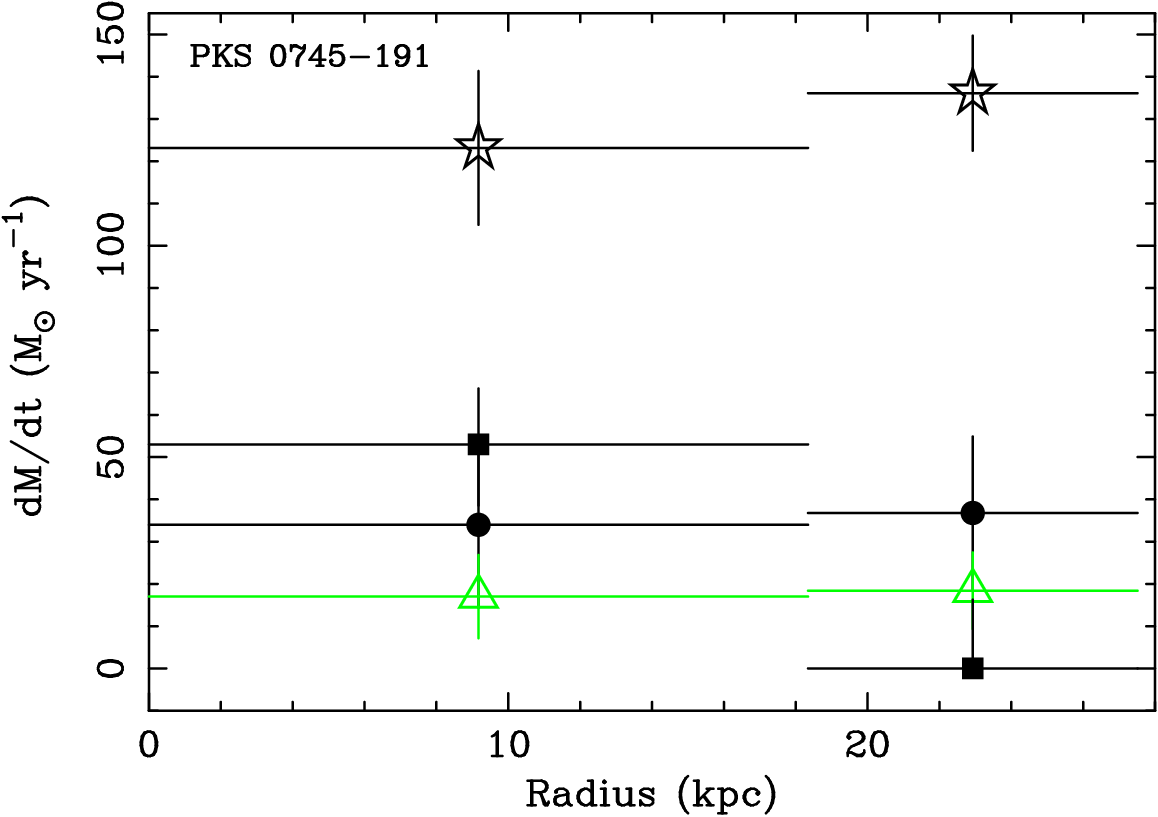}
\vspace{1.0mm}
\includegraphics[angle=-90,width=0.95\columnwidth]{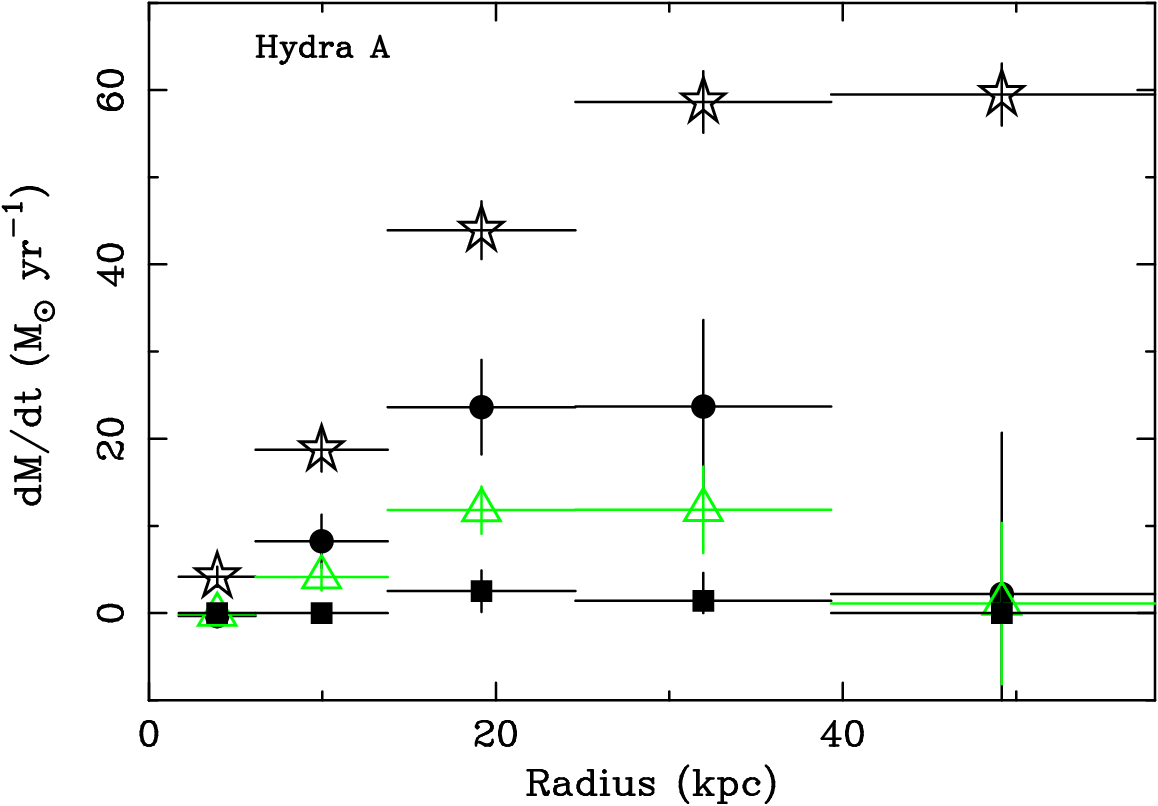}
\vspace{1.0mm}
\includegraphics[angle=-90,width=0.95\columnwidth]{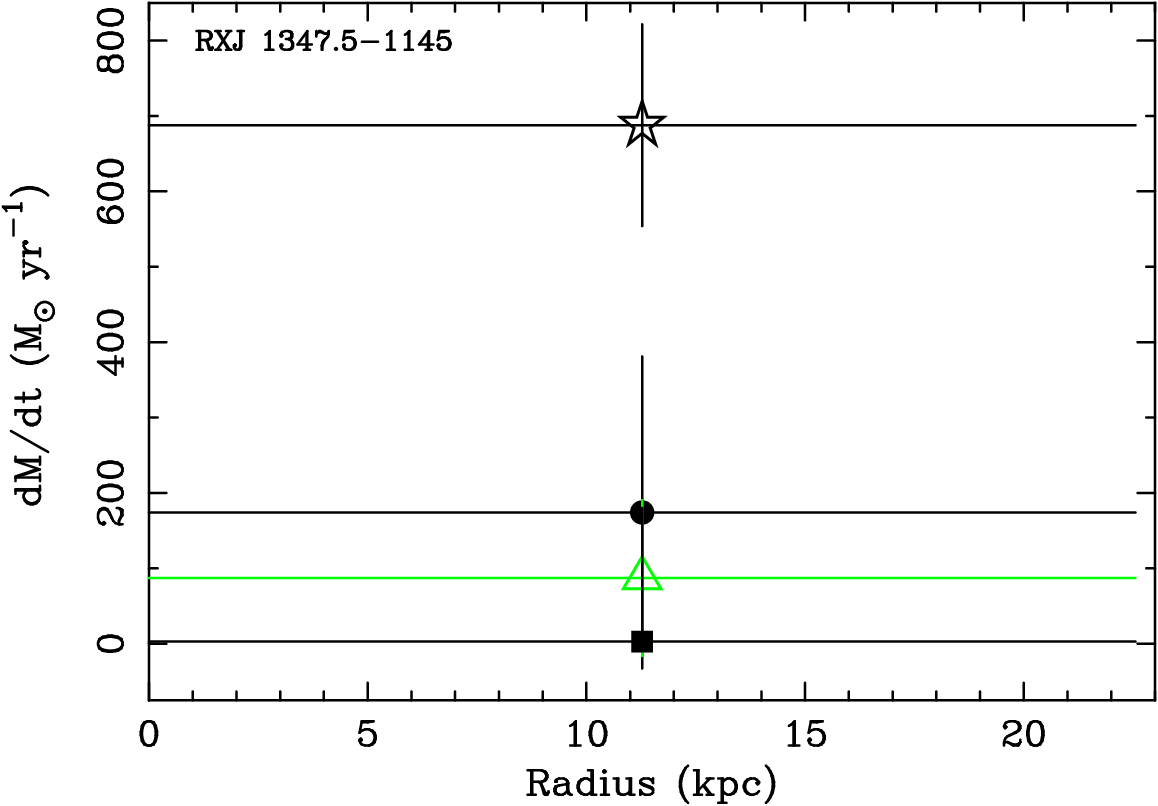}
\vspace{1.0mm}
\includegraphics[angle=-90,width=0.95\columnwidth]{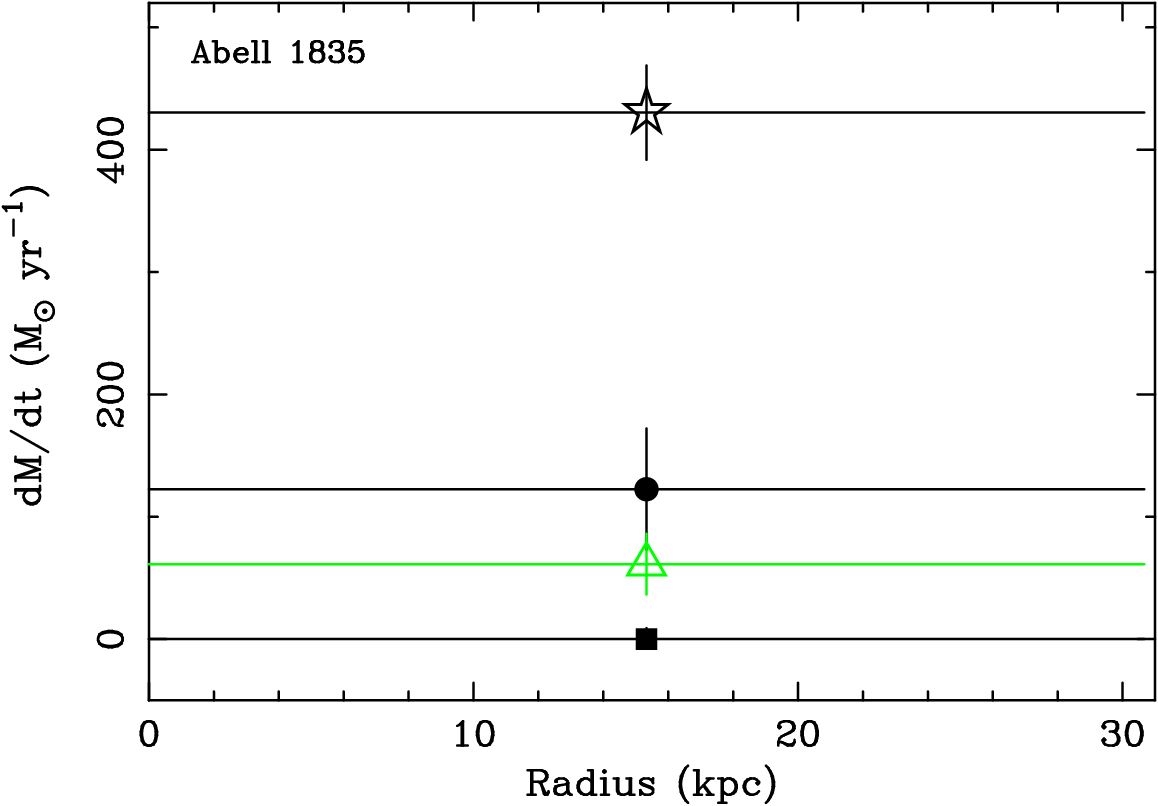}
\vspace{1.0mm}
\includegraphics[angle=-90,width=0.95\columnwidth]{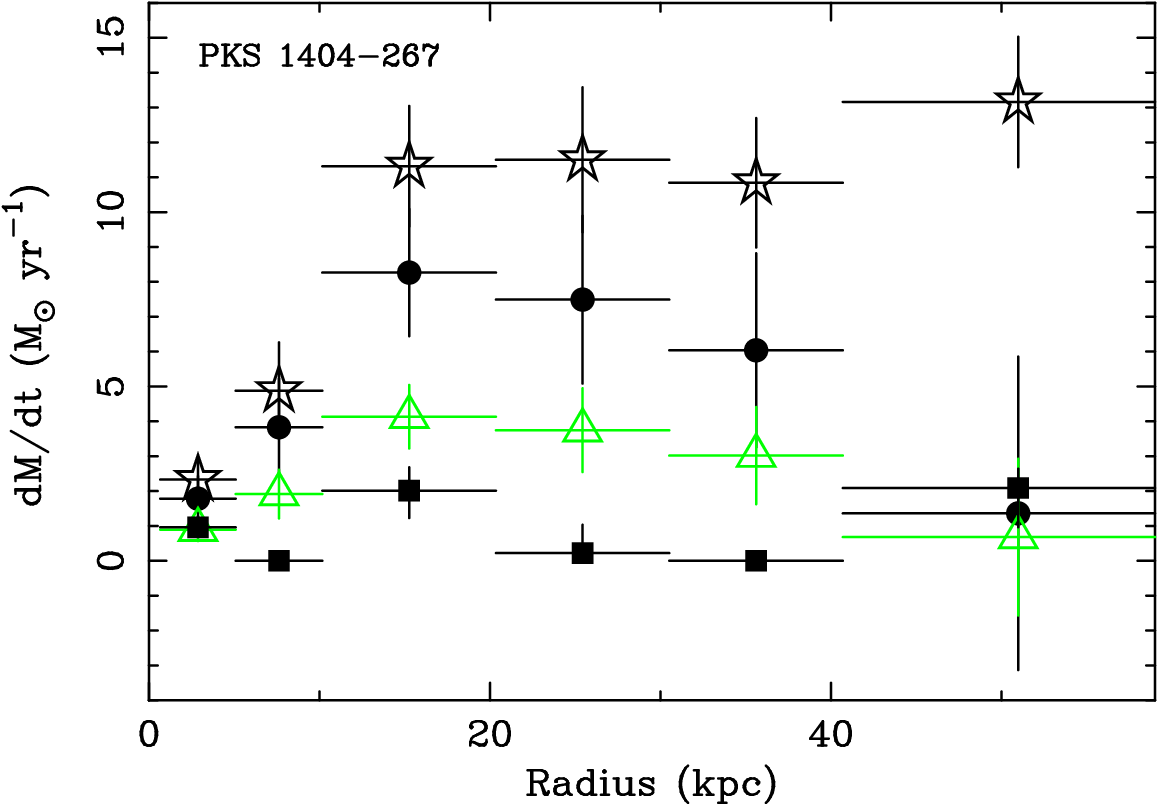}
\vspace{1.0mm}
\includegraphics[angle=-90,width=0.95\columnwidth]{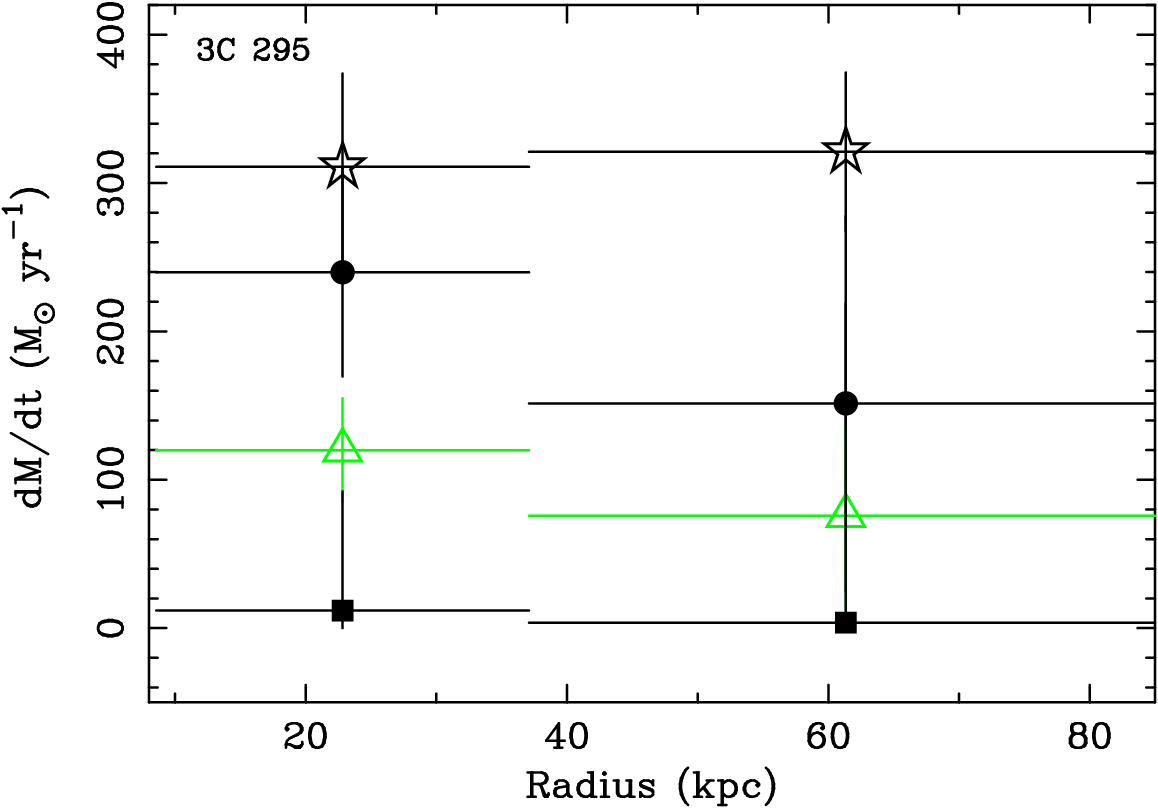}
\end{figure*}

\begin{figure*}
\includegraphics[angle=-90,width=0.95\columnwidth]{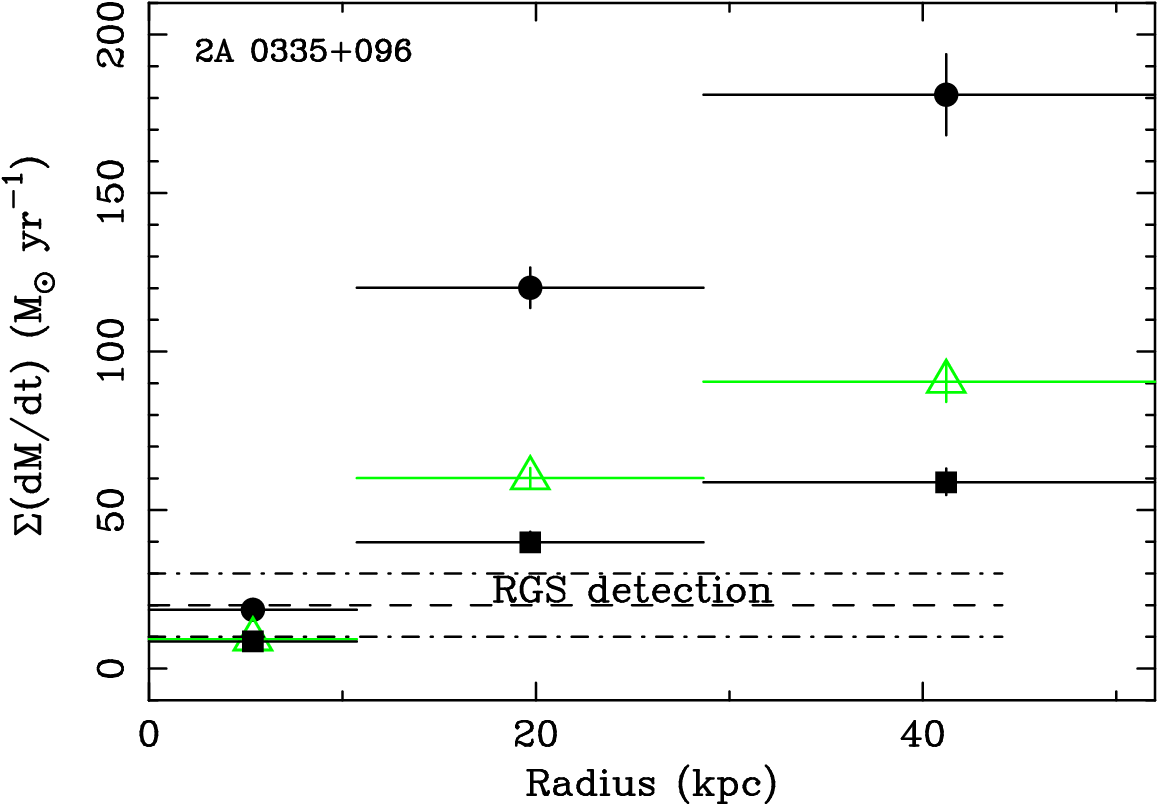}
\vspace{1.0mm}
\includegraphics[angle=-90,width=0.95\columnwidth]{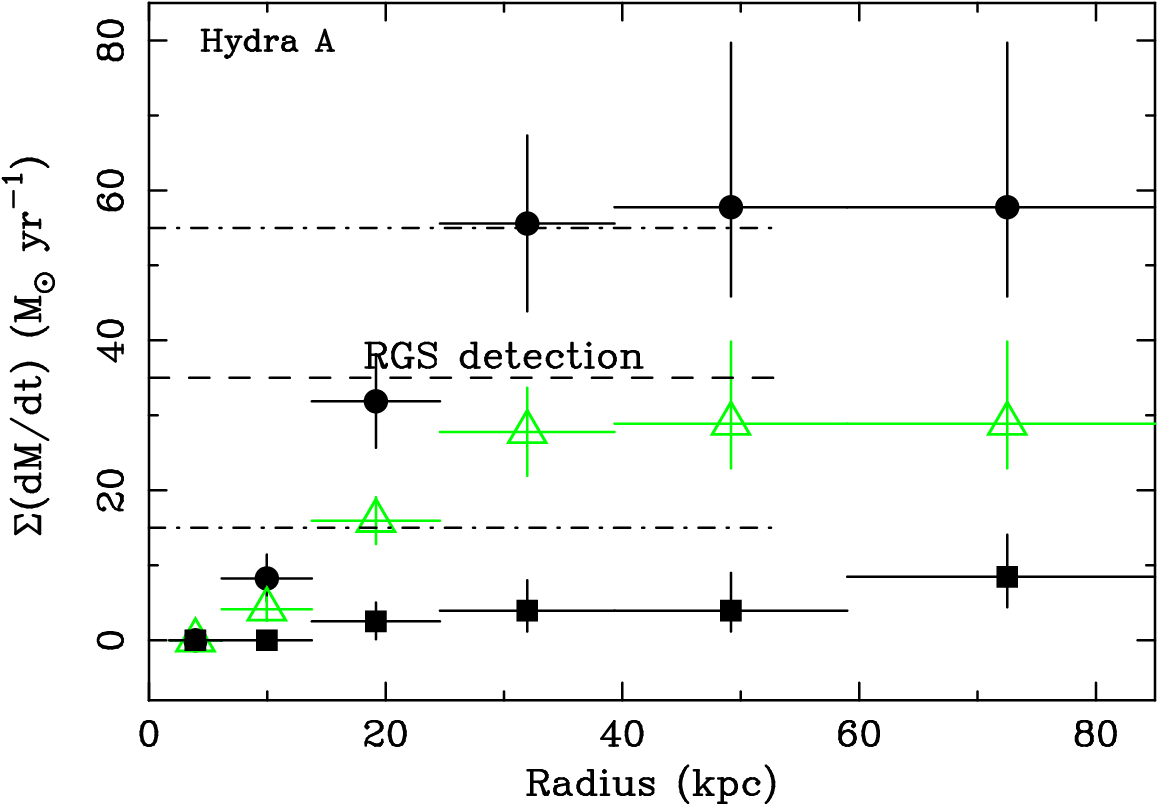}
\vspace{1.0mm}
\includegraphics[angle=-90,width=0.95\columnwidth]{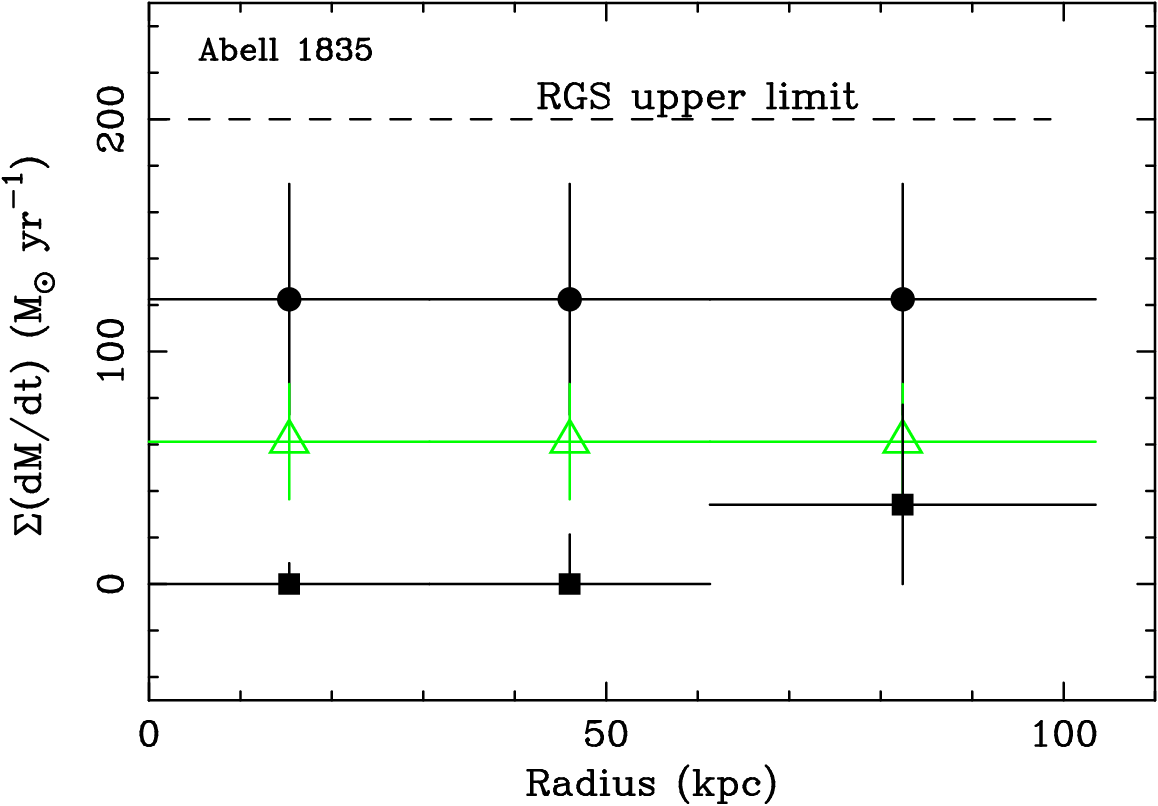}
\vspace{1.0mm}
\includegraphics[angle=-90,width=0.95\columnwidth]{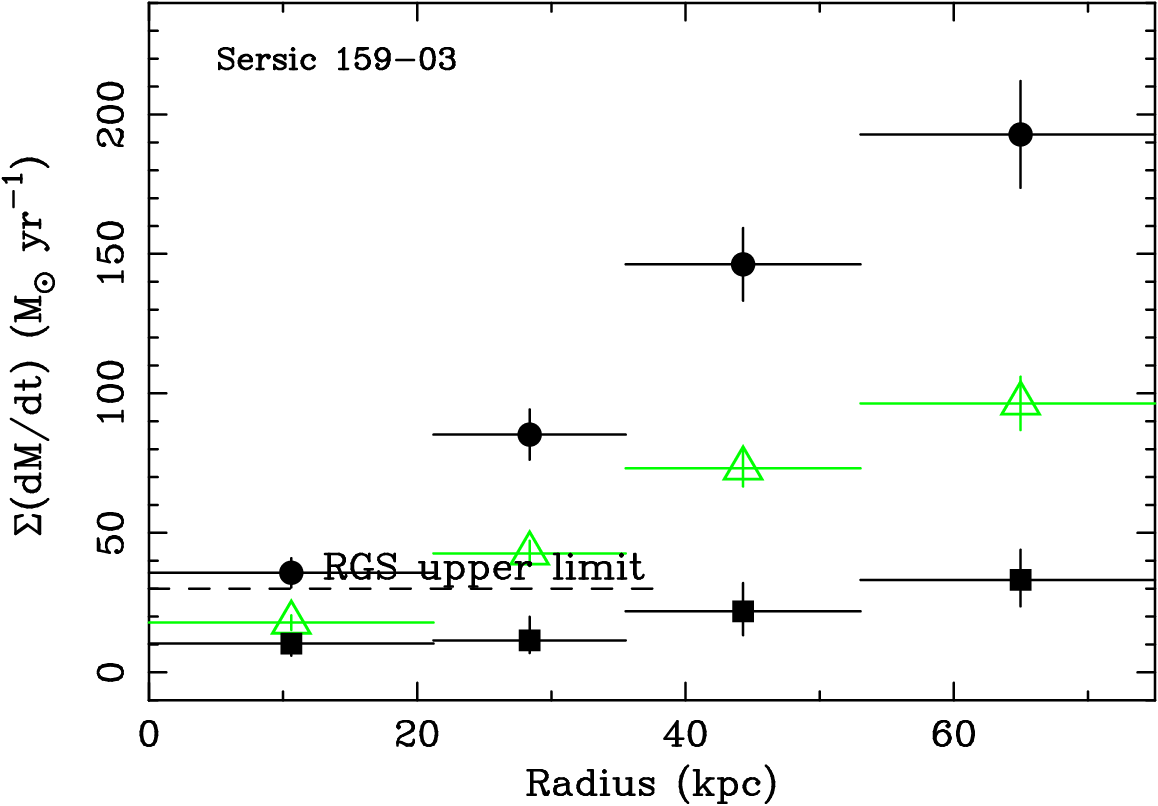}
\vspace{1.0mm}

\caption{As for Fig.~\ref{fig:mdot}, but with the mass deposition
  rates summed out to the radius shown. The upper limits and
  detections of cool gas from RGS data (Peterson et al.\ 2002) are
  shown by dashed lines (with the error bars shown by the dash-dot
  lines). The minimum $\dot M$ value from the three temperature bands
  used by Peterson et al.\ (2002) is plotted in each case.}
\label{fig:peterson}
\end{figure*}

For clusters which have also been studied by Peterson et al.\ (2002),
we plot $\Delta \dot M_{j}^{\mathrm{cool}}$ and $\Delta \dot
M_{j}^{\mathrm{spec}}$ summed out to the cooling radius (see
Fig.~\ref{fig:peterson}), together with the RGS upper limits or
detections of cool gas within the radius shown by the dashed line. For
Abell 1835 and Sersic 159-03 the spectral mass deposition rates found
in this study are consistent with the RGS upper limits. There is some
discrepancy between the Chandra and RGS data for Hydra A and 2A
0335+096, with a larger mass deposition rate found in this study for
2A 0335+096, and a smaller one for Hydra A.

\subsubsection{Conclusions}

Conduction at the Spitzer rate, together with a reduced, unabsorbed
cooling flow equal to the spectral mass deposition rates found here is
unable to account for the entire cooling luminosity in at least five
clusters (Hydra A, Sersic 159-03, 2A 0335+096, Abell 1835 and
PKS1404-267). Conduction at a lower rate will fail to prevent cooling
in even more clusters.

Unhindered conduction is, however, able to \emph{reduce} the mass
deposition rates required for a steady-state energy balance by a
significant factor (2--3 in most clusters) and the addition of a
modest amount of intrinsic absorption may bring the spectral mass
deposition rates into agreement with what is required. (We note too
that the Chandra detector contamination, affecting counts at low
energies, places an instrumental limitation on the accuracy of soft
X-ray emission measurements).

\begin{figure}
\includegraphics[angle=-90,width=0.95\columnwidth]{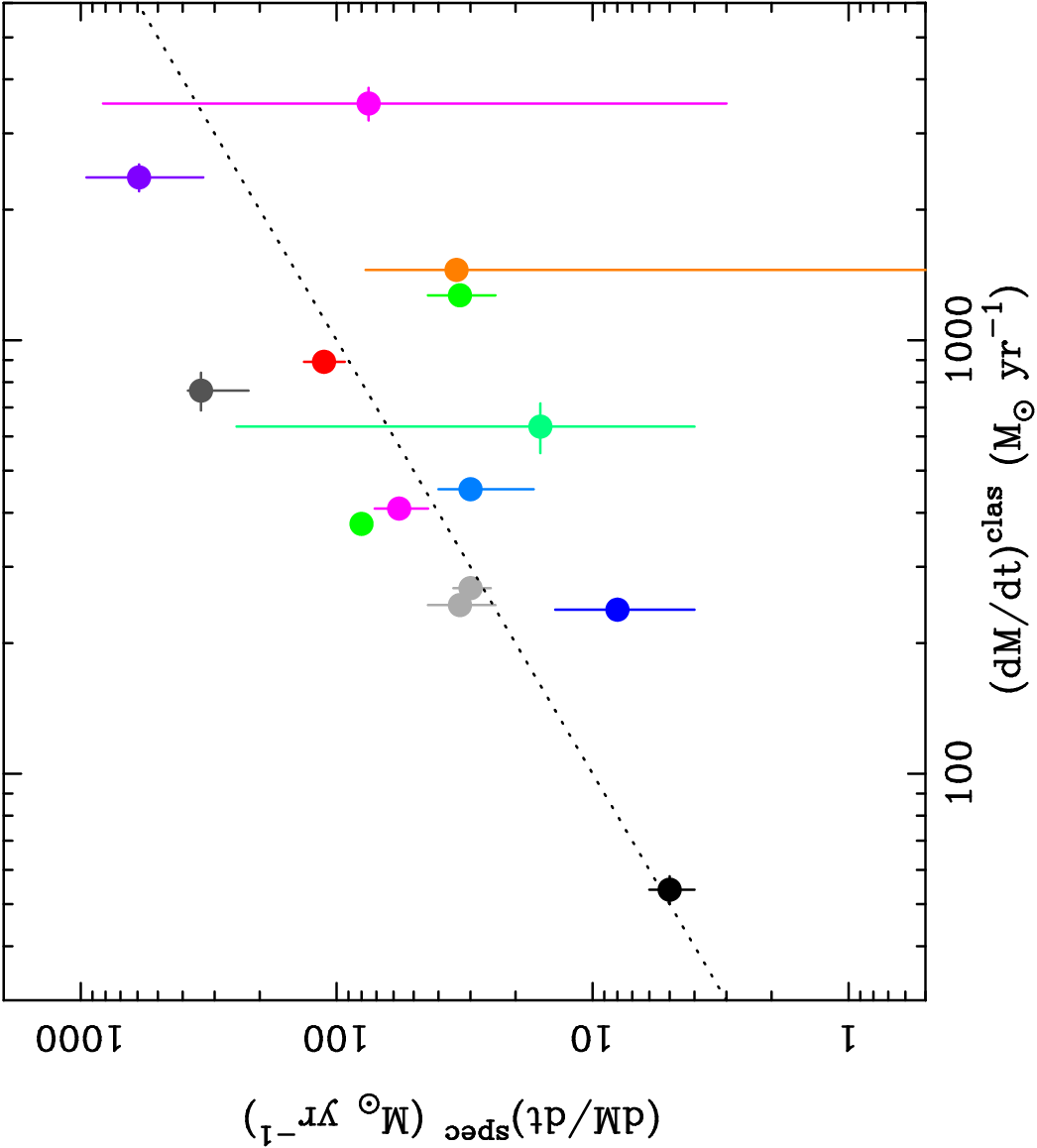}
\caption{The total spectral mass deposition rate
  within the cooling radius plotted against the classic mass
  deposition rate.  The dashed line indicates where the
  spectrally-determined mass deposition rate would be one-tenth of the
  classic mass deposition rate.}
\label{fig:totmdot}
\end{figure}

As a final comment on reduced cooling flows, we plot in
Fig.~\ref{fig:totmdot} the predicted classic and observed spectral
mass deposition rates obtained for the whole sample (tabulated in
Table~\ref{kappas}). The observed values do not assume any conduction
or excess absorption.  We find that $\dot M^{\mathrm{spec}} \sim 0.1
\dot M^{\mathrm{clas}}$. This correlation indicates that there is no
relation between the spectral mass deposition rates found and the
effective conductivity.

\section{Turbulent heat transport}
Cho et al.\ (2003) suggest that turbulent heat transport by mixing may
produce conductivities at least as high as those advocated by Narayan
\& Medvedev\ (2001). They derive a coefficient of turbulent diffusion
\(\kappa_{\mathrm{diff}}=C_{\mathrm{dyn}} L V_{\mathrm{L}}\), where
$V_{\mathrm{L}}$ is the amplitude of the r.m.s. turbulent velocity,
$L$ is the scale length of the turbulent motions and
$C_{\mathrm{dyn}}$ is a constant of the order unity. The turbulent
conductivity is then given by \(\kappa_{\mathrm{turb}}= n_{\rm e} k
\kappa_{\mathrm{diff}}\).

Assuming \(V_{\mathrm{L}} = \alpha c_{\rm s}\), where $c_{\rm s}$ is
the gas sound speed, and \(L = \beta r\), then

\begin{equation}
\kappa_{\rm turb}=\alpha \beta  n_{\rm e} k^{\frac{3}{2}} r \left(\frac{\gamma}{\mu
    m_{\rm H}}\right)^{\frac{1}{2}} T^{\frac{1}{2}}
\label{eqn:kappaturb}
\end{equation}
where $\alpha, \beta \leq 1$. 

Since the dependence of electron number density on temperature varies
from cluster to cluster, we find a general expression for $\kappa_{\rm
  turb}$ by calculating $n_{\rm e}$ as a function of temperature at a
fixed cooling time.
 
For $T \gtrsim 3 \times 10^{7}$ K, thermal bremsstrahlung is the main
emission mechanism and $\epsilon \approx 3.0 \times 10^{-27}
n_{\mathrm h}^{2} T^{\frac{1}{2}} $ erg cm$^{-3}$ s$^{-1}$. For $T
\lesssim 3 \times 10^{7}$ line cooling becomes important and $\epsilon
\approx 6.2 \times 10^{-19} n_{\mathrm h}^{2} T^{-\frac{3}{5}} $ erg
cm$^{-3}$ s$^{-1}$ (McKee \& Cowie\ 1977).  Substituting $\epsilon$
into Equation~\ref{eqn:tcool} we can re-write $t^{\rm cool}$ in terms
of the temperature and electron number density in the two temperature
regimes

\begin{equation}
t^{\rm cool}_{j} \approx 2.0 \times 10^{10}
\left(\frac{n_{\rm e}}{10^{-3}\hspace{0.1cm} \rm cm^{-3}}\right)^{-1}
\left(\frac{T}
{10^{7}\hspace{0.1cm}  \rm K}\right)^{\frac{1}{2}}\rm yr, 
\label{eqn:tcool2}
\end{equation}
\hspace{6.0cm} $T \gtrsim 3 \times 10^{7}$ K \\
and
\begin{equation}
t^{\rm cool}_{j} \approx 0.5 \times 10^{10}
\left(\frac{n_{\rm e}}{10^{-3}\hspace{0.1cm} \rm cm^{-3}}\right)^{-1}
\left(\frac{T}
{10^{7}\hspace{0.1cm}  \rm K}\right)^{\frac{8}{5}}\rm yr,
\label{eqn:tcool3}
\end{equation}
\hspace{6.0cm} $T \lesssim 3 \times 10^{7}$ K.\\

Substituting Equations~\ref{eqn:tcool2} and~\ref{eqn:tcool3} into
Equation~\ref{eqn:kappaturb} we find the following expressions for the
turbulent conductivity at a fixed cooling time within each cluster

\begin{equation}
\kappa_{\rm turb} \approx 8 \times 10^{12}
\left(\frac{T}{10^{7} \hspace{0.1cm} \rm K}\right) \ergpspcmpK,
\end{equation}
\hspace{6.0cm} $T \gtrsim 3 \times 10^{7}$ K \\
and
\begin{equation}
\kappa_{\rm turb} \approx 2 \times 10^{12}
\left(\frac{T}{10^{7} \hspace{0.1cm} \rm K}\right)^{2.1} \ergpspcmpK,
\label{eqn:turb}
\end{equation}
\hspace{6.0cm} $T \lesssim 3 \times 10^{7}$ K \\
\\
for $\alpha\beta$=1, $\gamma=\frac{5}{3}$, $t^{\rm cool}$= 1Gyr and
$L$ = 20 kpc.  (In the lower plot in Fig.~\ref{fig:sample} we see that
$L \approx$ 20 kpc at $t^{\rm cool}$=1Gyr for the clusters in the
sample). We plot this curve on the effective conductivity plot in
Fig.~\ref{fig:sample}. Turbulent heat transport may be important,
depending on the actual values of $\alpha$ and $\beta$.

We also compute a $\kappa_{\rm turb}$ profile for each cluster
individually (shown in Fig.~\ref{fig:kappa}) using
Equation~\ref{eqn:kappaturb}. The $\kappa_{\rm turb}$ and $\kappa^{\rm
  eff}$ profiles are strikingly similar if \(\alpha \beta \sim
\frac{1}{2}-\frac{1}{10}\). This may be explained by the scaling
involved. We have that
\begin{equation}
\frac{\kappa^{\rm eff}}{\kappa_{\rm turb}}
\propto \frac{n_{\rm e}}{\rm{d}T/\rm{d}r} = \frac{n_{\rm e} r}{T}
\end{equation}
for power-law temperature profiles and bremsstrahlung emission.
Typically $n_{\rm e} r \sim {\rm constant}$ and $T$ varies at most by
a factor of $\sim$ 3. The similarity could therefore be a coincidence.
More interestingly, the scaling (i.e. the temperature and density
profiles observed) could result from turbulence {\it being} the heat
transport process operating in cluster cores.

One problem with the last possibility is that the quasi-linear optical
filaments seen in the Perseus cluster (Conselice et al.\ 2001) argue
in that case against the gas being strongly turbulent (Fabian et al.\
2003c). A mild circulation pattern with eddies of a few kpc driven by
buoyant radio bubbles and oscillatory motion of the central galaxy may
however be consistent.

\section{Radio and optical data}
\begin{figure}
\centering
\includegraphics[angle=-90,width=0.95\columnwidth]{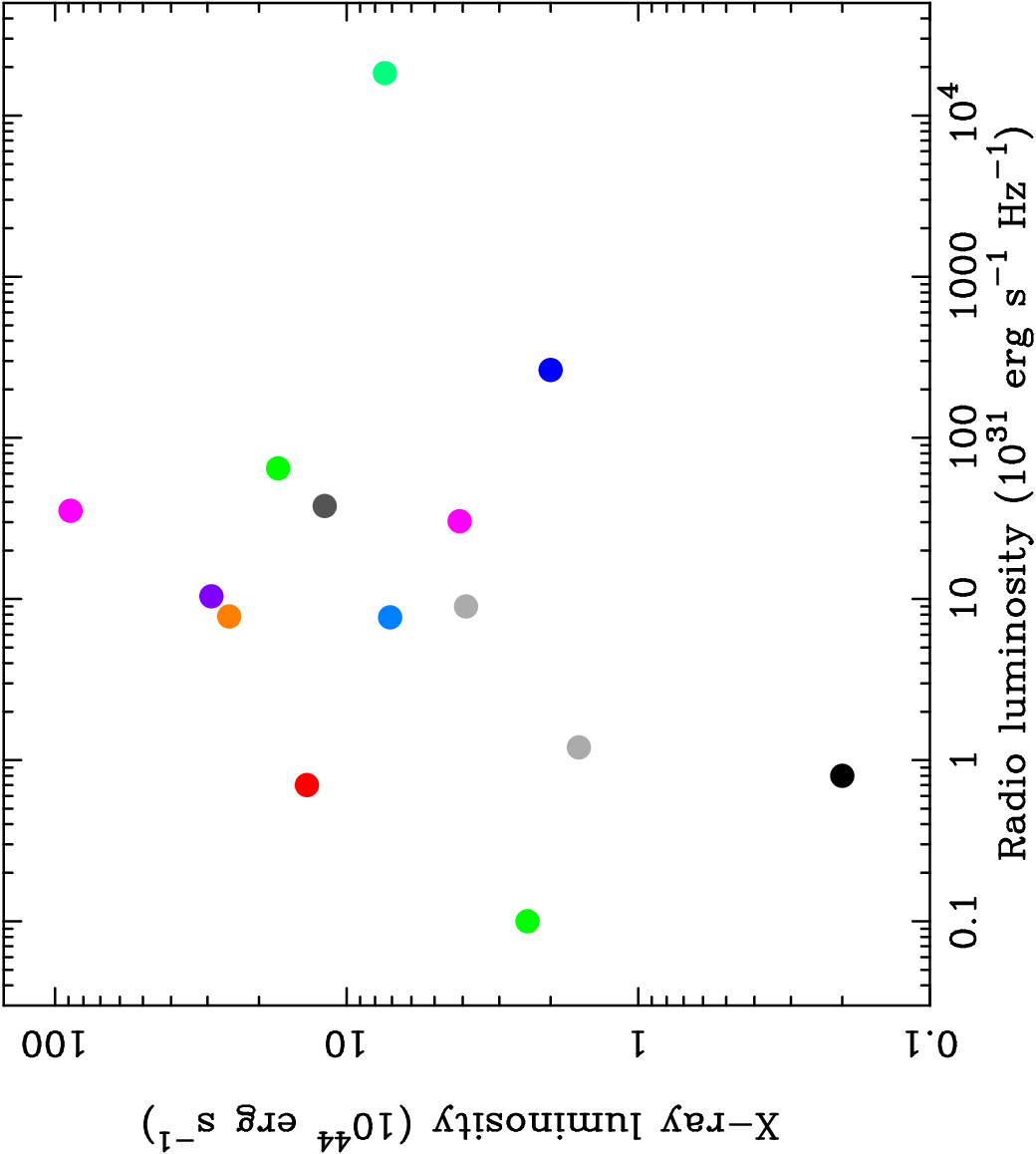} 
\caption{The total X-ray luminosity
  emitted from within the cooling radius of each cluster against the
  total 1.4 GHz radio power within 15 arsec of the cluster X-ray
  centre.}
\label{fig:radiolum}
\end{figure}

If the intracluster gas is heated by the central radio source, then we
would expect to see some degree of correlation between the heating
rate required to prevent cooling and the radio luminosity of the
central source. In Table~\ref{radio} we show the radio luminosities
measured at 20 cm (1.4 GHz), obtained from the NVSS survey (Condon et
al.\ 1998). In each case the radio sources detected within 15 arcsec
of the X-ray centre (see Table~\ref{clusters}) are tabulated. We plot
the rate of energy loss from the cluster within the cooling radius
against the total 1.4 GHz radio luminosity in Fig.~\ref{fig:radiolum}.

We see that there is no obvious correlation between the radio power
observed and the radio power required to prevent cooling in these
clusters. Although the radio luminosity is not directly related to the
mechanical power of the radio lobes, it is unlikely that the same
$P$d$V$ work can be done on clusters with similar X-ray luminosities
when their radio source luminosities vary by over 5 orders of
magnitude.

Observations in the UV, optical and infrared suggest that a small
cooling flow may be required to fuel the observed star formation
rates, emission line nebulosities and molecular gas masses (Donahue
2000; Edge 2001; Edge et al.\ 2002) in the centres of cooling flow
galaxy clusters.  In Table~\ref{radio} we indicate whether or not
optical emission lines have been observed in the object (Crawford \&
Fabian\ 1992; Crawford et al.\ 1999; Peres et al.\ 1998).

It is interesting that there is no evidence for optical emission lines
in Abell 2029 (Johnstone et al.\ 1987; McNamara \& O'Connell 1989),
given that conduction at $\kappa \sim$ (0.3--0.4) $\kappa_{\rm S}$,
the range in plasma conductivity advocated by Narayan \& Medvedev
2001, can prevent cooling from taking place right into the inner few
kpc of this object.

We also note that molecular gas is generally found only within the
central $\sim$ 20 kpc of clusters (Donahue \& Voit 2003); indicating
that if we are indeed detecting the cool sink produced by a reduced
cooling flow, then any heat process operating in these objects fails
to stop cooling only in the innermost region. We find that, in
general, the effective conductivity profiles rise above the Spitzer
curve towards the centres of the clusters.

\begin{table*}
\centering
\begin{tabular}{lccccc} 
\hline\hline
Cluster & Position (J2000)& Flux& Total Radio Luminosity &X-ray Luminosity & Optical lines \\
&RA\hspace{0.5cm} Dec &(mJy)&(10$^{31}$ erg s$^{-1}$ Hz$^{-1}$)& (10$^{44}$ erg s$^{-1}$)& \\
\hline
2A0335+096 & (03 38 40.62) (+09 58 12.2)    & 36.7  & 0.1  & 2.3  &  YES\\
& (03 38 40.06) (+09 58 11.0) & 1.8 &  & & \\
A478 & (04 13 25.18) (+10 27 54.1) & 36.9  & 0.7  & 13.6  &  YES\\
& (04 13 20.04)(+10 27 50.7) &  1.5 & & &\\
PKS 0745-191 & (07 47 31.35) (07 47 30.0) & 2372.0 & 64.7  & 17.2 & YES \\
& (07 47 30.03) (-19 17 30.6) & 83.7 & & & \\
Hydra A & (09 18 05.78) (-12 05 41.3) & 40849.9 & 263.9  & 2.0  & YES\\
& (09 18 00.03) (-12 05 40.6)&  1278.8& && \\
M87 & (12 30 49.46) (+12 23 21.6) & 138487.0 & 5.7 &-- & YES\\
& (12 30 00.03) (+12 23 20.3) & 4858.7  & & & \\
RXJ 1347.5-1145 & (13 47 30.67) (-11 45 08.6) & 45.9 & 35.3   & 89.0 & -- \\ 
& (13 47 30.04) (-11 45 00.7) & 1.5 & & & \\
A1795 & (13 48 52.43) (+26 35 33.6) & 924.5 & 9.0  &  3.7  & YES   \\
& (13 48 50.03) (+26 35 30.6) & 27.7 & && \\
A1835 & (14 01 02.05) (+02 52 41.0) & 39.3& 7.8   & 26.0 & YES\\
& (14 01 00.04) (+02 52 40.7) & 1.6 & & & \\
PKS 1404-267 & (14 07 29.83) (-27 01 04.2) & 645.5 & 0.8  & 0.17  & --\\
& (14 07 20.03) (-27 01 00.6) & 22.8 & & &\\
3C295 & (14 11 20.63) (+52 12 09.0) & 22720.1 & 1.83$\times10^{4}$ & 7.2 & --\\
& (14 11 20.05) (+52 12 00.6) & 681.6 & & & \\
A2029 & (15 10 55.87) (+05 44 39.3) & 527.8 & 7.7  & 6.9 & NO\\
& (15 10 50.03) (+05 44 30.6) & 18.2 & & & \\
RXJ 1532.9+3021 & (15 32 53.77) (+30 20 59.8) & 22.8 & 10.4  & 27.5 & YES \\ 
& (15 32 50.06) (+30 20 50.8) & 0.8 & & & \\
Cygnus A & (19 59 28.1) (40 44 0.5.00)  & 5.7$\times10^{5}$ & 4.2$\times10^{3}$ & -- & YES \\
& & & & & \\
A2390 & (21 53 36.81) (+17 41 44.8) & 235.3 & 37.8   & 11.4  & YES \\
& (21 53 30.06) (+17 41 40.6) & 8.3& & & \\
Sersic 159-03 &(23 13 58.60) (-42 43 38.0) &156.7 & 1.2  & 1.9 & YES\\
& & & & & \\
A2597 & (23 25 19.82) (-12 07 28.6) & 1874.6 & 30.4  & 4.0 & YES \\
& (23 25 10.03) (-12 07 20.6) & 56.2 & & & \\
\\
\hline
\end{tabular}
\caption{1.4 GHz radio fluxes from within 15 arcsec of the X-ray
  peak (from the NVSS survey; Condon et al.\ 1998), radio
  luminosities calculated from the radio fluxes and X-ray
  luminosity from within the cooling radius. The 1.4 GHz radio fluxes for Cygnus A and Sersic 159-03
  were found using 4.85
  GHz data from the Green Bank and PMN (Wright et al.\ 1994)
  surveys respectively, assuming $I_{\nu} \propto \nu^{-\alpha}$,
  with $\alpha$=0.8. In the last column we
  indicate the presence, absence or lack of information on optical
  emission lines in the cluster (Crawford \& Fabian\ 1992; Crawford et al.\ 1999; Peres et
  al.\ 1998). }
\label{radio}
\end{table*}

\section{Discussion}
We have carried out spatially-resolved, deprojected analyses of the
cooling flow regions in a sample of galaxy clusters and find effective
conductivity profiles which lie both above and below the Spitzer
curve.  If the plasma conductivity in galaxy clusters is close to the
unhindered Spitzer value, then conduction will play a important role
in reducing cooling flows.

For clusters with effective conductivity values above the Spitzer
curve, we calculated the rate at which gas must cool out of the
ambient plasma, assuming unhindered conduction, and found spectral
mass deposition rates at least a factor of two lower than is required
for energy balance in five objects.

The possibility that the plasma conductivity is as high as 0.1--1.0
$\kappa_{\rm S}$ continues to be debated in the literature, both from
an observational viewpoint (Ettori \& Fabian; Vikhlinin et al.\ 2001;
Markevitch et al.\ 2003; Nath 2003) and theoretically (Chandran \&
Cowley 1998; Narayan \& Medvedev 2001; Chandran \& Maron 2003).  We
also point out that a certain amount of fine-tuning is required for
the conduction model to be successful. For an energy balance between
conduction and cooling to exist in each object, the suppression factor
must vary from cluster to cluster. However, it is not unreasonable to
expect there to be some variation in the suppression of conduction in
different objects; in particular since rotation measures show that the
length over which fields are ordered varies between clusters (Taylor
et al.\ 2002).  Also, mass inflow itself may pull the field lines
radially, increasing the conductivity (Bregman \& David\ 1988; Fabian
et al.\ 2002b).

Since the model requires some finite $\dot M$ in most clusters, we
must consider whether or not it is physically plausible for an
inhomogeneous cooling flow to exist in a conducting medium. If the gas
which cools out of the flow `pulls' magnetic field lines around it as
it condenses then the cool blobs may be magnetically isolated from the
surrounding hotter, conducting medium. We note that a certain amount
of cool gas may need be deposited in the centres of galaxy clusters in
order to account for the enhanced star formation rates (Crawford et
al.\ 1999) and molecular gas masses detected (Edge\ 2001; Edge et al.\ 
2002).

In this paper we concentrate on the energetic feasibility of balancing
bremsstrahlung emission with conductive heat transfer. For the
conduction model to be successful, this balance must be stable; or at
least quasi-stable over the timescales considered. The conduction
scenario may be a self-regulating process. Radiative cooling occurs
isobarically at each radius and so a perturbation causing the
temperature to drop leads to an increase in the cooling rate. We might
then expect the gas to cool dramatically since conductivity decreases
with temperature. However, the decline in temperature leads to an
increase in the temperature gradient, which is beneficial for
conductive heat transfer. The situation is complex and time-dependent
simulations are needed to understand how the intracluster gas may
evolve. Past work has shown that steady state solutions to the energy
equation are unable to reproduce the large temperature gradients
observed (Bregman \& David 1988; Bertschinger \& Meiksin 1986; Meiksin
1988).  Zakamska \& Narayan (2002), on the other hand, argue that the
balance is stable with the observed temperature and density profiles.
Kim \& Narayan (2003) suggest that thermal instabilities are
unimportant on timescales $\sim$ 2--5 Gyr.

\begin{figure*}
  \vspace{1.0mm}
  \includegraphics[angle=-90,width=0.95\columnwidth]{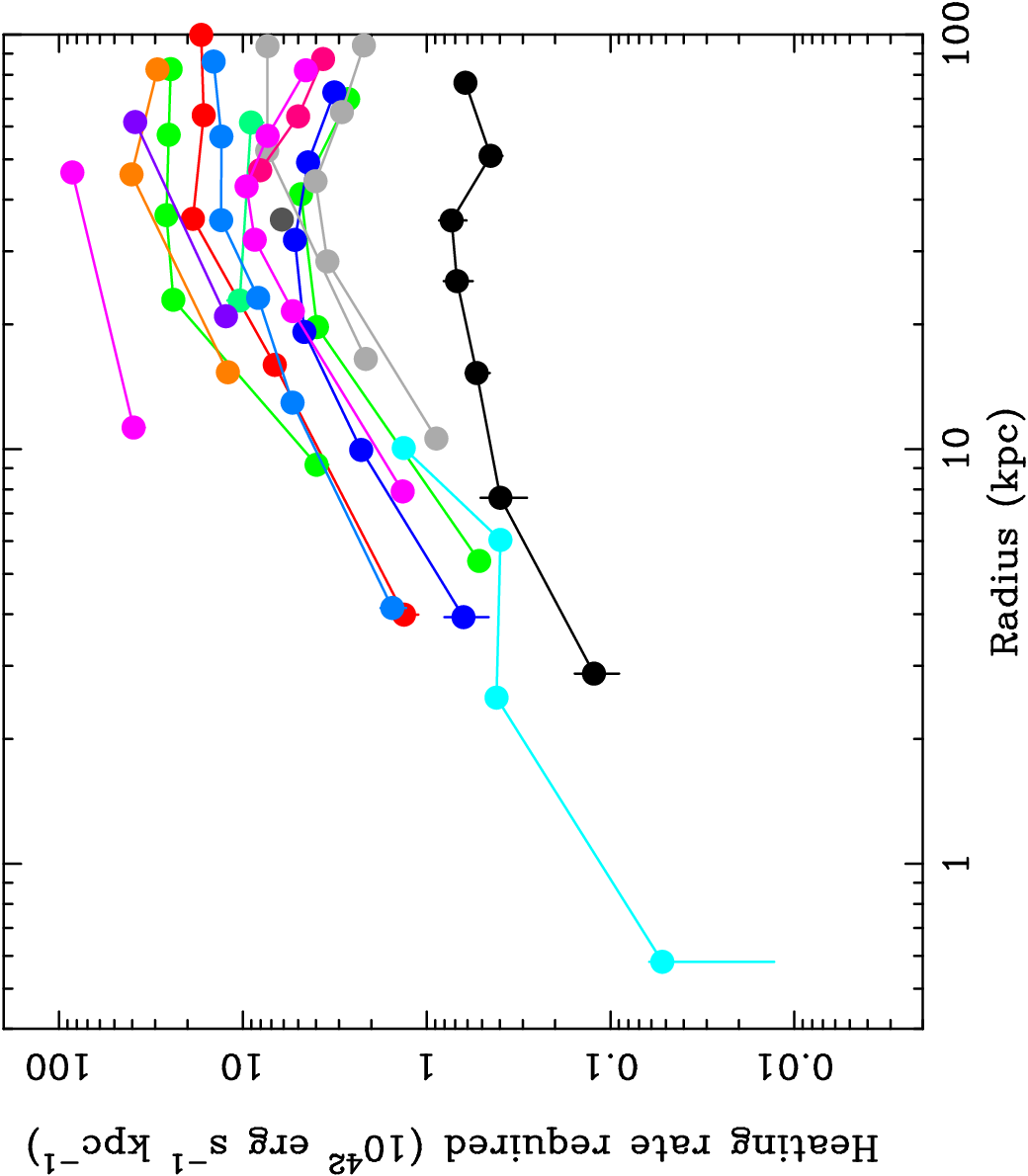}
\vspace{1.0mm}
  \includegraphics[angle=-90,width=0.95\columnwidth]{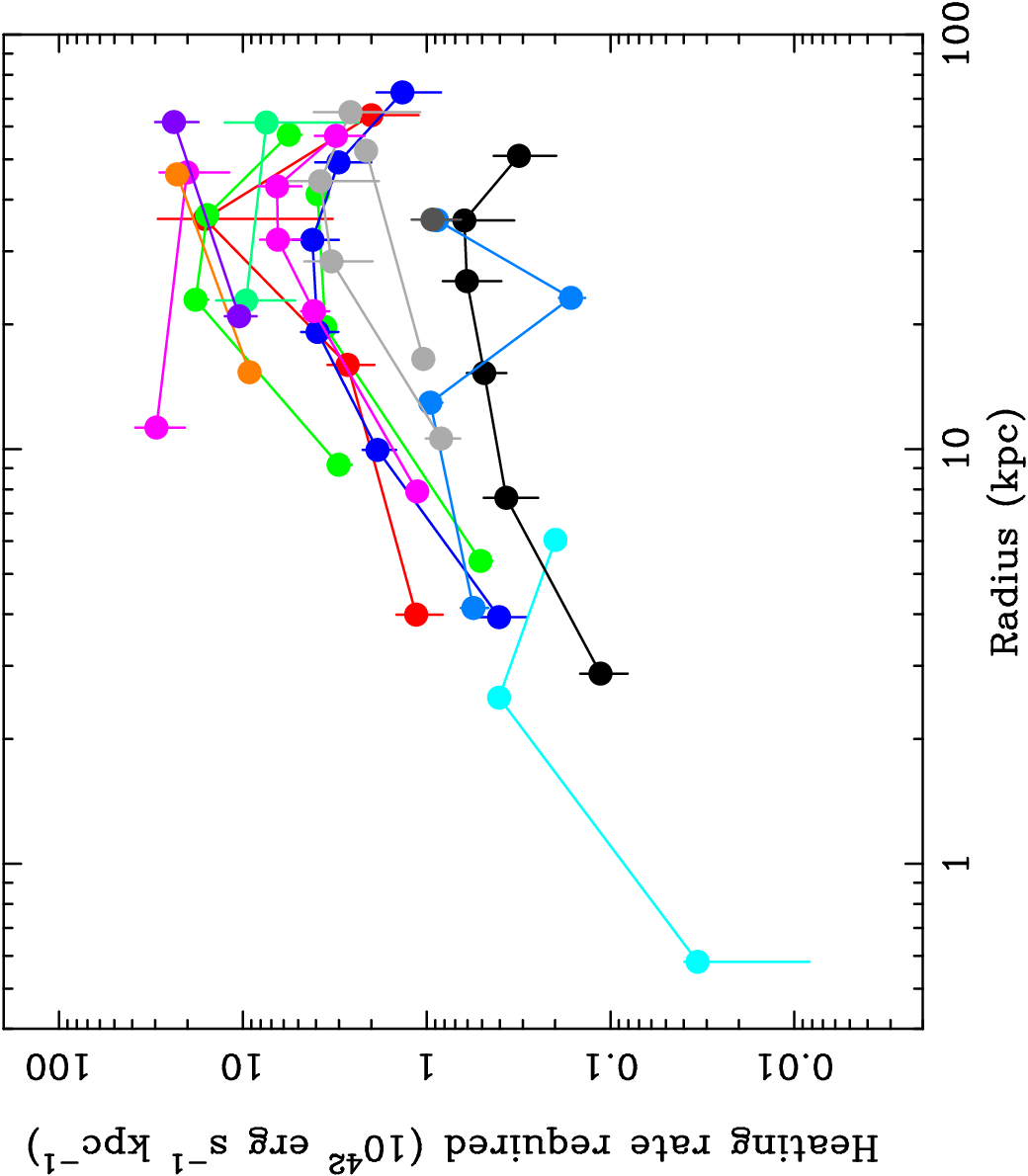}
\caption{Heating rate required per kpc at increasing radii within the
  central 100 kpc, with $\kappa=0$ (left) and $\kappa=\kappa_{\rm
    S}/3$ (right). Cygnus A does not appear in the plot on the right
  since conduction below one-third the Spitzer rate offsets cooling
  completely in this object.}
\label{fig:heatrateperkpc}
\end{figure*}

If there is some heating process preventing cooling flows from
developing then it must be distributed. We see from
Fig.~\ref{fig:heatrateperkpc} (left plot) that the heat which must be
transferred per kpc to shells at increasing radii is approximately
constant between 20 and 100 kpc, with a rise in the innermost region.
If the cluster is heated by the central AGN then the $P$d$V$ work done
in blowing bubbles through the intracluster medium must lead to the
widespread deposition of energy.  How this heat is transported is
unclear (viscous dissipation of the sound waves produced by the
bubbles is a possibility; Fabian et al.\ 2003b). It has been suggested
that conduction and AGN heating work together in galaxy clusters
(Brighenti \& Matthews 2003, Ruszkowski \& Begelman 2002), and, in
particular, that conduction prevents cooling in the outer regions,
with AGN heating dominating in the inner regions (Kim \& Narayan
2003). Since we find $\kappa^{\rm eff}$ decreases with radius in most
clusters it would be worth pursuing this scenario. We plot the heating
rate required per kpc with conduction at $\kappa=\kappa_{\rm S}/3$
(right plot in Fig.~\ref{fig:heatrateperkpc}). The required heating is
little changed for the cooler clusters in our sample.

\section{Summary}
Recent \emph{XMM-Newton} observations using the RGS have shown that
although cooling may have occurred in the past, it is likely that some
heating mechanism is preventing net heat loss at the observed epochs
($z=0.0-0.5$). We have shown that, in general, conduction at close to
the Spitzer rate is able to completely offset cooling in the hotter
parts of clusters ($T$ $\gtrsim$ 5 keV). But in the inner, cooler
parts of hot clusters and most of cool clusters conduction is
insufficient. The observed spectral mass deposition rates found here
are less than the amounts expected from the radiative cooling rate and
some additional heating and/or absorption of the cooling gas is
therefore required.

It has been suggested that turbulent heat diffusion can provide a high
effective heat conductivity (Cho et al.\ 2003). We have shown that
this process may supply adequate heat to balance radiative cooling in
the observed clusters. In detail, it provides remarkably similar
profiles to those required.

\section{Acknowledgements}
We thank Roderick Johnstone for help with the analysis of PKS
1404-267 and the use of his program for evaluating the cooling
function. We are also grateful to Robert Schmidt and Jeremy Sanders
for the use of their scripts in the Monte Carlo simulations and
cooling flow models, respectively. We thank Steve Allen and Martin
Laming for helpful discussions. ACF and LMV acknowledge support from
The Royal Society and PPARC, respectively.

\begin{figure*}
\includegraphics[angle=-90,width=0.95\columnwidth]{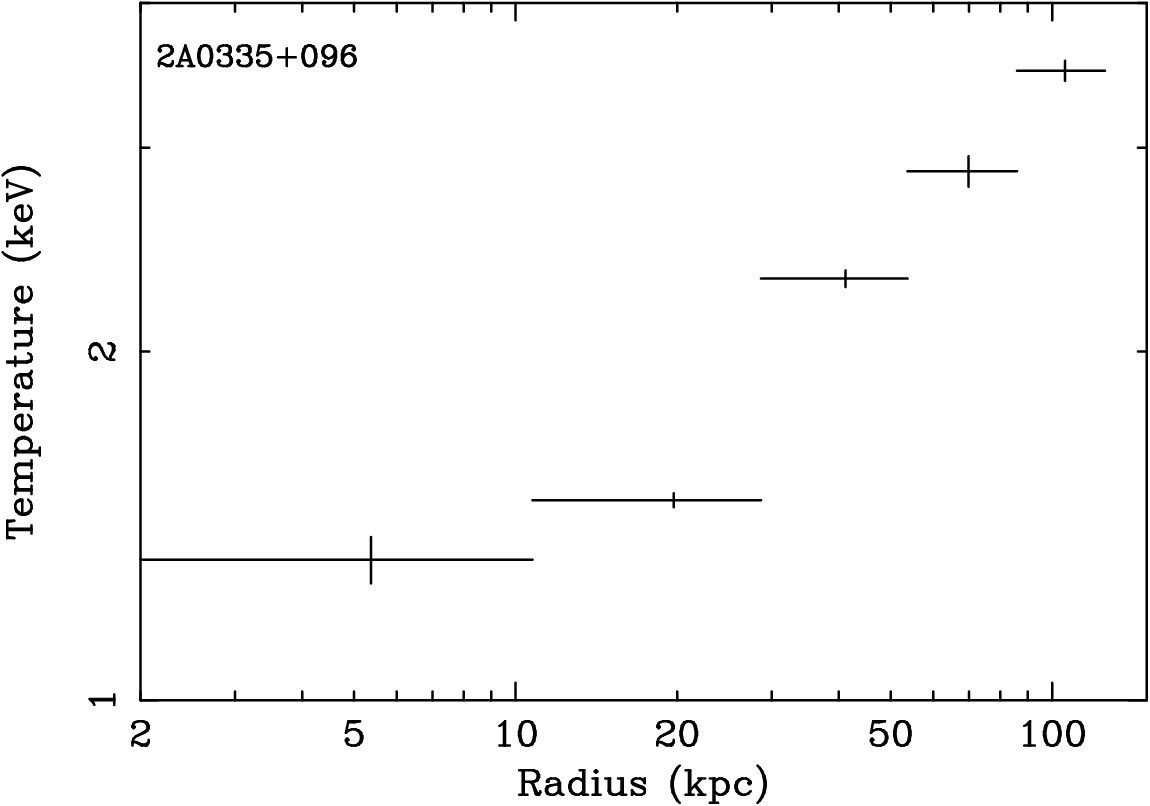}
\vspace{1.0mm}
\includegraphics[angle=-90,width=0.95\columnwidth]{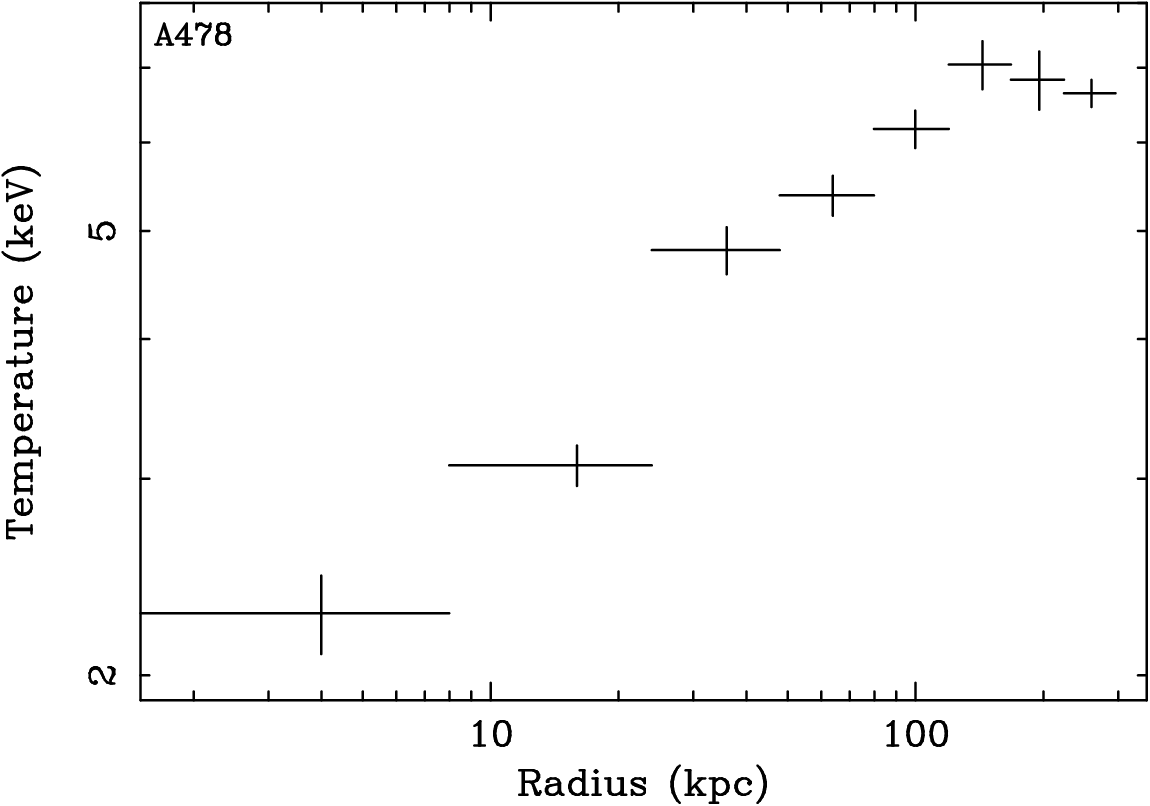}
\vspace{1.0mm}
\includegraphics[angle=-90,width=0.95\columnwidth]{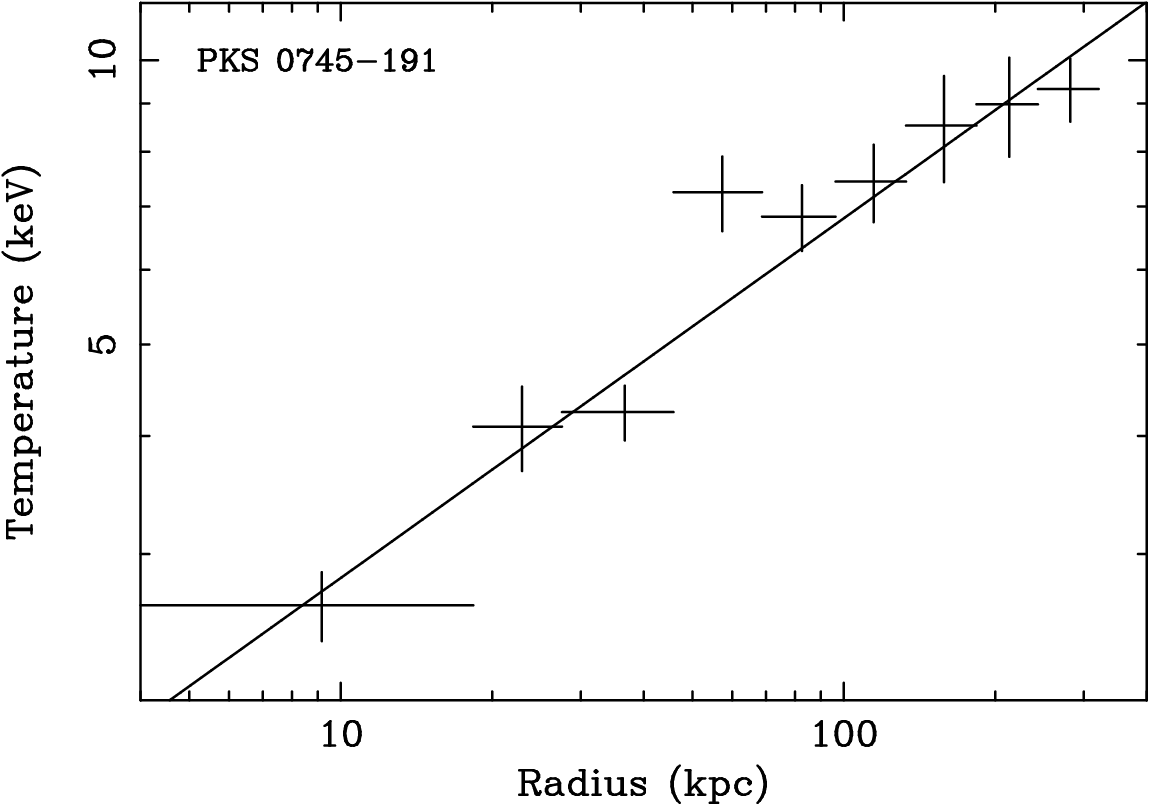}
\vspace{1.0mm}
\includegraphics[angle=-90,width=0.95\columnwidth]{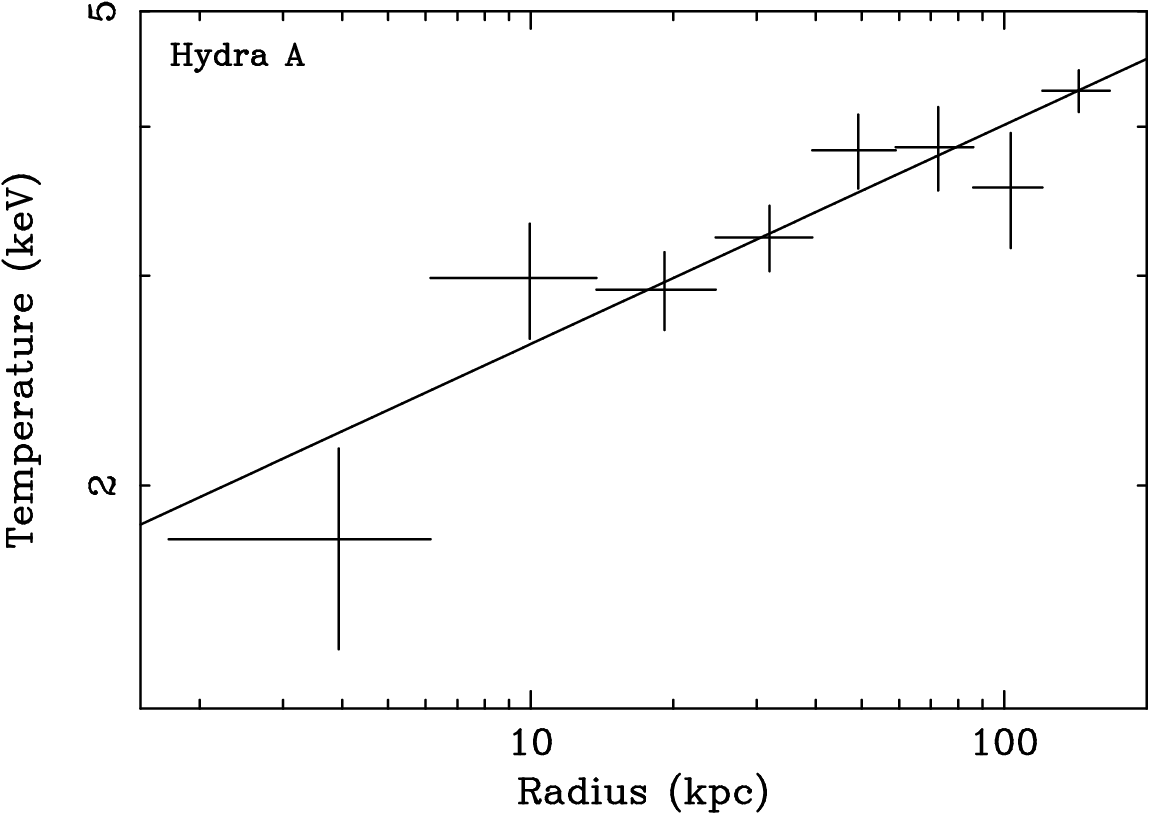}
\vspace{1.0mm}
\includegraphics[angle=-90,width=0.95\columnwidth]{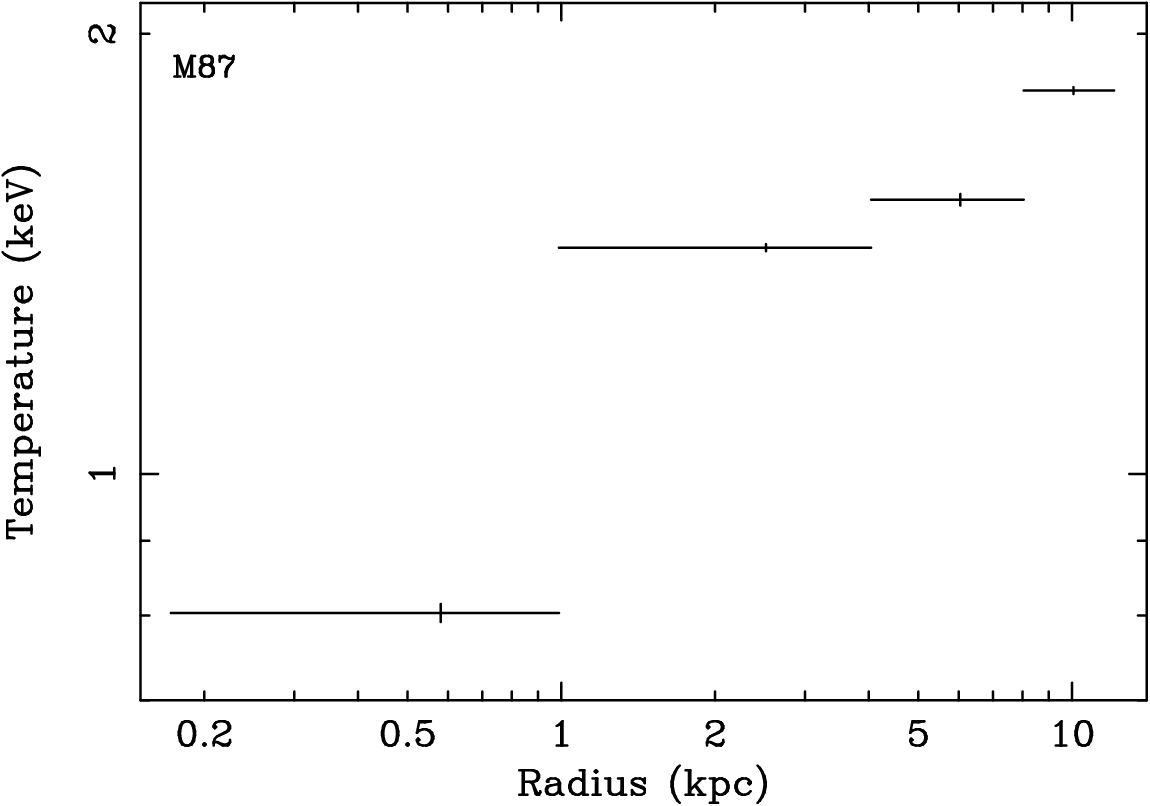}
\vspace{1.0mm}
\includegraphics[angle=-90,width=0.95\columnwidth]{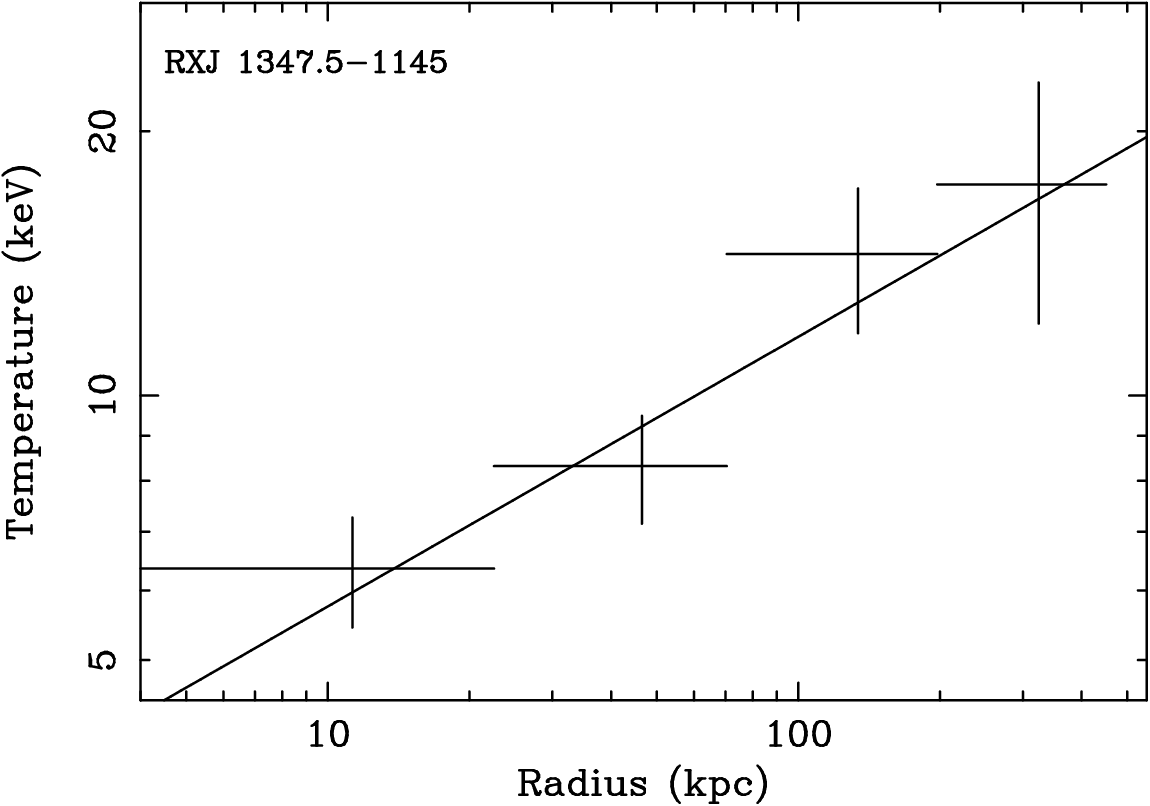}
\vspace{1.0mm}
\includegraphics[angle=-90,width=0.95\columnwidth]{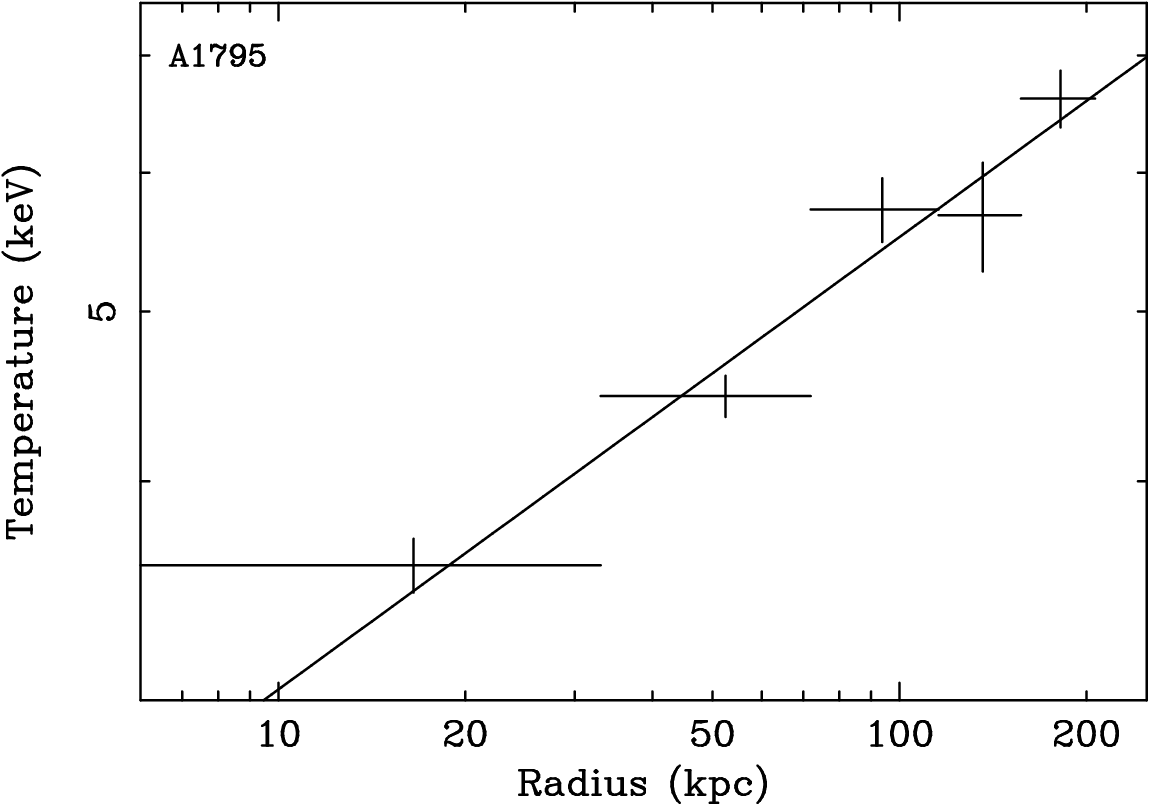}
\vspace{1.0mm}
\includegraphics[angle=-90,width=0.95\columnwidth]{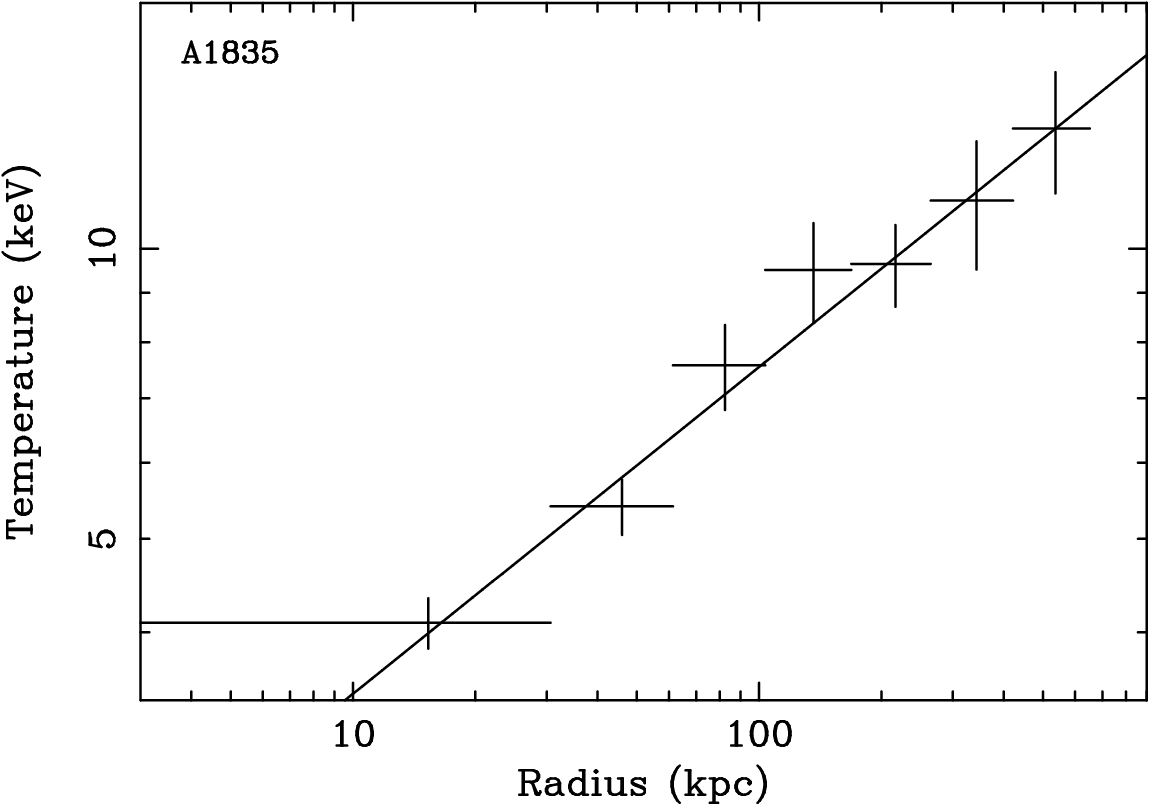}
\end{figure*}

\begin{figure*}
\includegraphics[angle=-90,width=0.95\columnwidth]{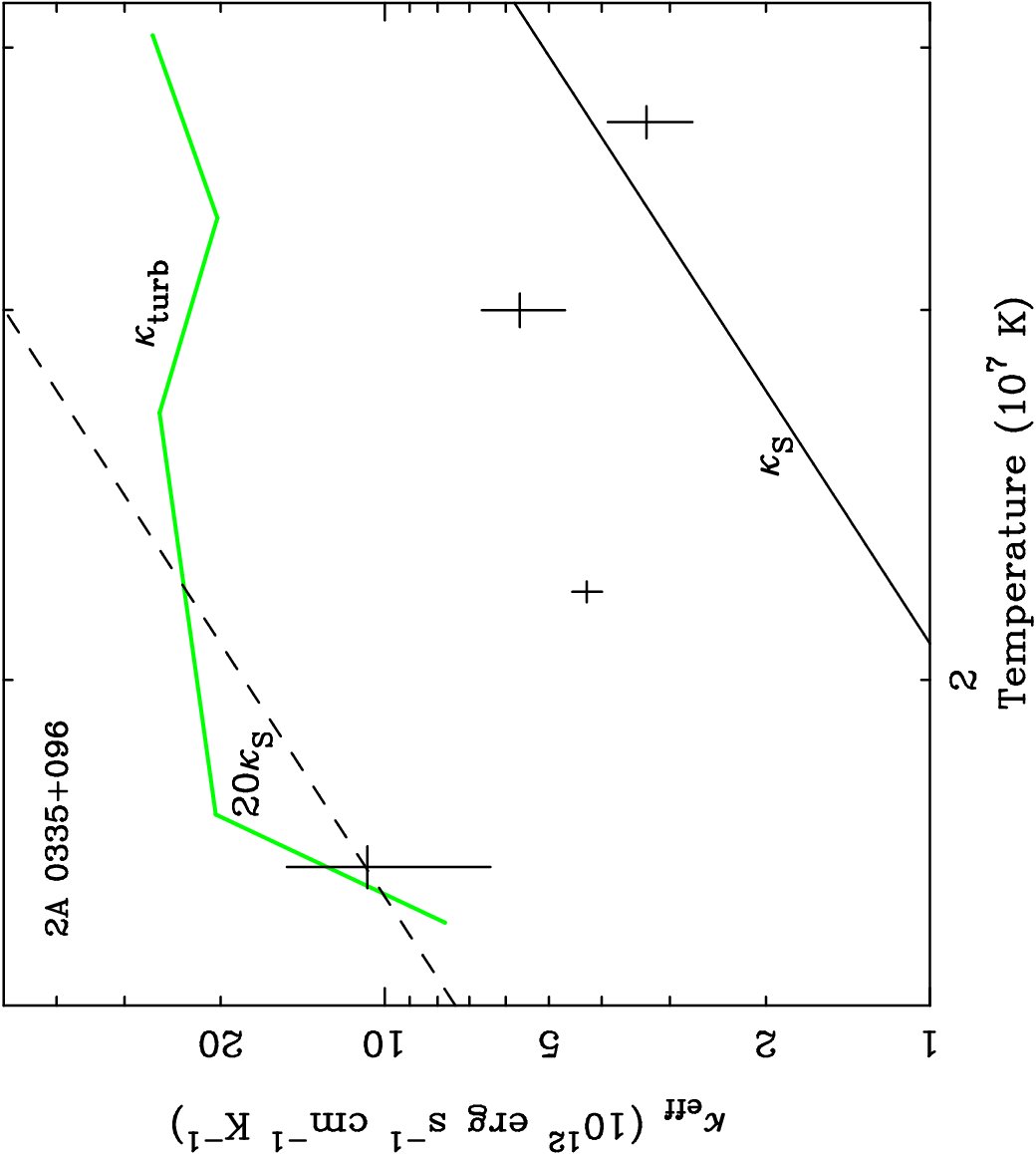}
\vspace{1.0mm}
\includegraphics[angle=-90,width=0.95\columnwidth]{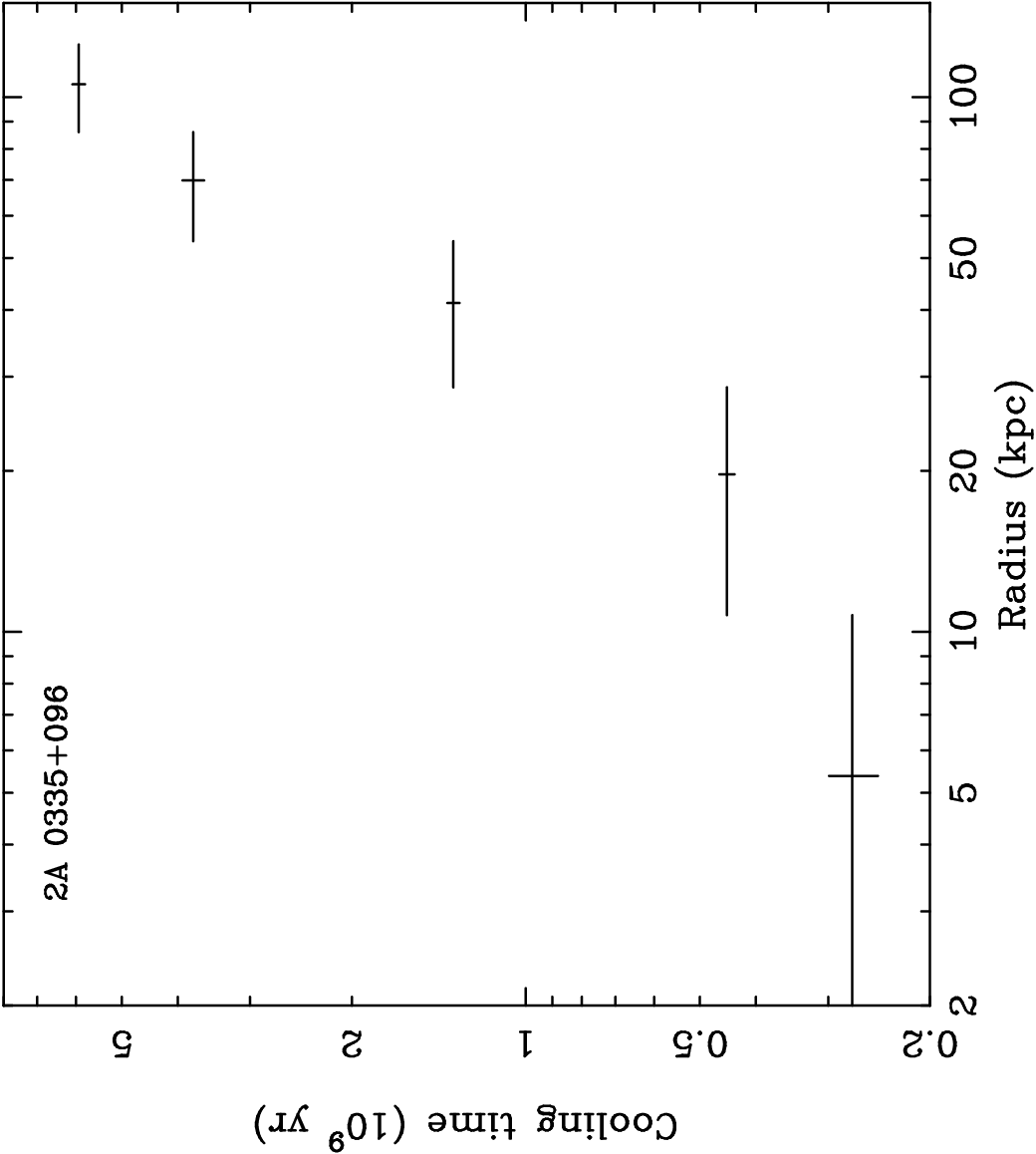}
\vspace{1.0mm}
\includegraphics[angle=-90,width=0.95\columnwidth]{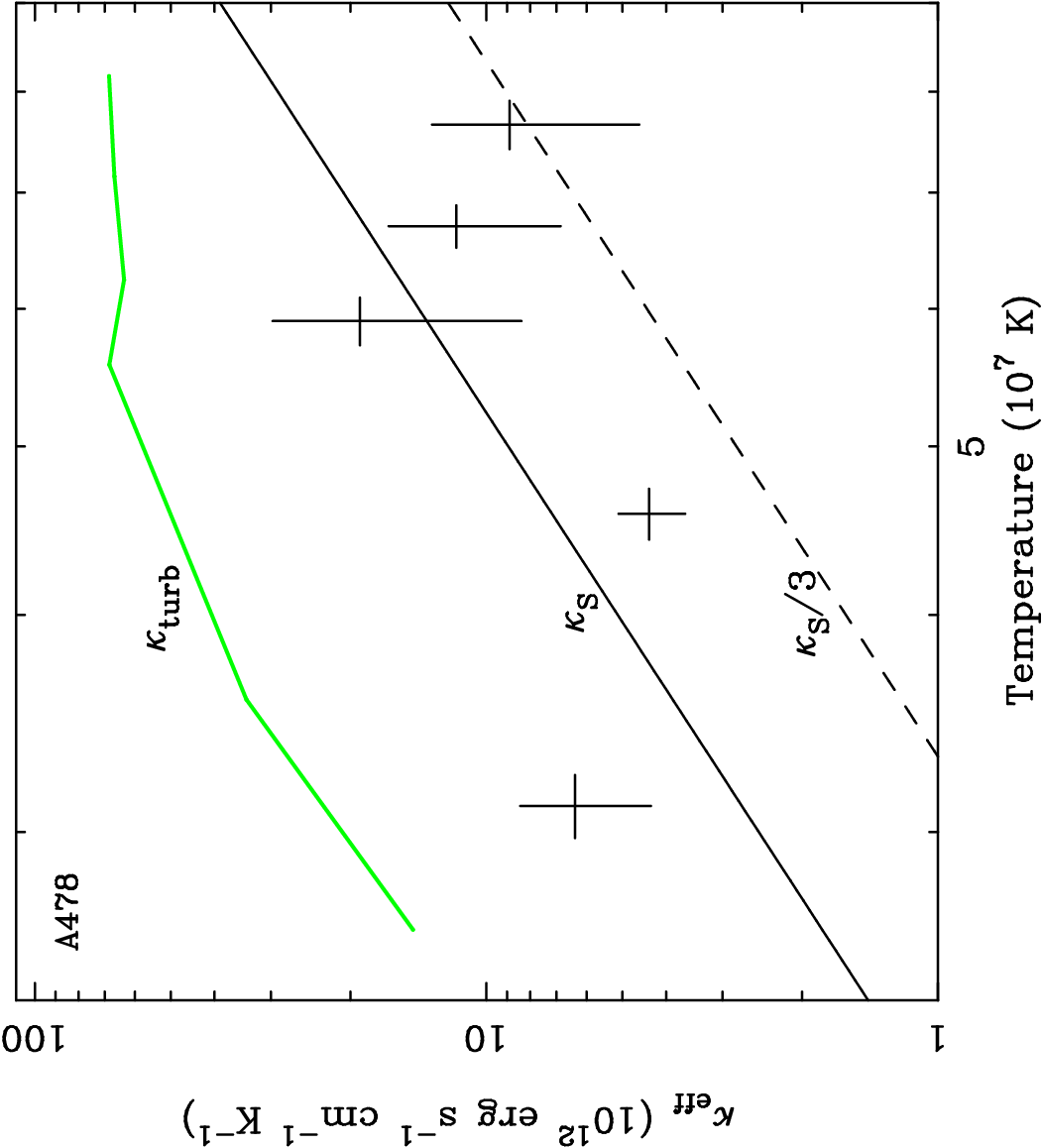}
\vspace{1.0mm}
\includegraphics[angle=-90,width=0.95\columnwidth]{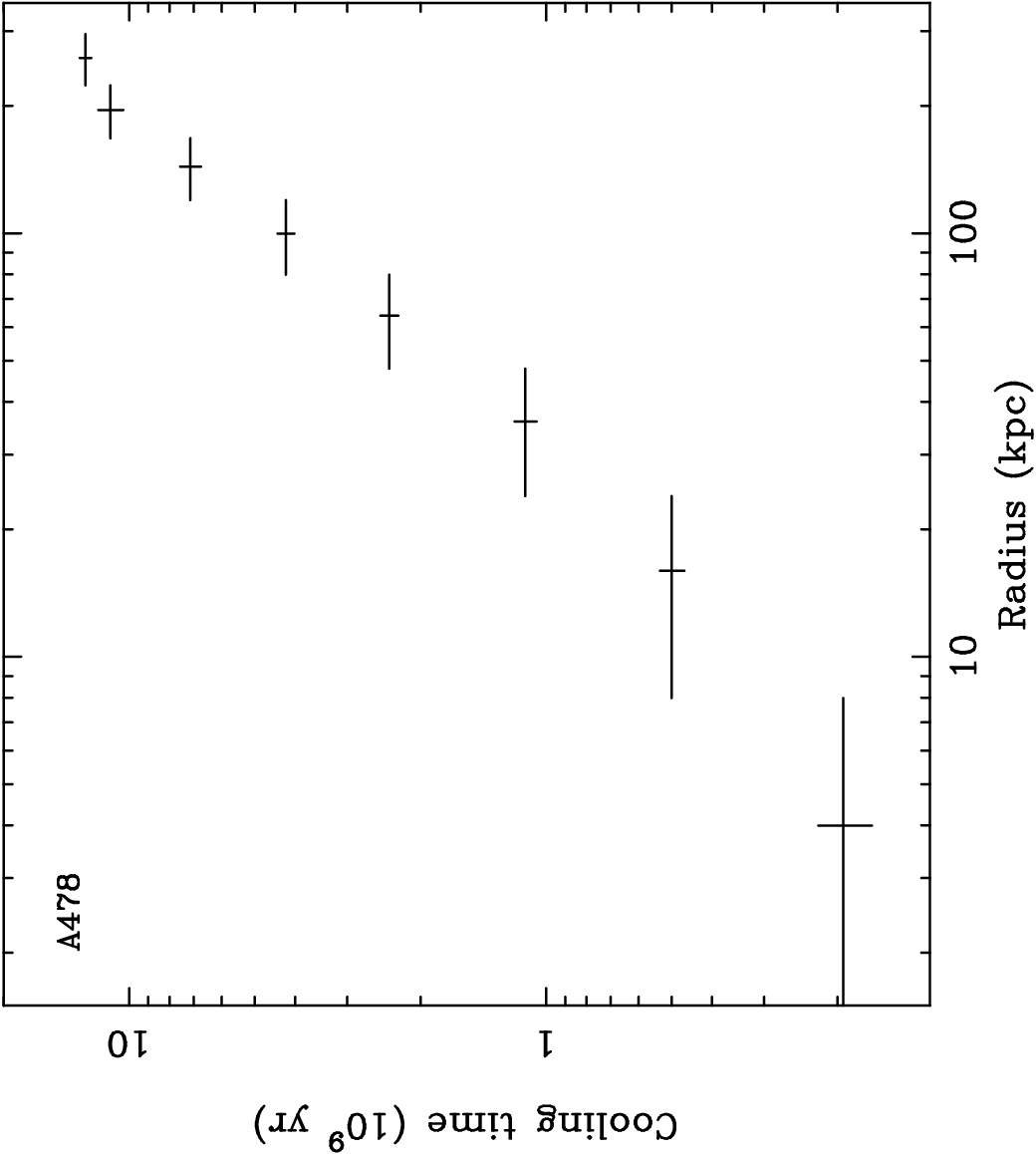}
\vspace{1.0mm}
\includegraphics[angle=-90,width=0.95\columnwidth]{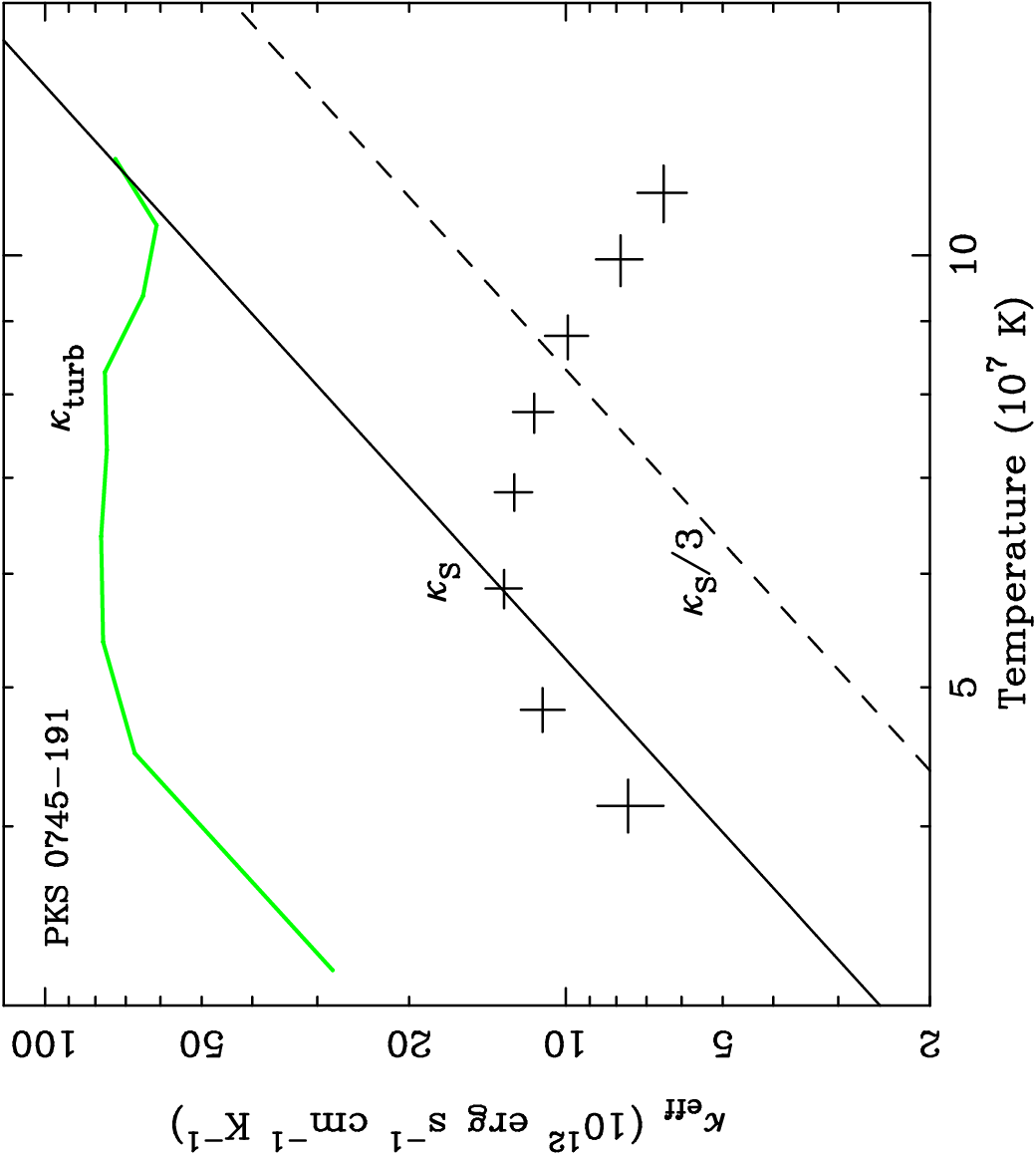}
\vspace{1.0mm}
\includegraphics[angle=-90,width=0.95\columnwidth]{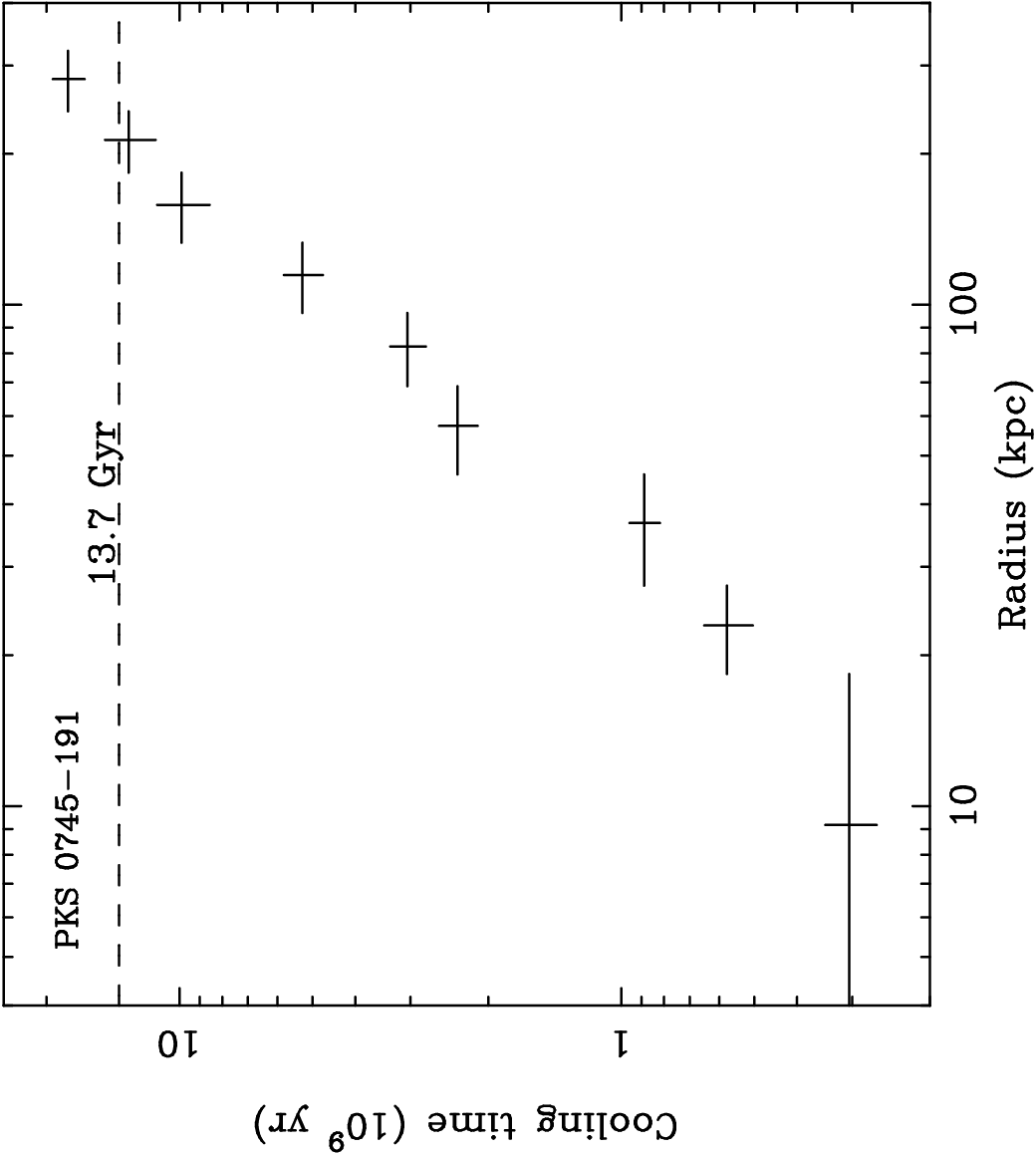}
\vspace{1.0mm}
\end{figure*}

\begin{figure*}
\includegraphics[angle=-90,width=0.95\columnwidth]{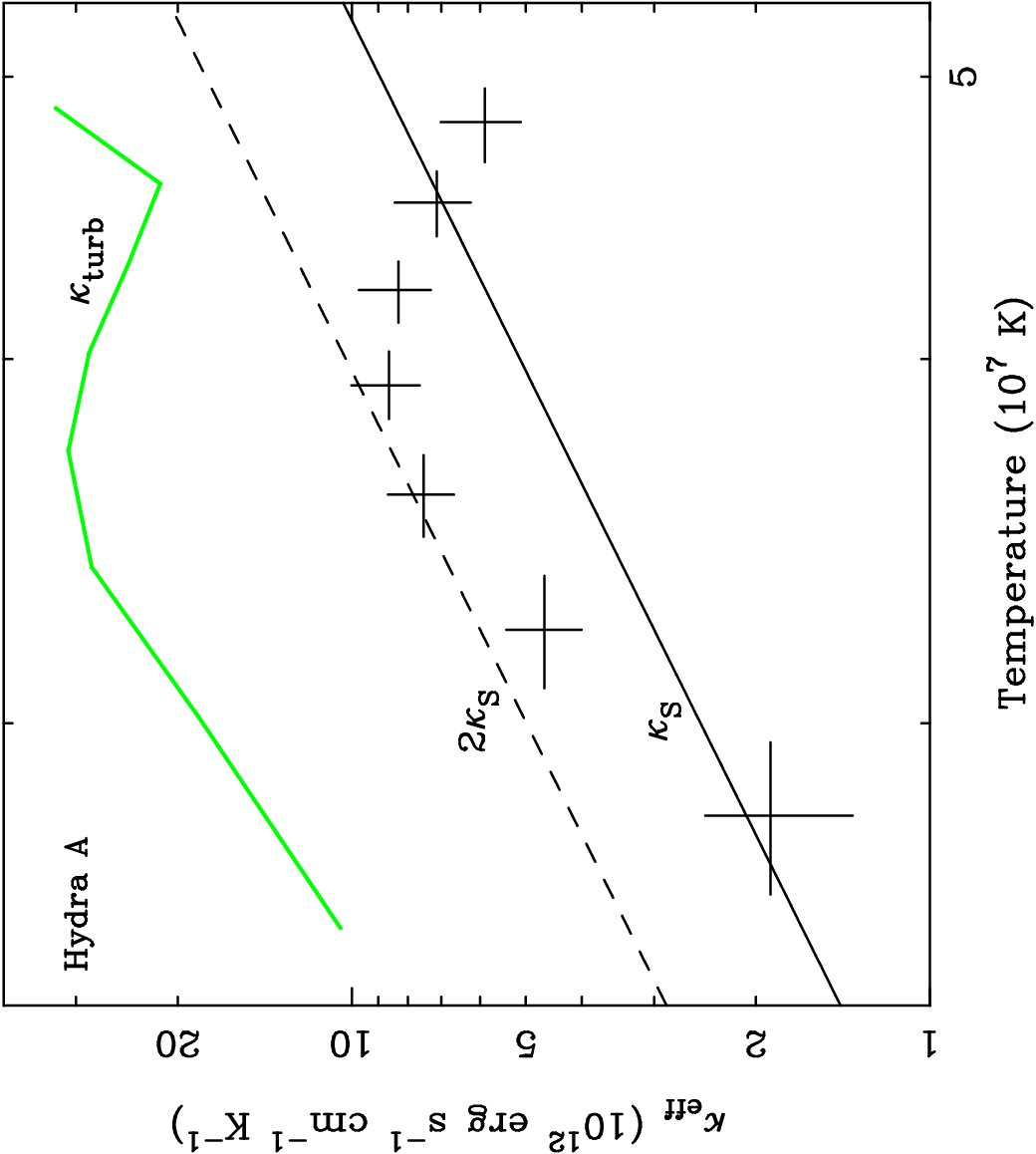}
\vspace{1.0mm}
\includegraphics[angle=-90,width=0.95\columnwidth]{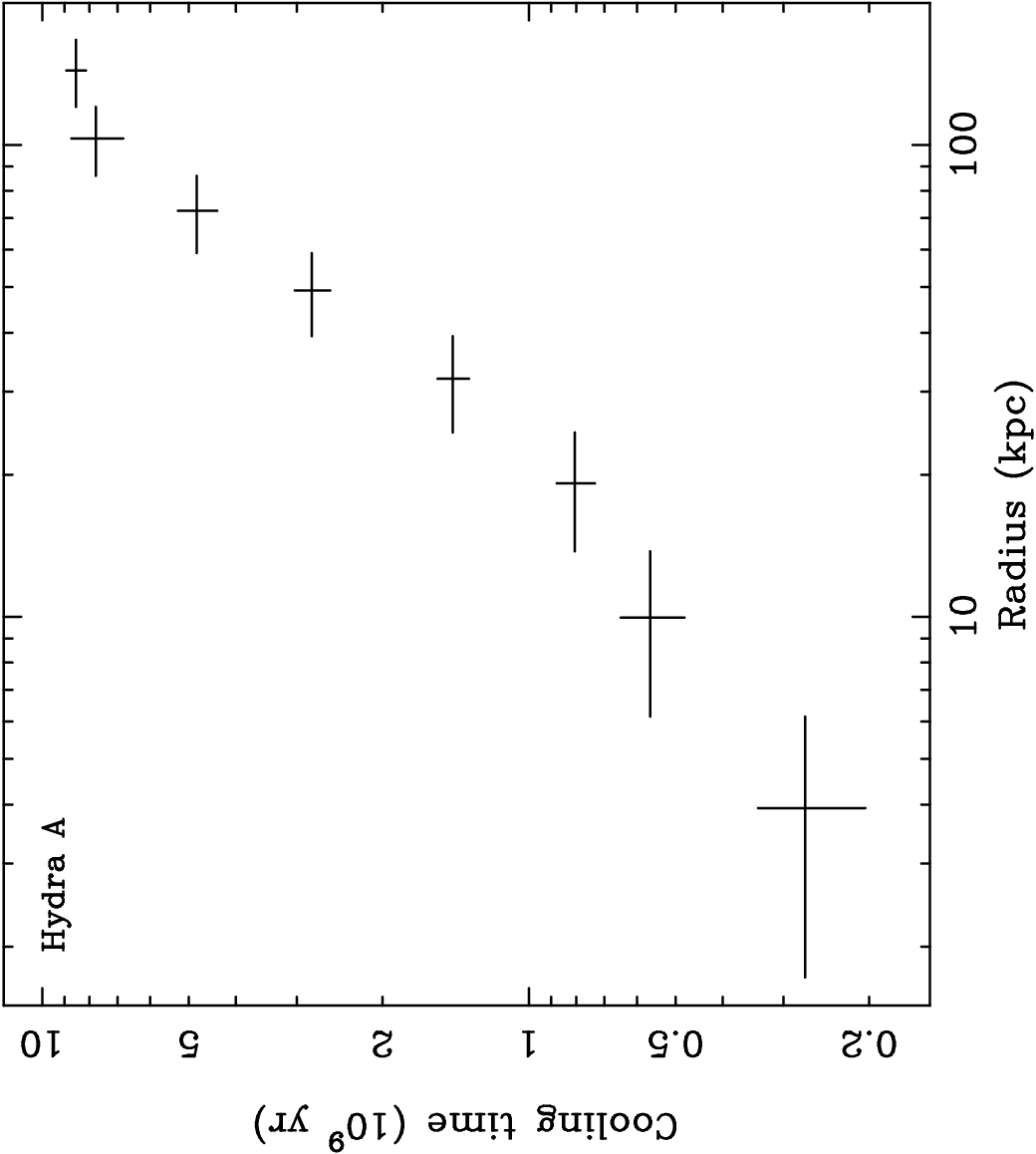}
\vspace{1.0mm}
\includegraphics[angle=-90,width=0.95\columnwidth]{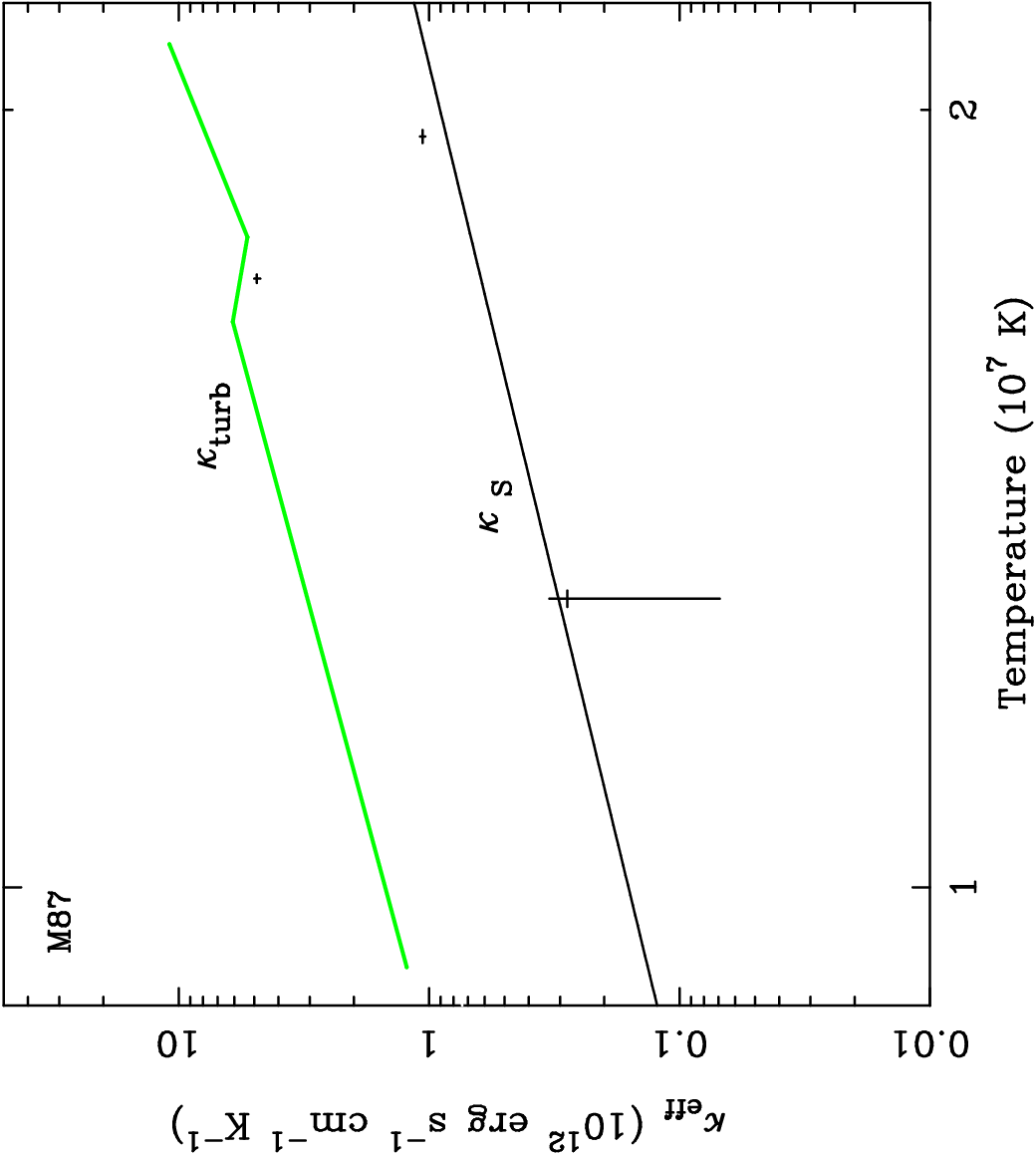}
\vspace{1.0mm}
\includegraphics[angle=-90,width=0.95\columnwidth]{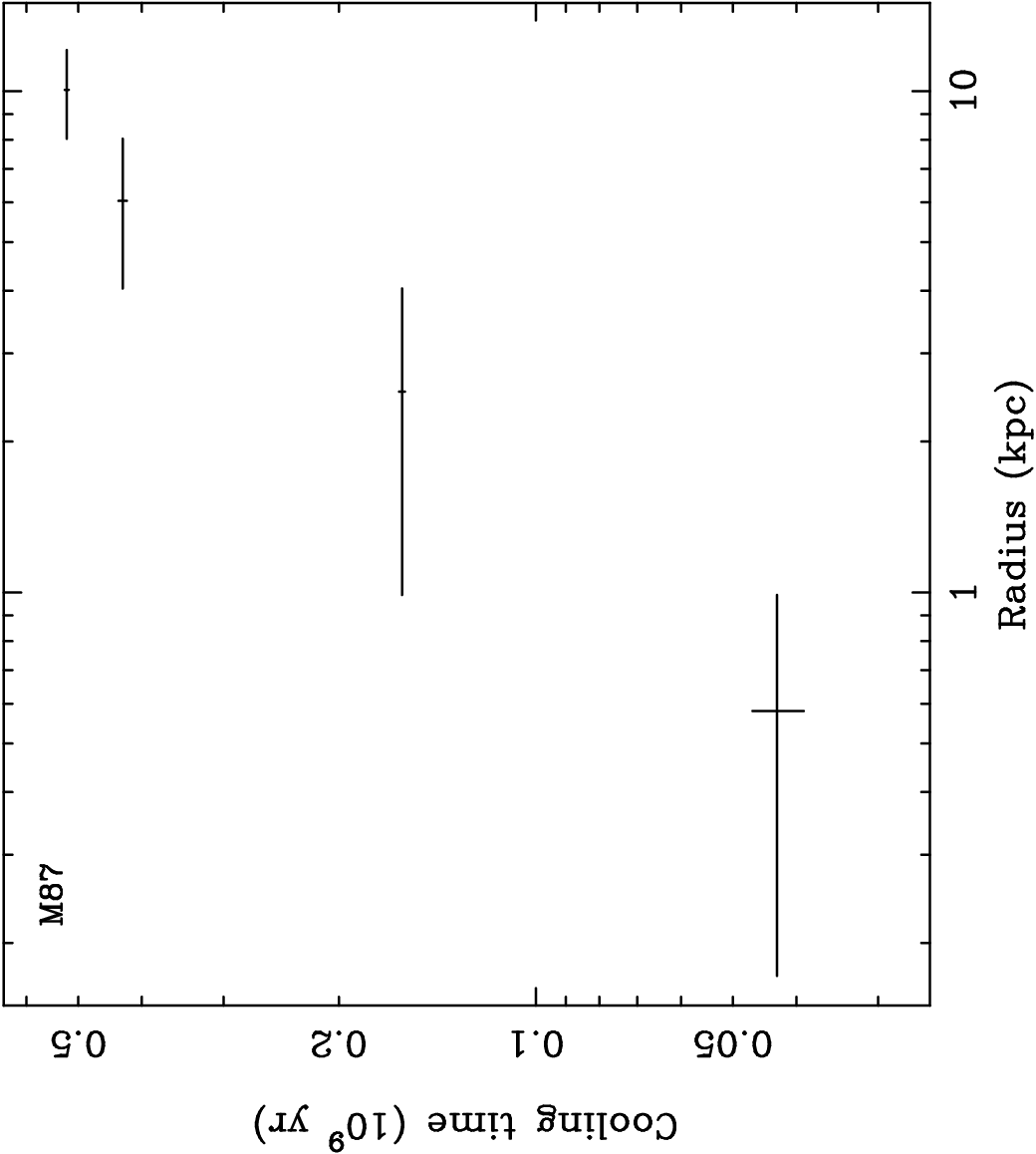}
\vspace{1.0mm}
\includegraphics[angle=-90,width=0.95\columnwidth]{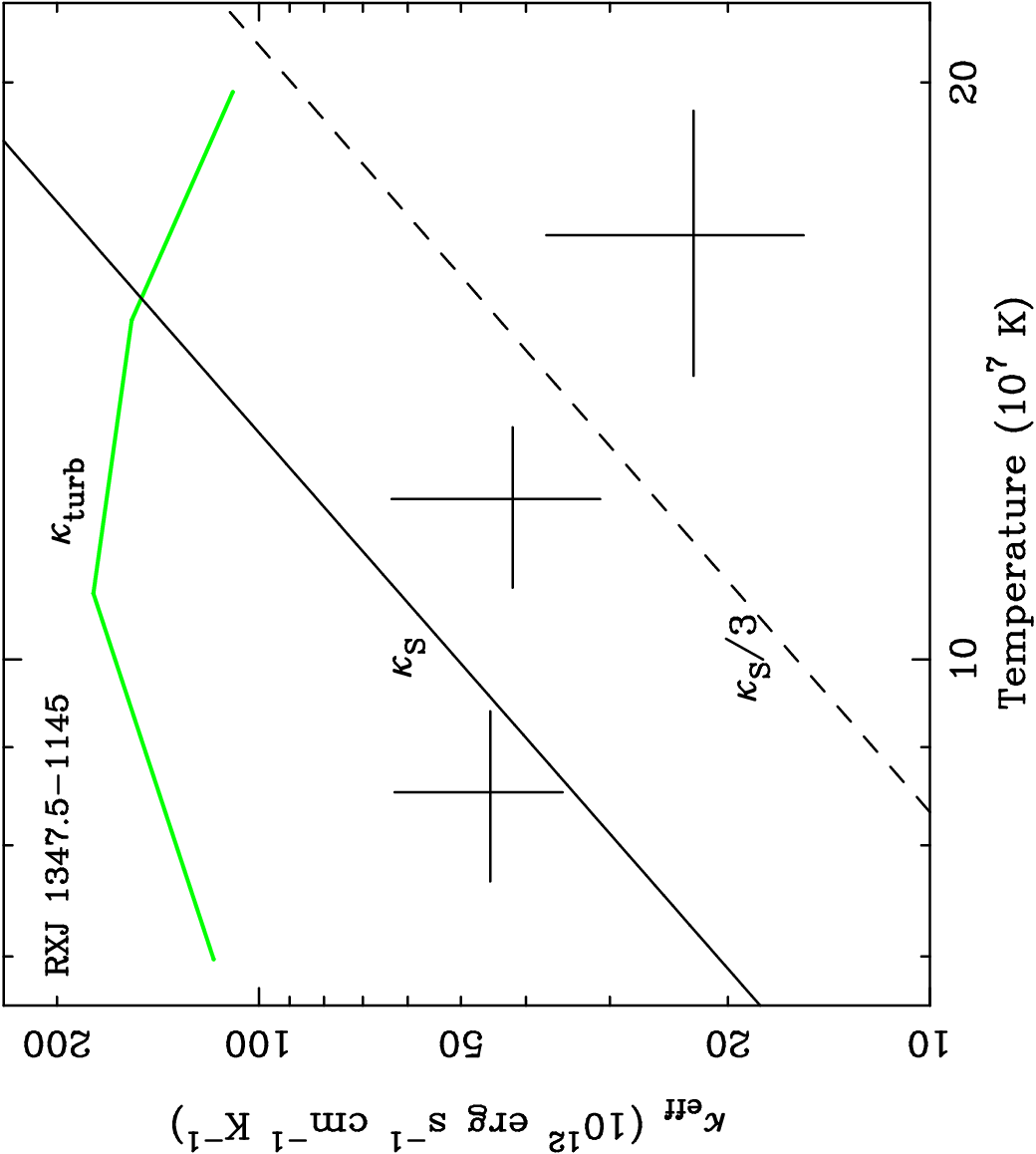}
\vspace{1.0mm}
\includegraphics[angle=-90,width=0.95\columnwidth]{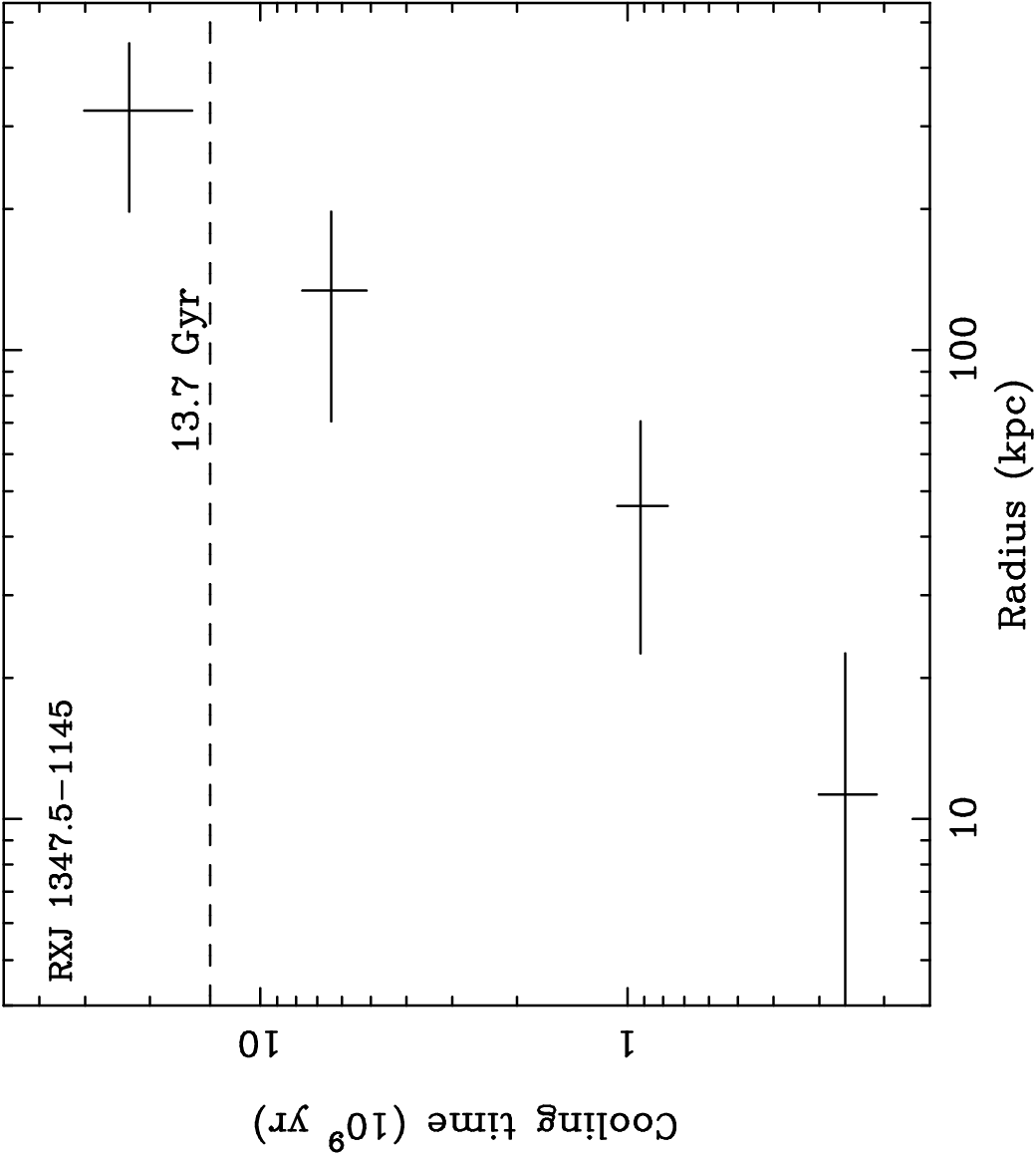}
\end{figure*}

\begin{figure*}
\includegraphics[angle=-90,width=0.95\columnwidth]{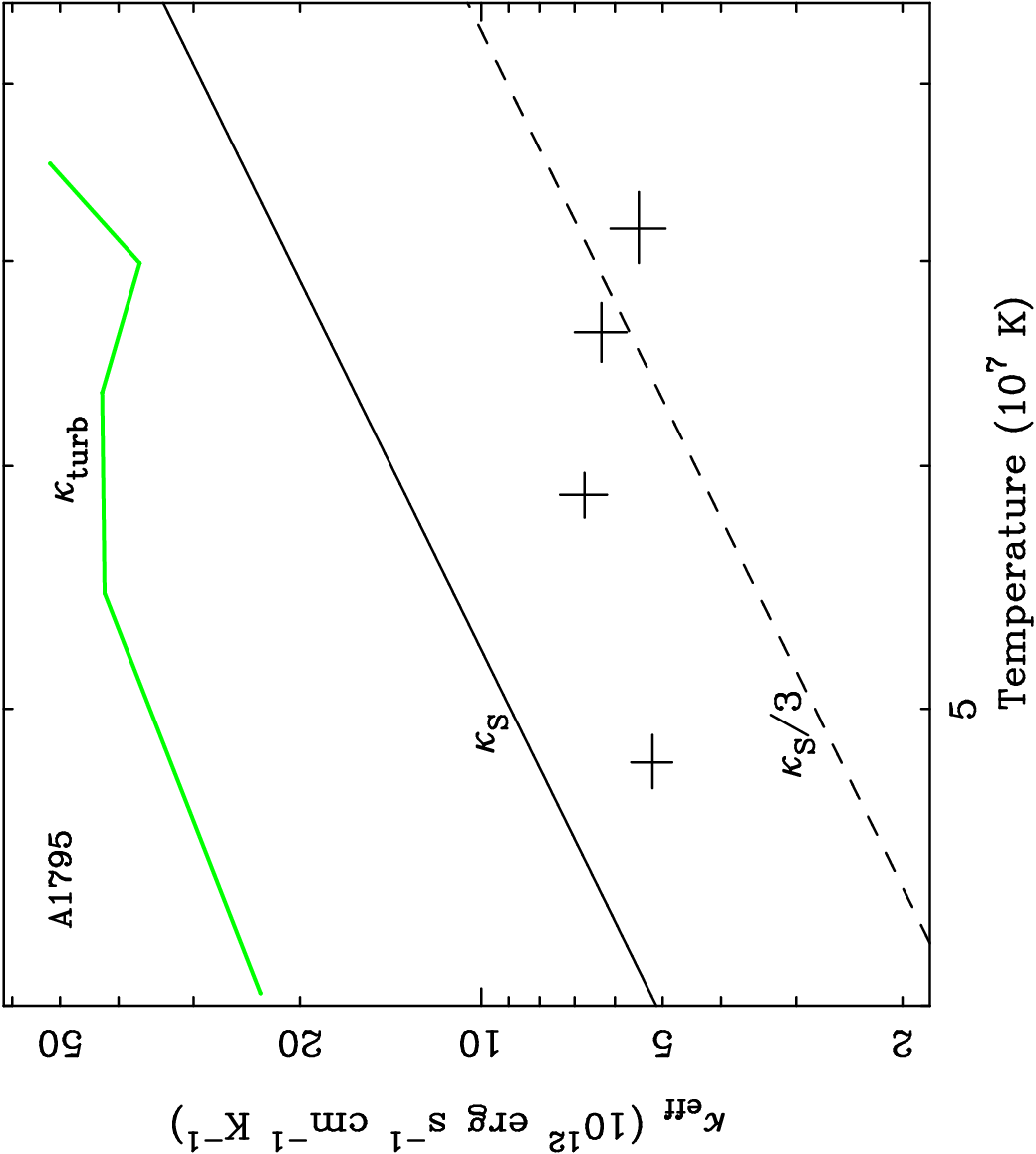}
\vspace{1.0mm}
\includegraphics[angle=-90,width=0.95\columnwidth]{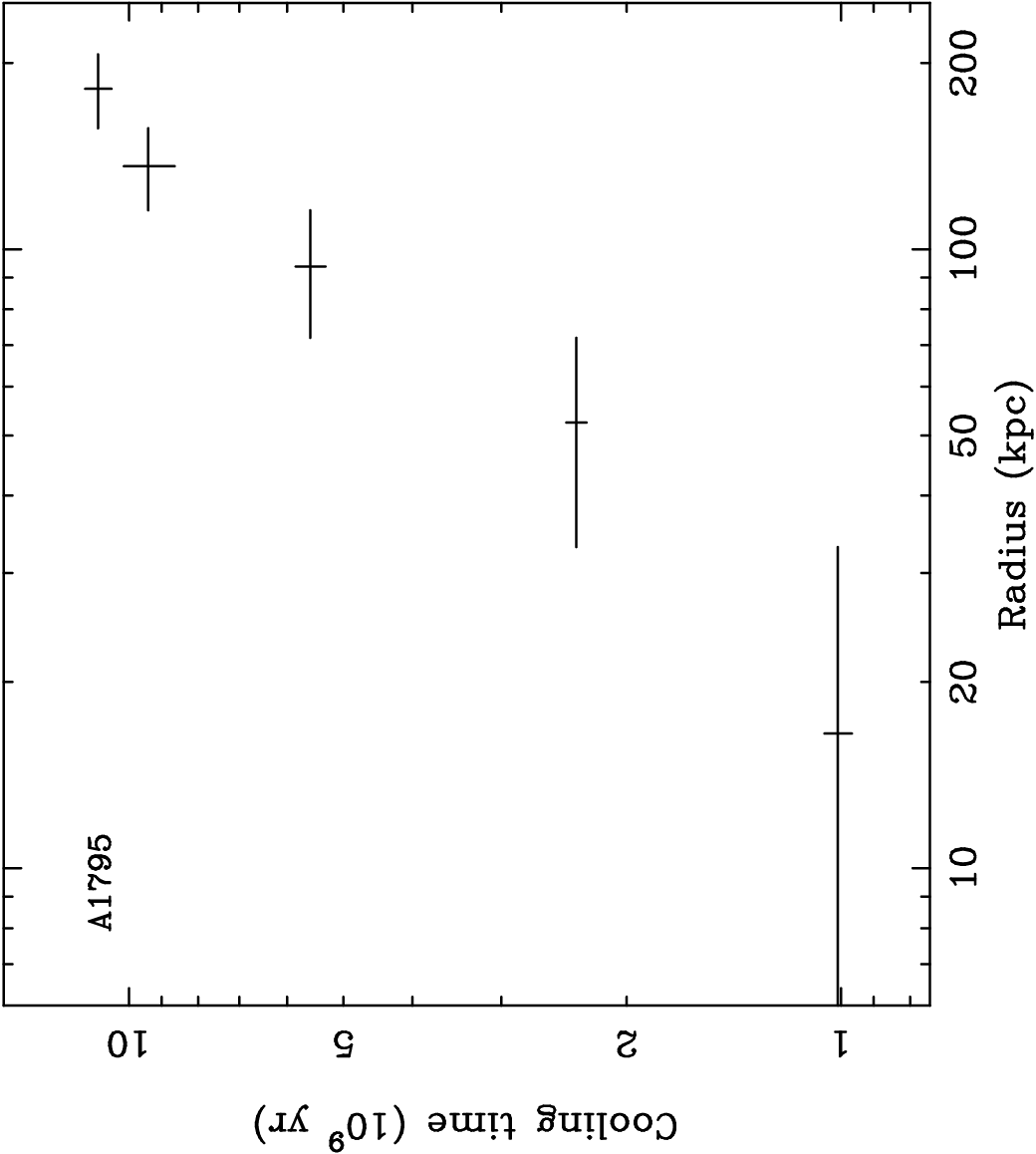}
\vspace{1.0mm}
\includegraphics[angle=-90,width=0.95\columnwidth]{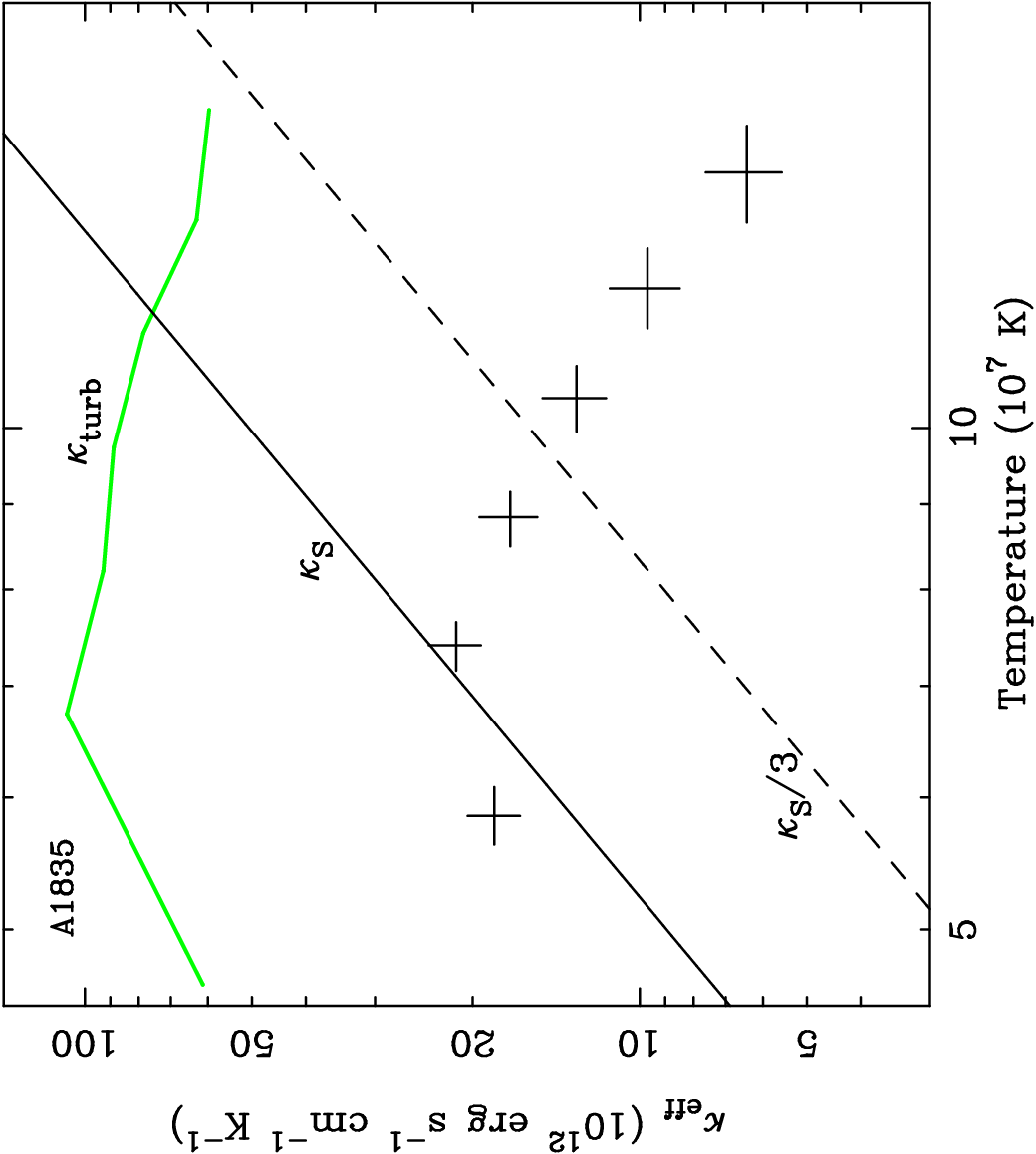}
\vspace{1.0mm}
\includegraphics[angle=-90,width=0.95\columnwidth]{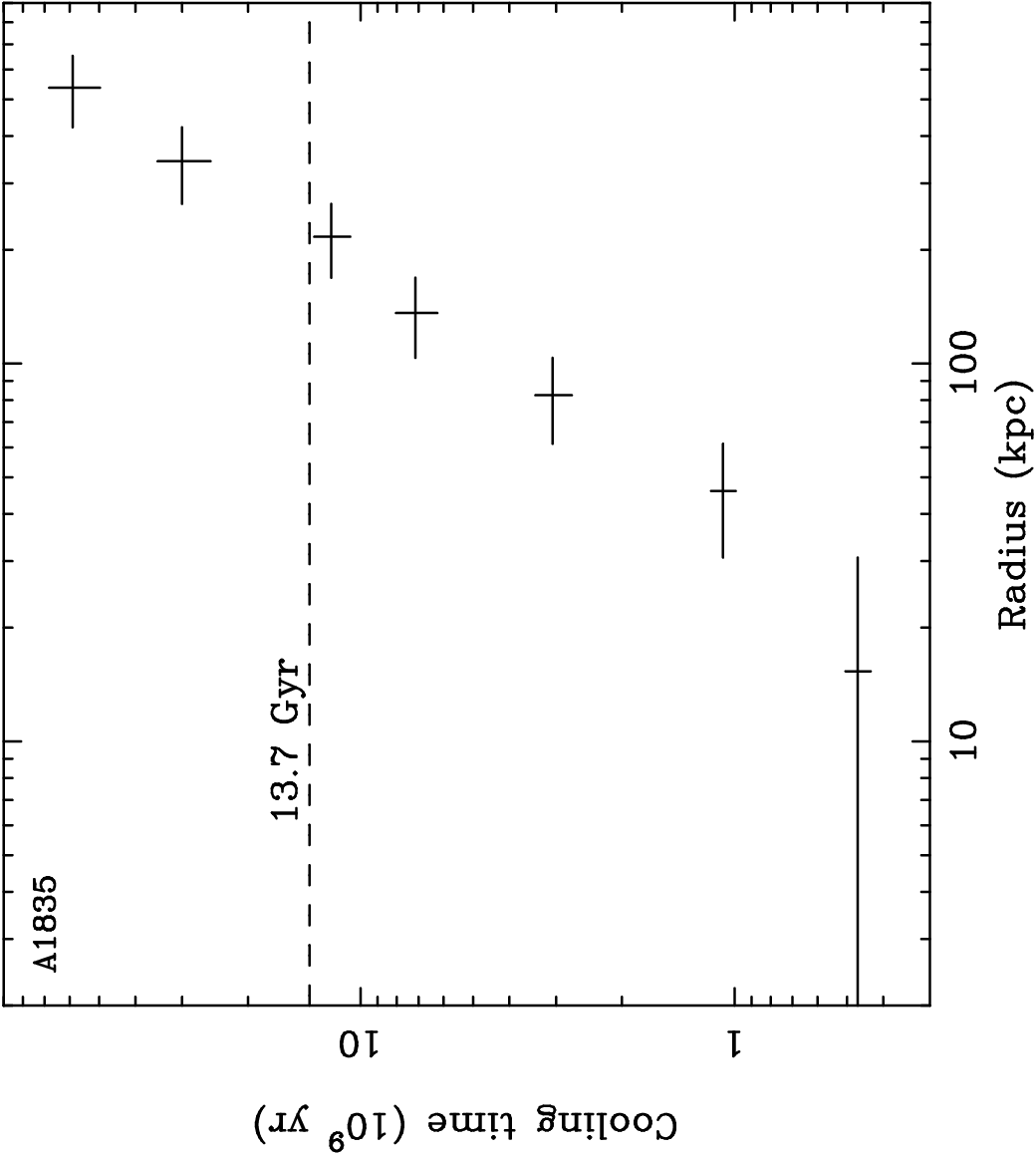}
\vspace{1.0mm}
\includegraphics[angle=-90,width=0.95\columnwidth]{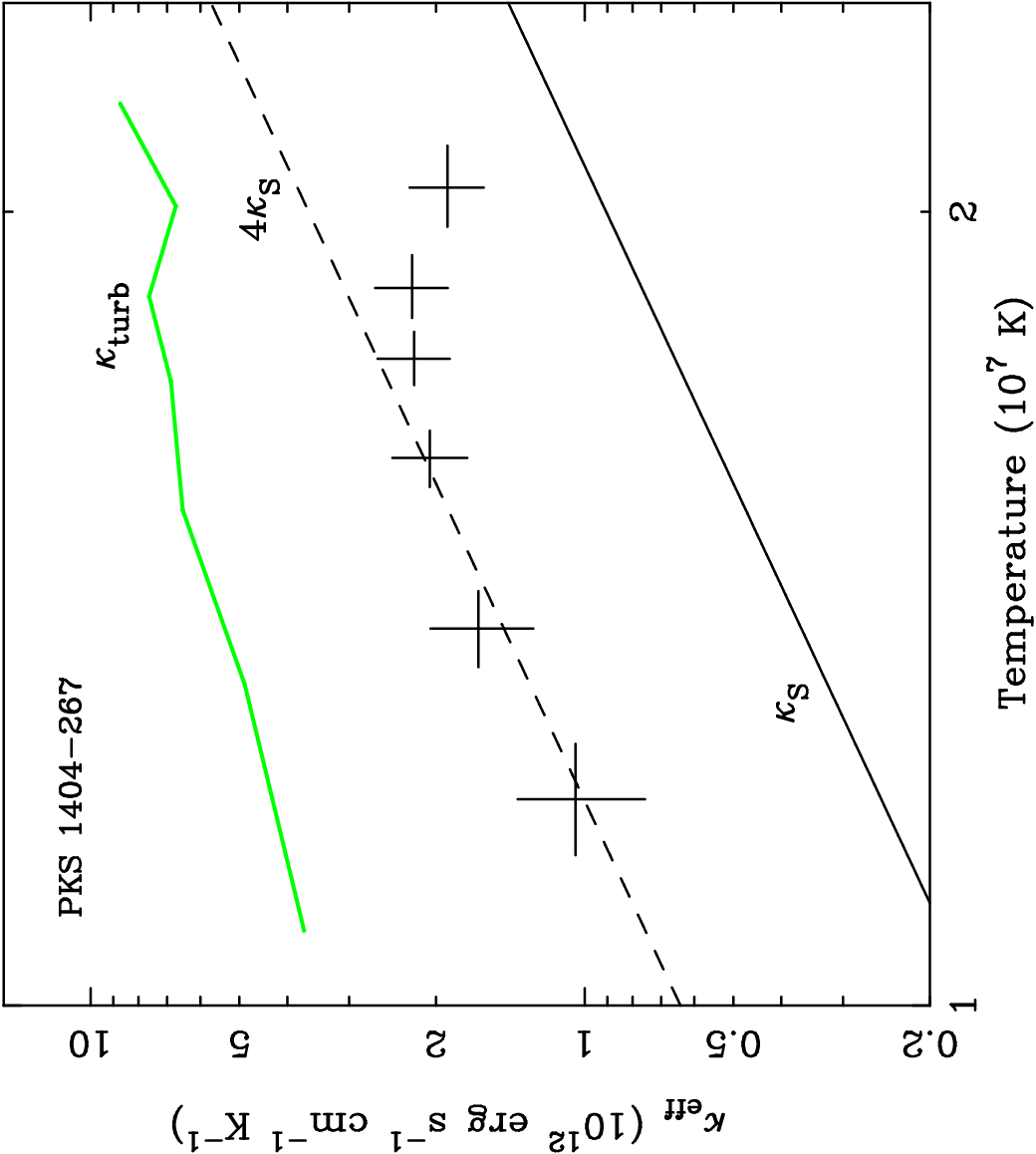}
\vspace{1.0mm}
\includegraphics[angle=-90,width=0.95\columnwidth]{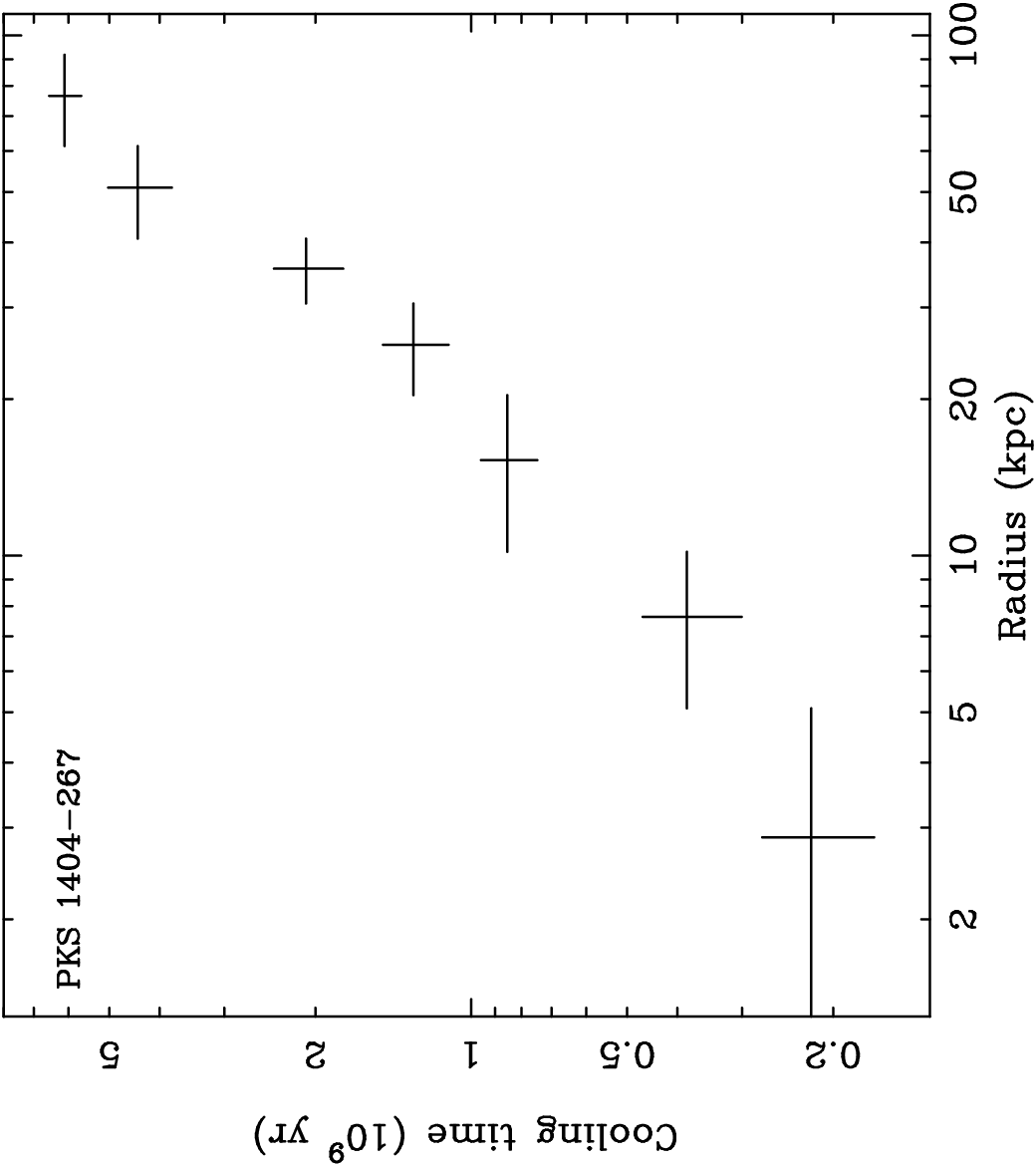}
\end{figure*}

\begin{figure*}
\includegraphics[angle=-90,width=0.95\columnwidth]{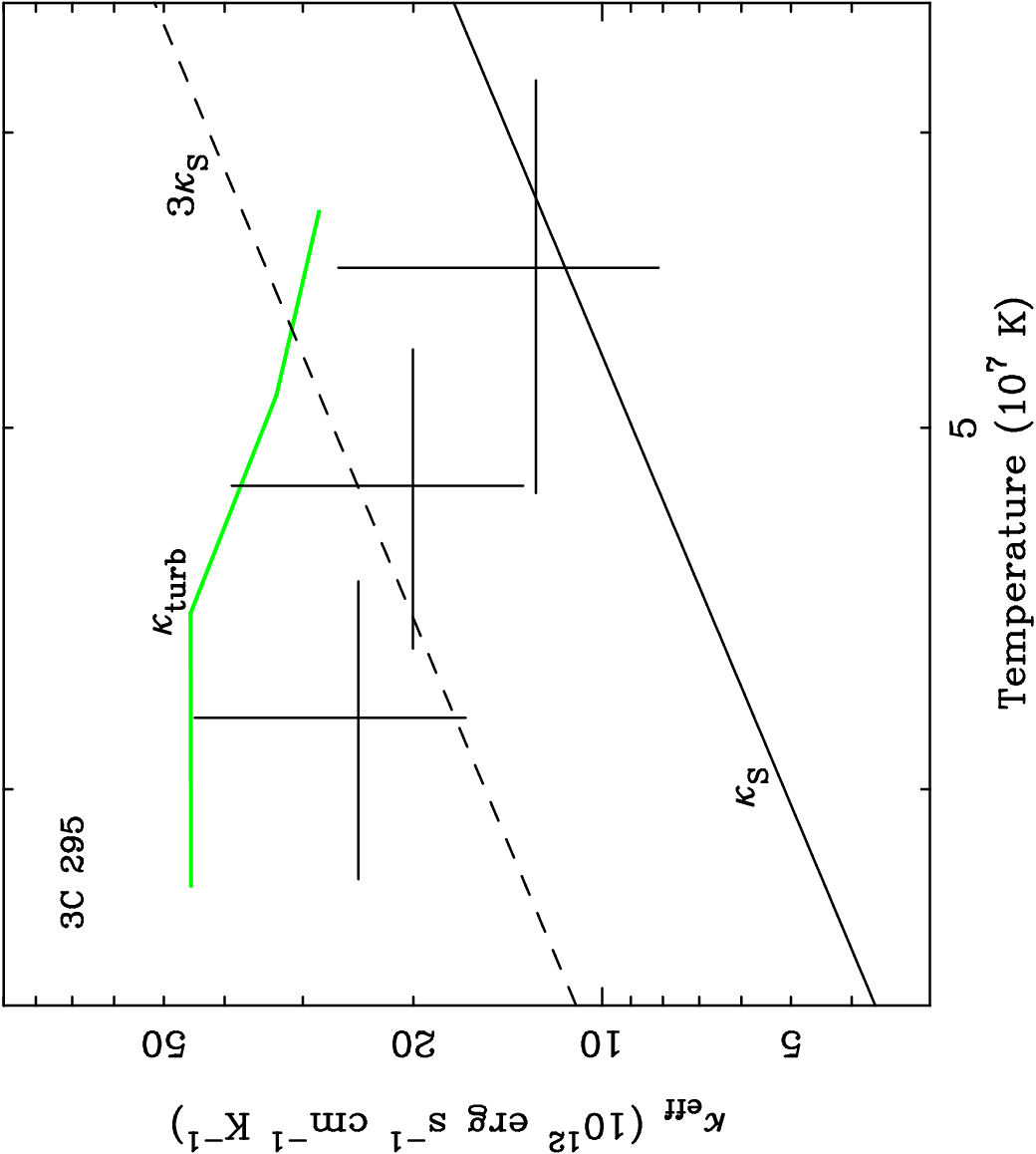}
\vspace{1.0mm}
\includegraphics[angle=-90,width=0.95\columnwidth]{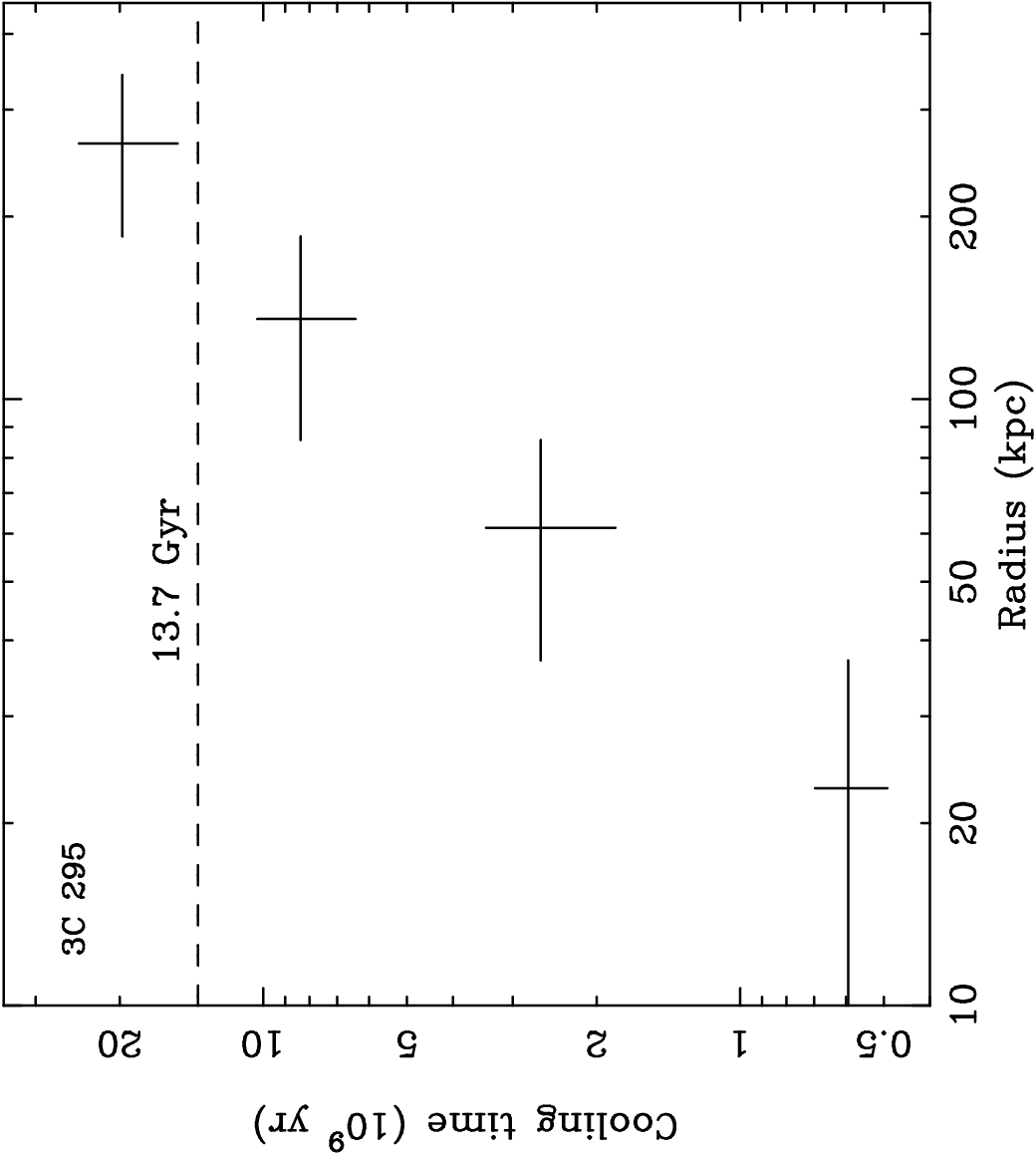}
\vspace{1.0mm}
\includegraphics[angle=-90,width=0.95\columnwidth]{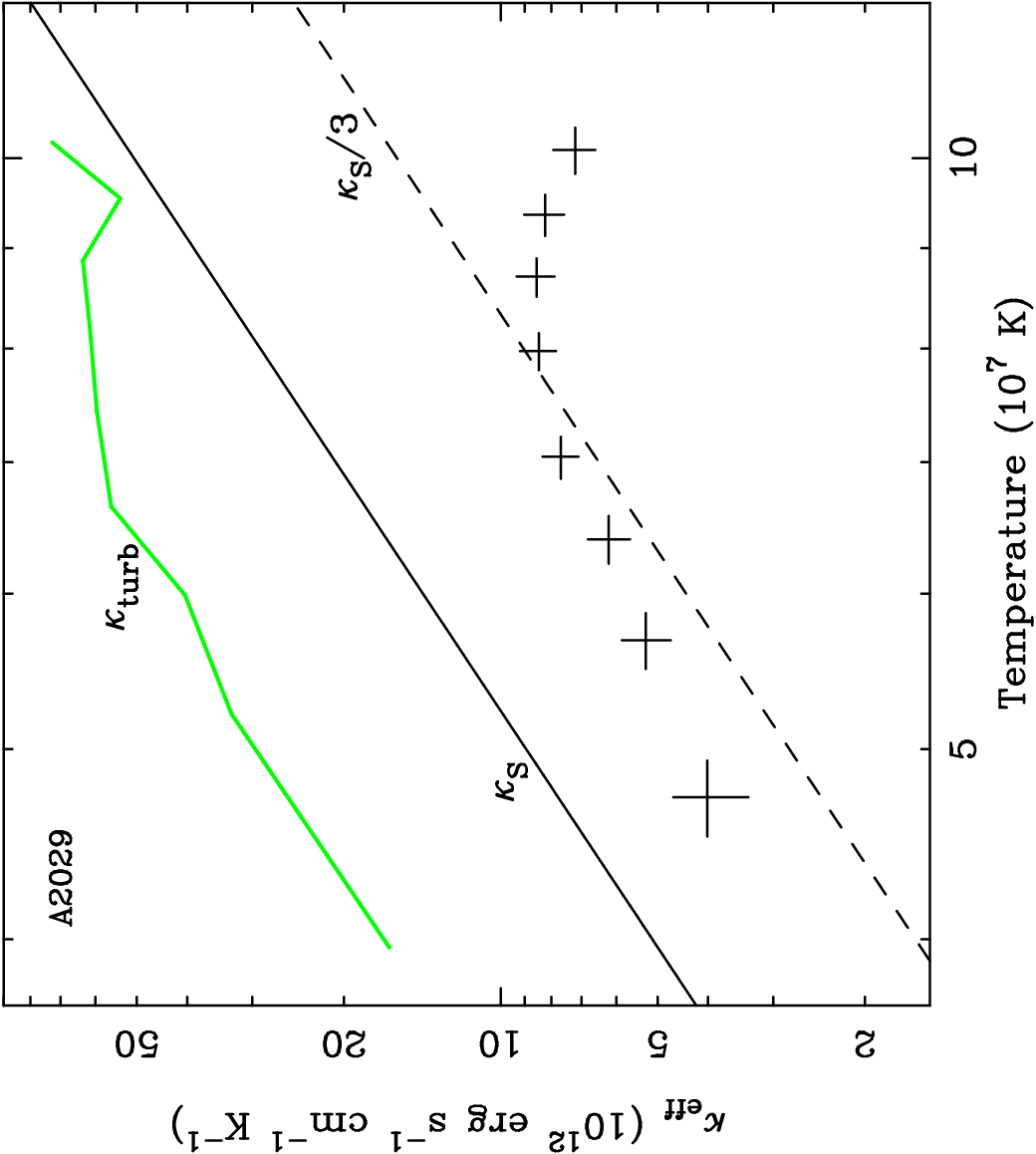}
\vspace{1.0mm}
\includegraphics[angle=-90,width=0.95\columnwidth]{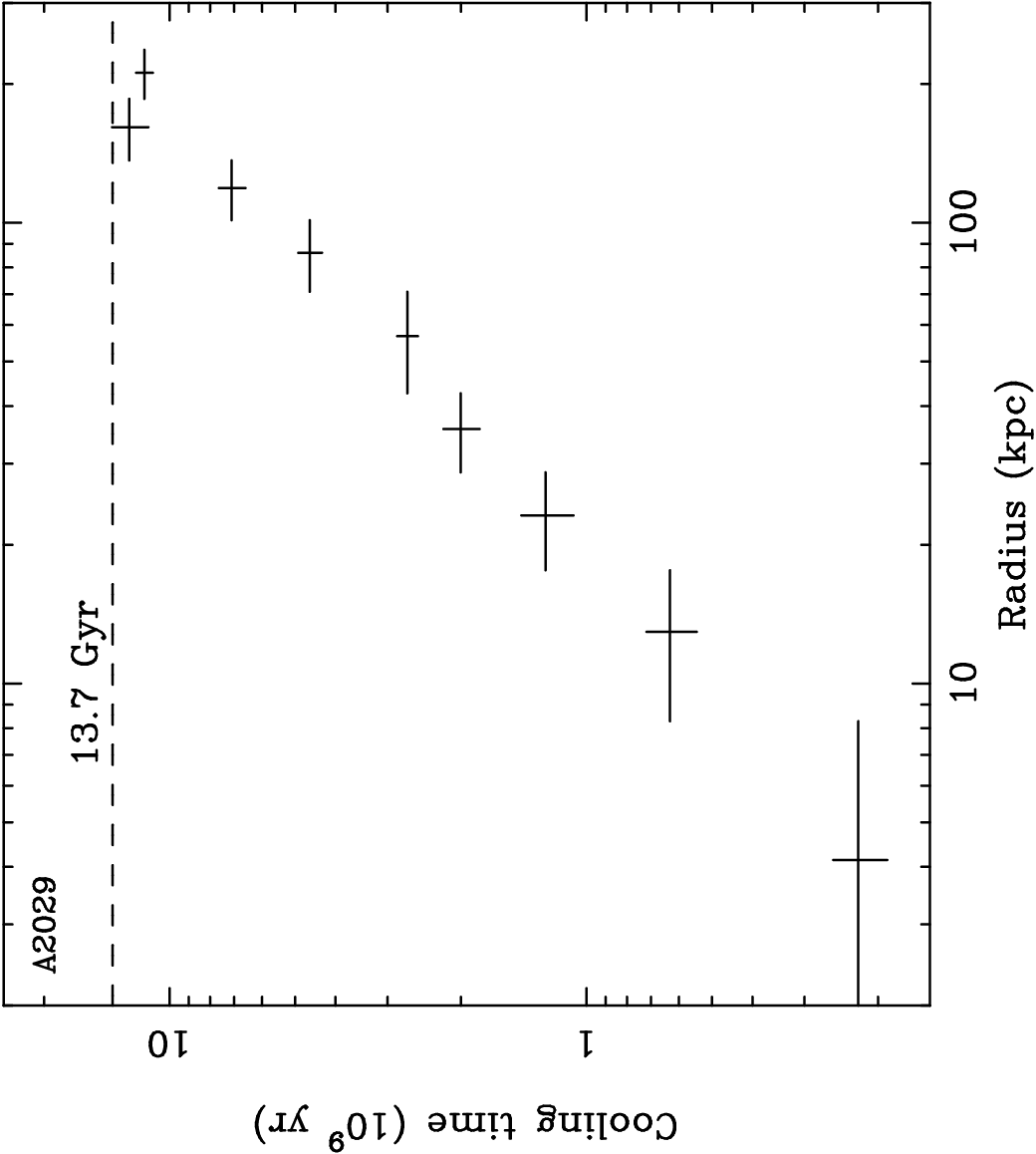}
\vspace{1.0mm}
\includegraphics[angle=-90,width=0.95\columnwidth]{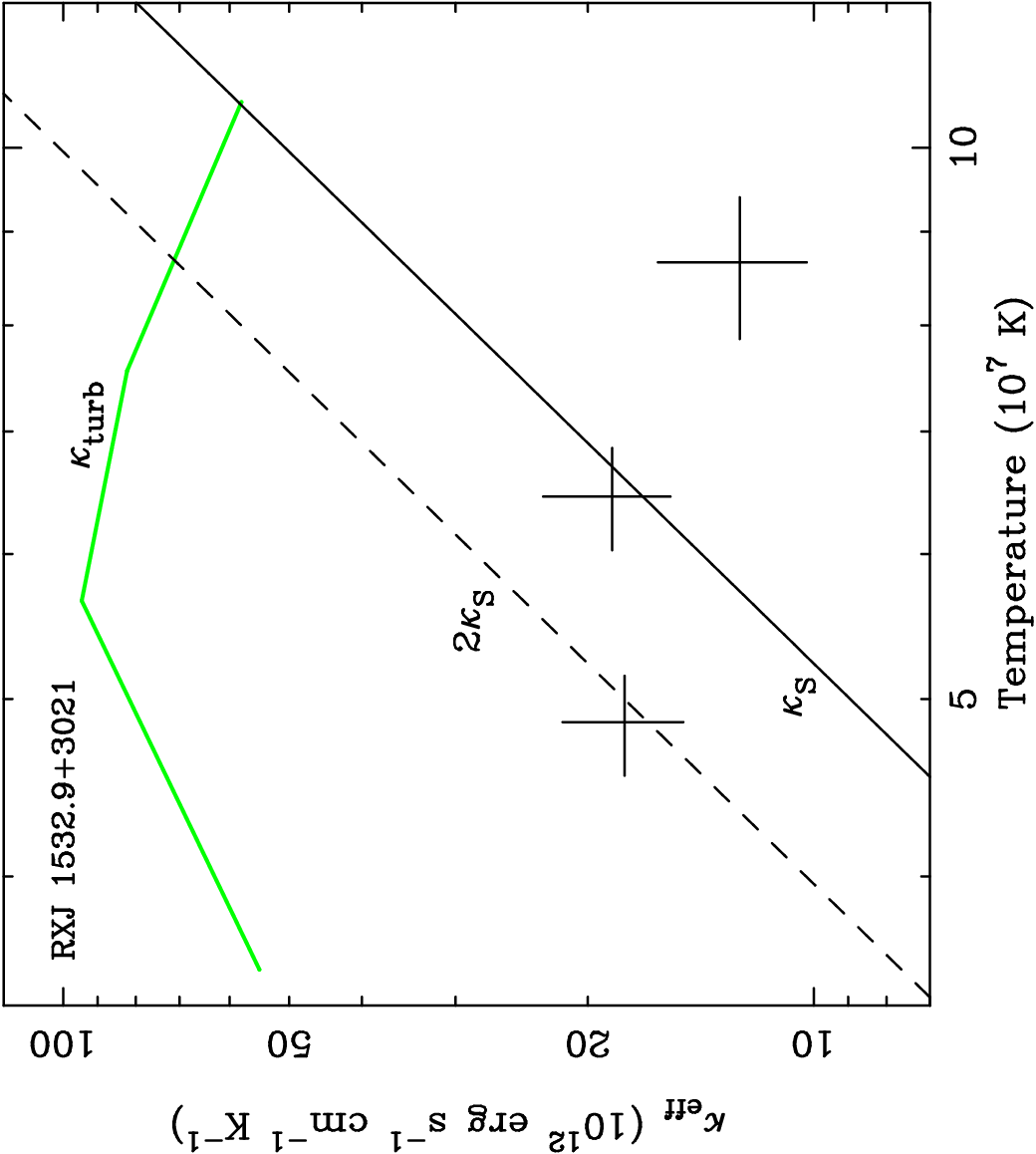}
\vspace{1.0mm}
\includegraphics[angle=-90,width=0.95\columnwidth]{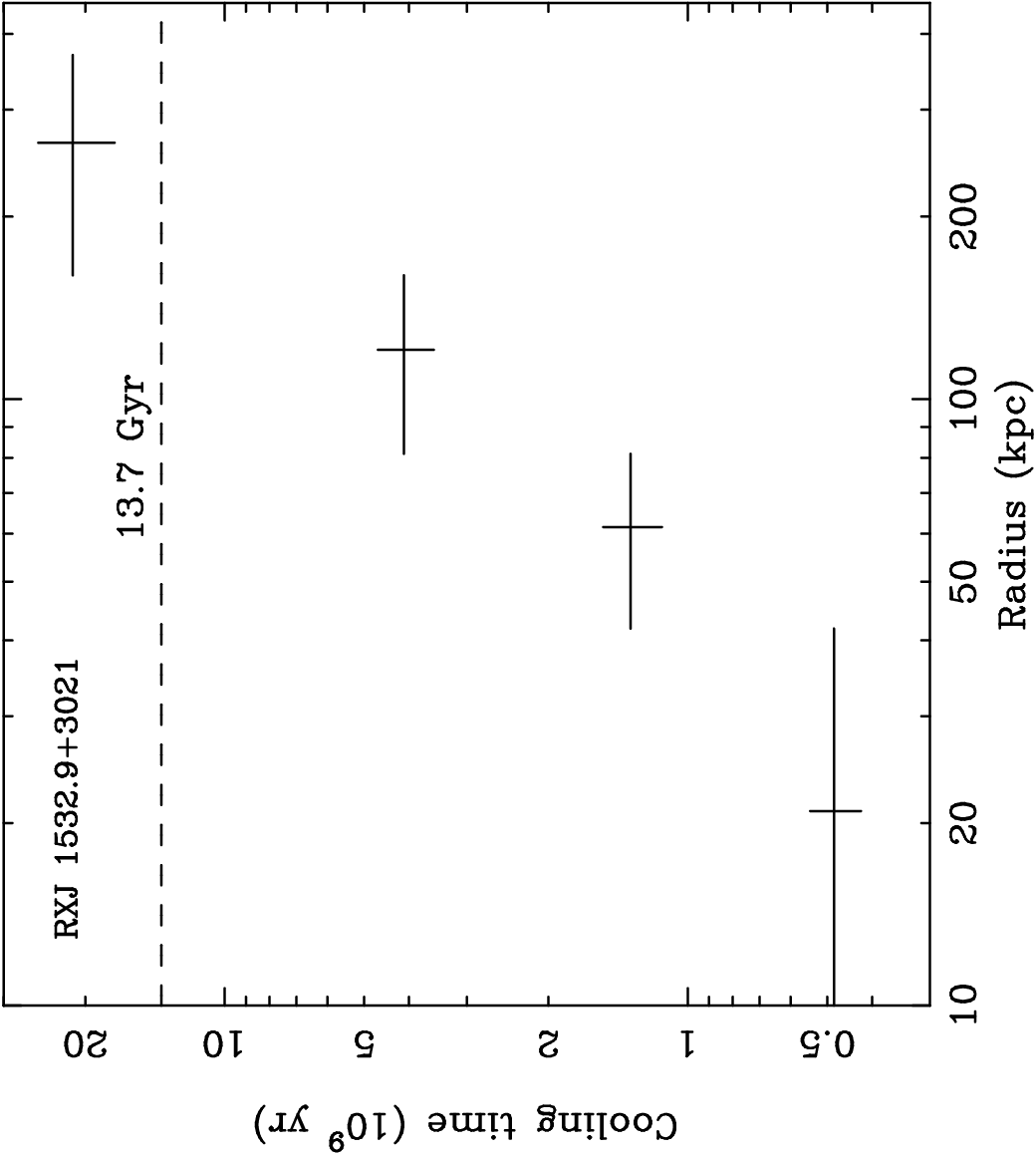}
\end{figure*}

\begin{figure*}
\includegraphics[angle=-90,width=0.95\columnwidth]{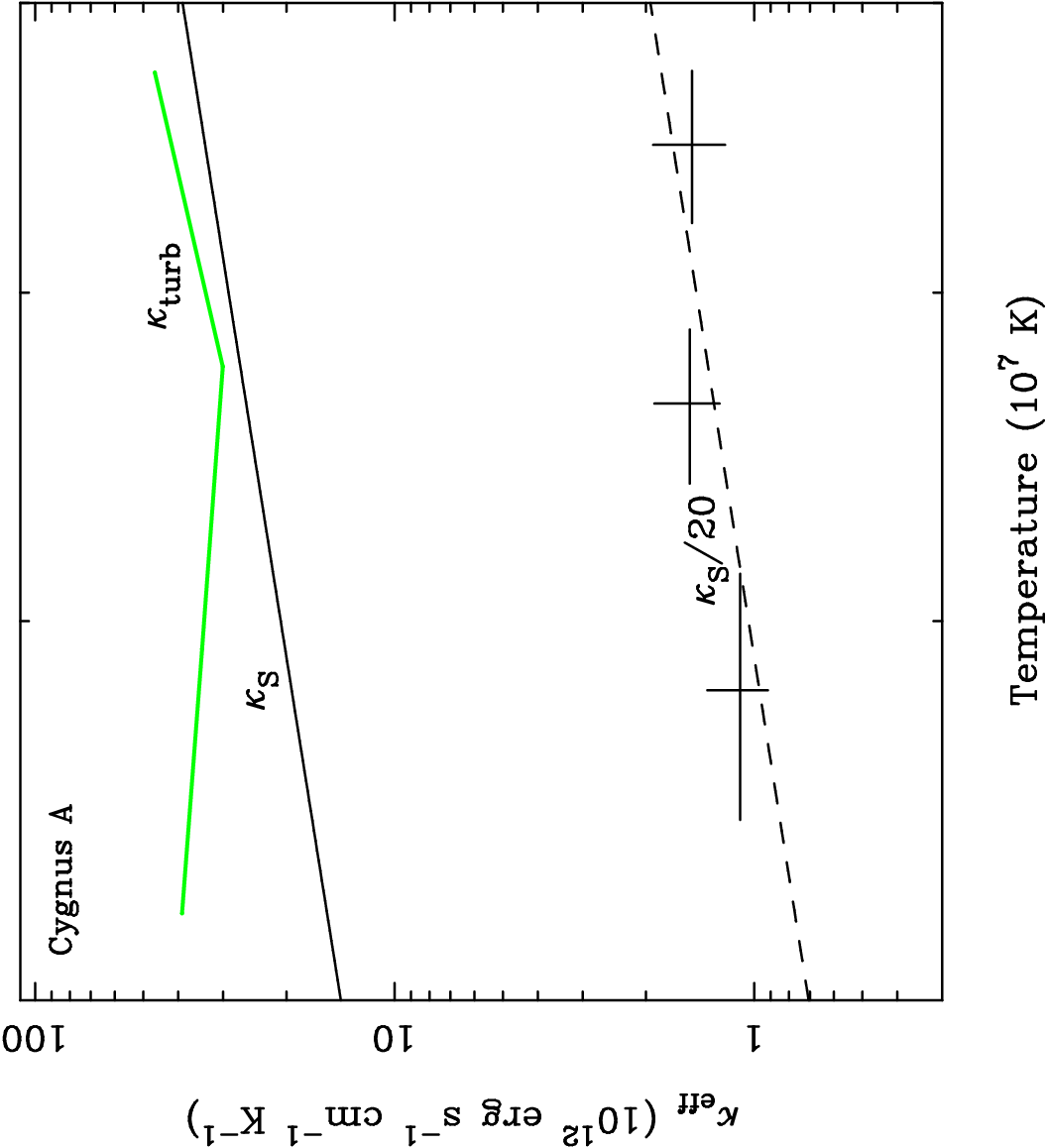}
\vspace{1.0mm}
\includegraphics[angle=-90,width=0.95\columnwidth]{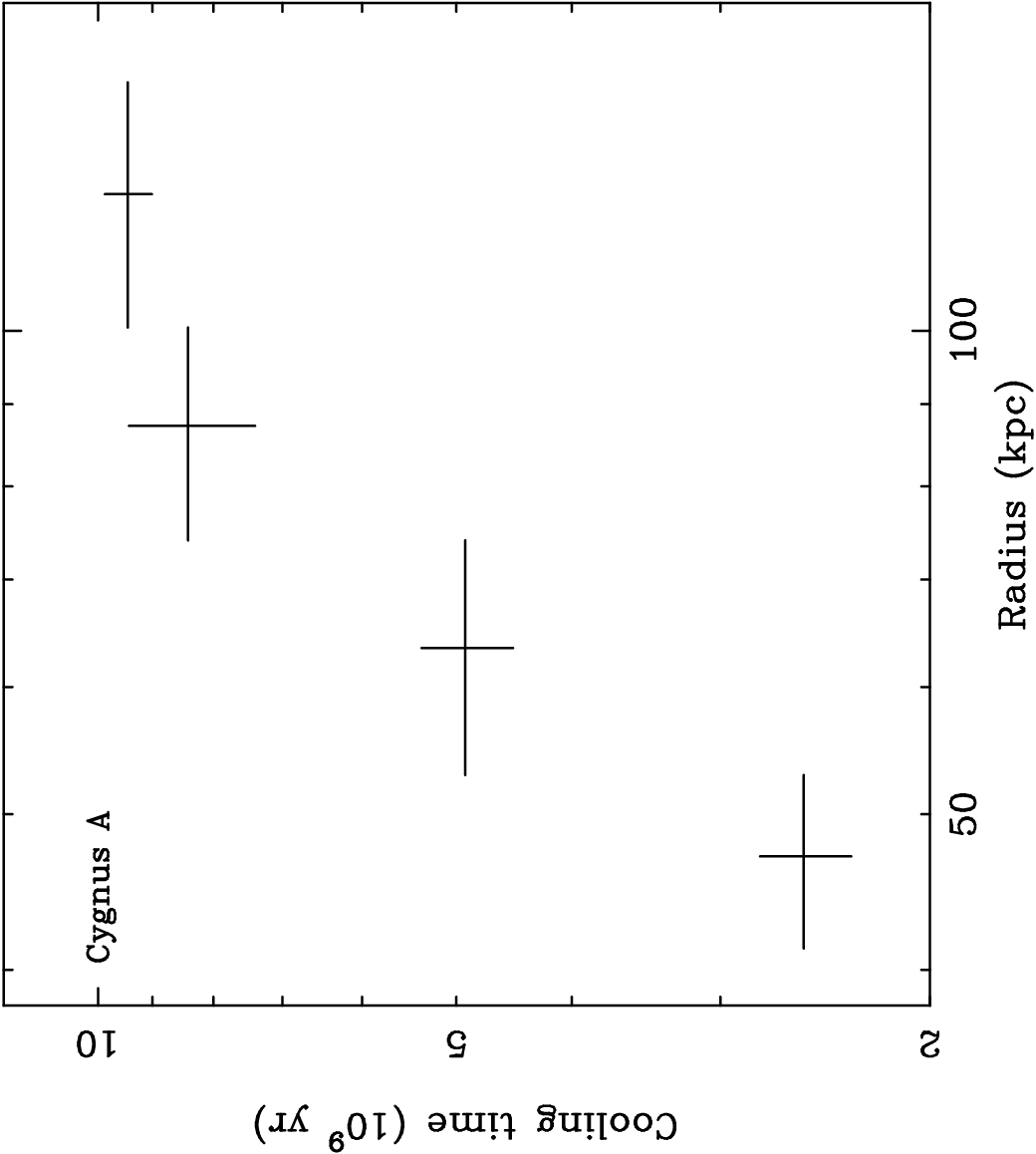}
\vspace{1.0mm}
\includegraphics[angle=-90,width=0.95\columnwidth]{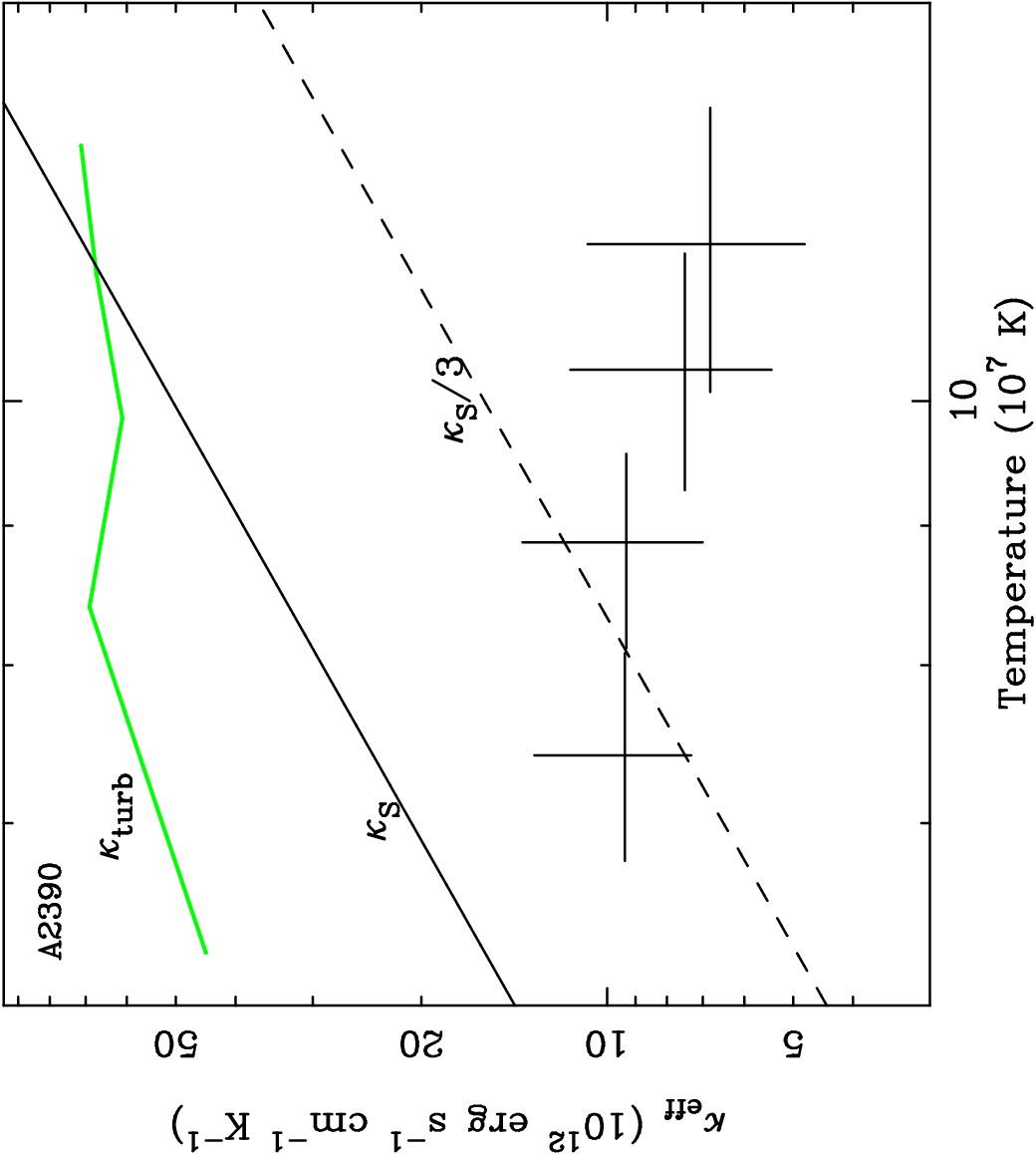}
\vspace{1.0mm}
\includegraphics[angle=-90,width=0.95\columnwidth]{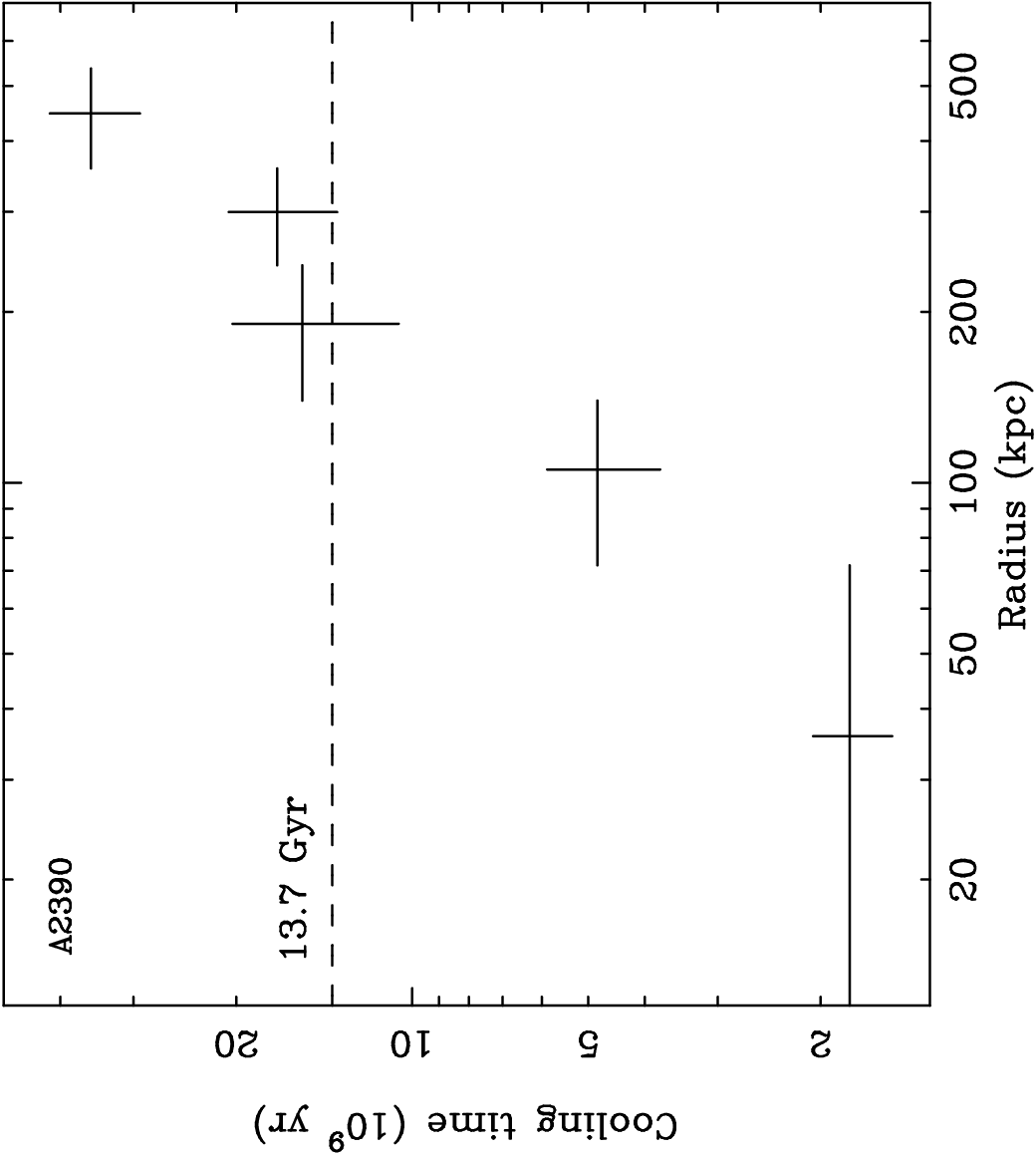}
\vspace{1.0mm}
\includegraphics[angle=-90,width=0.95\columnwidth]{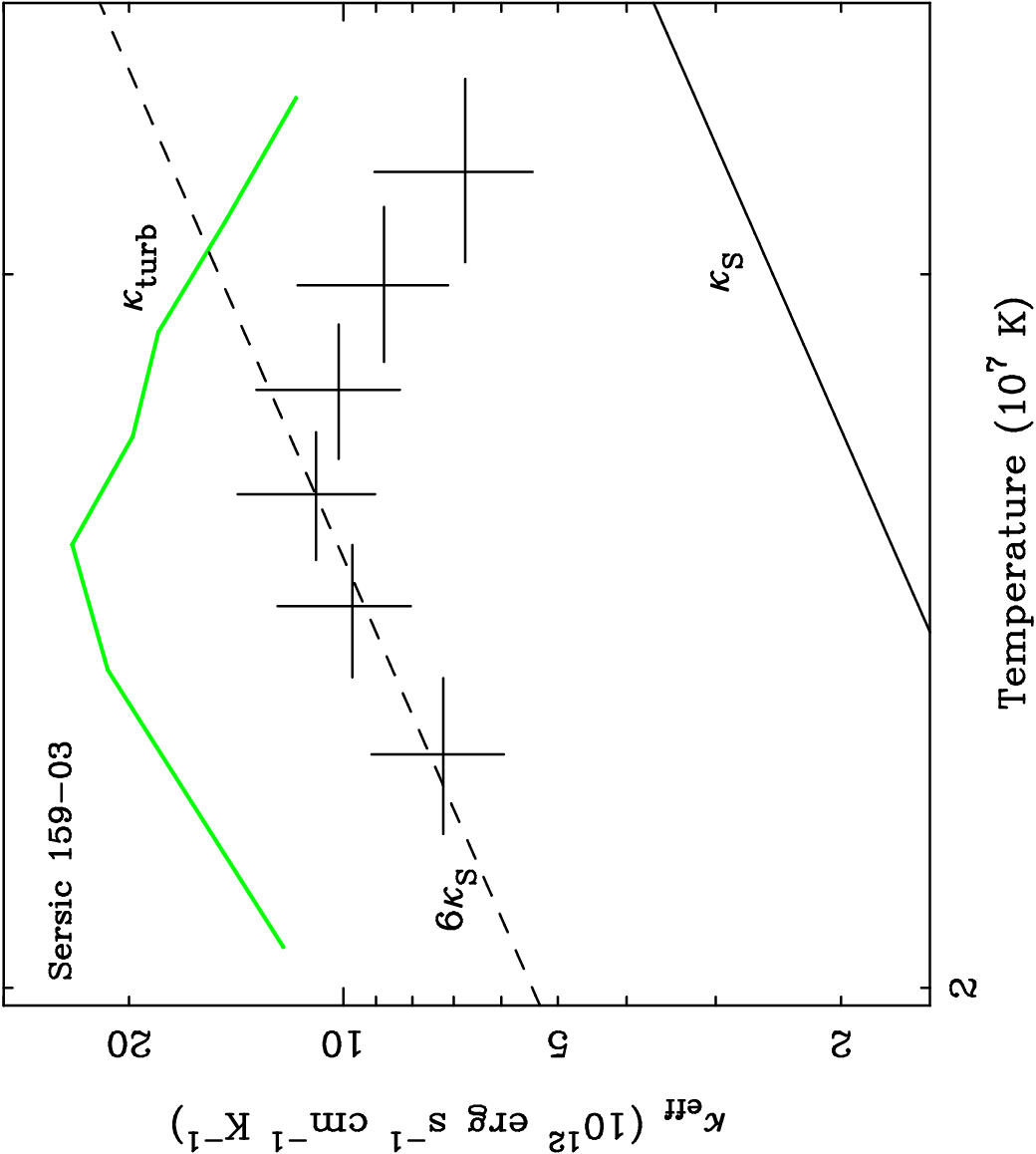}
\vspace{1.0mm}
\includegraphics[angle=-90,width=0.95\columnwidth]{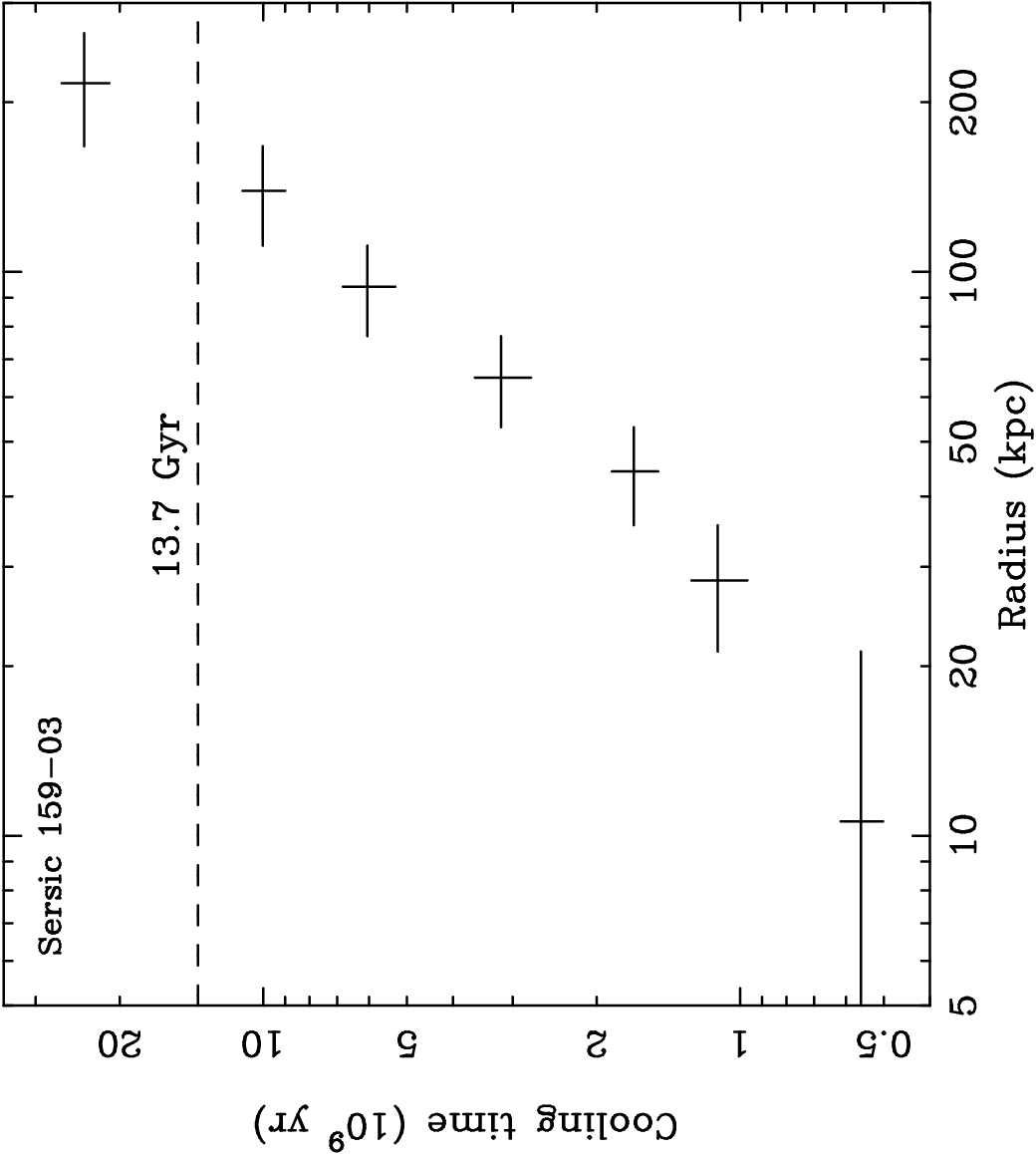}
\end{figure*}

\end{document}